\documentclass[aps,prd,twocolumn,showpacs,groupedaddress,preprintnumbers,superscriptaddress,amsmath,amssymb,floatfix,longbibliography]{revtex4-1} 

\def\nn{\nonumber}

\newcommand{\vect}[1]{\boldsymbol{#1}}

\newcommand{\DM}[0]{ \rm DM }

\newcommand{\mMED}[0]{ m_\phi }

\newcommand{\TCS}[0]{ \sigma }

\newcommand{\LSDM}[0]{ \chi_1 }

\newcommand{\DBWformula}[0]{ D_{\rm  DM} }

\newcommand{\Avrg}[1]{\left\langle #1 \right\rangle}
\newcommand{\TAvrgCS}[0]{\left\langle \TCS v \right\rangle}

\usepackage{color}	
\usepackage{graphicx}
\usepackage{bm}
\usepackage{slashed}

\usepackage[colorlinks=true, filecolor=blue, citecolor=blue,urlcolor=blue]{hyperref}

\usepackage{enumitem}

\usepackage{tikz-feynhand}

\date{\today}

\begin{document}

\abovedisplayskip = 4pt
\belowdisplayskip = 4pt
\abovedisplayshortskip = 4pt
\belowdisplayshortskip= 4pt

\title{Examining scalar portal inelastic dark matter with lepton fixed target experiments}

\author{I.~V.~Voronchikhin}
\email[\textbf{e-mail}: ]{i.v.voronchikhin@gmail.com}
\affiliation{ Tomsk Polytechnic University, 634050 Tomsk, Russia}
\affiliation{Institute for Nuclear Research, 117312 Moscow, Russia}

\author{D.~V.~Kirpichnikov}
\email[\textbf{e-mail}: ]{dmbrick@gmail.com}
\affiliation{Institute for Nuclear Research, 117312 Moscow, Russia}

\begin{abstract}
Inelastic dark matter scenarios have attracted considerable 
attention in contemporary particle physics. In this study, we 
investigate the phenomenology of sub-GeV inelastic dark matter 
interacting via a lepton-specific scalar portal. By solving the Boltzmann 
equations, we obtain thermal target curves for several inelastic DM 
mass splittings in the sub-GeV mediator-mass range. We study the 
discovery potential of lepton fixed-target experiments, particularly 
NA64e, LDMX, and NA64$\mu$, via their missing-energy signatures. Our 
analysis focuses on the $\phi$-strahlung process, $l N \to l N \phi$, 
followed by the invisible decay of the scalar mediator into Majorana 
dark matter particles $\phi \to \chi_1 \chi_2$.  We use this channel to probe the 
mediator coupling to charged leptons of the Standard Model. For 
phenomenologically viable parameters of the inelastic dark matter scenario, we derive 
projected sensitivities for NA64e, LDMX, and NA64$\mu$, assuming 
that the boosted state $\chi_2$ decays visibly via $\chi_2 \to \chi_1 
e^+ e^-$ outside the detector acceptance. Our results demonstrate 
the complementary roles of electron- and muon-beam experiments in 
exploring the sub-GeV inelastic dark matter sector interacting via a 
scalar portal.
\end{abstract}

\maketitle

\section{Introduction}

In recent decades, astrophysical observations have provided strong evidence for the 
existence of dark matter (DM)~\cite{Bergstrom:2012fi,Bertone:2016nfn}. Empirical support 
for DM arises from a variety of astrophysical phenomena, including galactic rotation curves, 
anisotropies in the cosmic microwave background, and gravitational lensing 
observations~\cite{Cirelli:2024ssz,Bertone:2004pz,Gelmini:2015zpa}. Current cosmological 
measurements indicate that DM constitutes approximately 85\% of the total matter 
content of the Universe, a component not explained within the framework of the 
Standard Model (SM) of particle physics~\cite{Planck:2015fie,Planck:2018vyg}. 

One can assume that thermalization between DM and SM sectors occurs 
via different portal interactions. Theoretical frameworks typically consider bosonic mediators 
with different spins: spin-0 (e.~g., light hidden Higgs bosons)~\cite{Ponten:2024grp,Kolay:2025jip,Kolay:2024wns,McDonald:1993ex,Burgess:2000yq,Wells:2008xg,Schabinger:2005ei,Bickendorf:2022buy,Boos:2022gtt,Sieber:2023nkq,Guo:2025qes,Voronchikhin:2023qig}, spin-1 (e.~g., sub-GeV dark photons)~\cite{Catena:2023use,Holdom:1985ag,Izaguirre:2015yja,Essig:2010xa,Kahn:2014sra,Batell:2014mga,Izaguirre:2013uxa,Kachanovich:2021eqa,Lyubovitskij:2022hna,Gorbunov:2022dgw,Claude:2022rho,Wang:2023wrx}, and spin-2 (e.~g., massive dark  gravitons)~\cite{Lee:2013bua,Kang:2020huh,Bernal:2018qlk,Folgado:2019gie,Kang:2020yul,Dutra:2019xet,Clery:2022wib,Gill:2023kyz,Wang:2019jtk,deGiorgi:2021xvm,deGiorgi:2022yha,Jodlowski:2023yne,Voronchikhin:2024ygo}.

The search for DM particles at accelerator-based facilities is one of the major efforts of contemporary high-energy physics, aimed at probing physics beyond the SM~\cite{Berlin:2018jbm,Dienes:2023uve,Mongillo:2023hbs,DallaValleGarcia:2025giv,Guo:2025qes,Jodlowski:2019ycu,Izaguirre:2015zva}. The portal-mediated 
light dark matter scenarios predict distinctive experimental signatures characterized by events with missing energy. These signatures 
originate from bremsstrahlung-like processes in which a charged lepton interacts with the nucleus of a target, producing a mediator that subsequently decays into a pair of DM particles. Fixed-target experiments 
are particularly effective for probing sub-GeV DM due to their combination of high-energy lepton beams and substantial beam 
intensities. Current experimental efforts include operational at CERN SPS
fixed-target facilities such as 
NA64e~\cite{Gninenko:2016kpg,NA64:2016oww,NA64:2017vtt,Gninenko:2018ter,Banerjee:2019pds,Dusaev:2020gxi,Andreev:2021fzd,NA64:2022yly,NA64:2022rme,Arefyeva:2022eba,Zhevlakov:2022vio,NA64:2021acr,NA64:2021xzo}, 
NA64$\mu$~\cite{NA64:2024klw,Sieber:2021fue,Kirpichnikov:2021jev}. In our analysis, we also discuss  planned  LDMX electron fixed-target experiment at SLAC~\cite{Berlin:2018bsc,Schuster:2021mlr,Akesson:2022vza,LDMX:2025bog}. 

One of extended dark sector models is the concept of inelastic dark matter~\cite{Tucker-Smith:2001myb,Tucker-Smith:2004mxa,Chang:2008gd}. The inelastic nature of the scenario implies a  mass splitting between DM states, which gives rise to novel experimental signatures. 
Originally proposed to explain the anomalous signals observed by the DAMA 
collaboration~\cite{Bernabei:2013xsa}, inelastic DM scenarios have 
emerged as a theoretically well-motivated framework for sub-GeV thermal DM
interacting through a vector 
mediator~\cite{Jordan:2018gcd,Berlin:2018pwi,Batell:2021ooj,Mongillo:2023hbs,Izaguirre:2015zva}. The key phenomenological 
feature of these models involves off-diagonal couplings between the DM ground state, 
$m_{\chi_1}$, and an excited state, $m_{\chi_2}$, with $m_{\chi_2} \gtrsim m_{\chi_1}$. This structure kinematically suppresses the DM-nucleon scattering cross sections for current direct-detection experiments. 

This work investigates the projected sensitivities of the NA64e, LDMX, and NA64$\mu$ 
experiments, where we use the simplified scenario of inelastic fermionic dark matter model with a 
scalar mediator predominantly coupled to the charged leptons  of the SM.
We focus on a scalar mediator with the mass from~$\mathcal{O}(1)$~MeV to~$\mathcal{O}(1)$~GeV and the mass splitting of inelastic dark matter,~$\Delta_2 \equiv (m_{\chi_2} -m_{\chi_1})/m_{\chi_1}$, in the range from $0.1$ to $0.5$. 

This paper is organized as follows. In  Sec.~\ref{sec:BenchModels} we discuss a simplified benchmark model for inelastic DM mediated by scalar portal. In Sec.~\ref{sec:RelAbCalcEq} we 
estimate the inelastic DM relic abundance by solving the Boltzmann equations. In Sec.~\ref{sec:ExperimentSetup}   we discuss the missing-energy signatures at NA64e, LDMX, and NA64$\mu$ experiments, arising from the process $l N \to l N \phi(\to \chi_1 \chi_2)$ followed by $\chi_2 \to \chi_1 e^+ e^-$ decays. In Sec.~\ref{sec:ExpectedReach} we discuss the thermal target curves for inelastic DM and regarding sensitivity at NA64e, LDMX, and NA64$\mu$ fixed-target facilities. Finally, conclusions are drawn in Sec.~\ref{sec:Conclusion}.  
In Appendices~\ref{sec:RelicAbundance}, \ref{SectionCScalculation}, \ref{sectionAveraging}, and~\ref{sec:DD} we provide some useful formulas and discuss direct-detection constraints.

\section{Simplified Benchmark scenarios\label{sec:BenchModels}}

In this section, we  discuss simplified benchmark scenarios for inelastic dark matter
with a focus on the lepton-specific scalar mediator interacting with Majorana dark matter. 

The simplified Lagrangian describing a lepton-philic spin-0 mediator reads  as follows~\cite{Chen:2018vkr,Berlin:2018bsc}
\begin{equation}
    \mathcal{L}_{\rm eff}^\phi  \supset  \frac{1}{2}(\partial_\mu \phi)^2 -\frac{1}{2} m_\phi^2 \phi^2 - \sum_{\l = e, \mu,\tau}c^{\phi}_{ll}  \overline{l}  l \phi,
    \label{GenMeDCoupl}
\end{equation}
the last term in Eq.~(\ref{GenMeDCoupl}) can be understood as originating from the effective gauge-invariant dimension-5 operators~\cite{Batell:2017kty}
\begin{equation} \label{eq:EFT}
\mathcal{L} \supset 
 \sum_{i =e,\mu,\tau}\frac{c_i}{\Lambda} \phi \overline{L}_iH E_i.
\end{equation}
In this framework, $\Lambda$ represents the characteristic scale of new physics, while $c_i$ 
denotes the Wilson coefficient corresponding to the flavor $i$. We adopt the assumption that these 
couplings remain diagonal in the mass basis.
Regarding the relative magnitudes of the Wilson coefficients $c_i$, a theoretically motivated 
ansatz suggests proportionality to the respective Yukawa couplings $y_i$. Consequently, following 
electroweak symmetry breaking, the effective couplings $c^\phi_{ll}$ scale with the 
corresponding lepton masses. This leads us to establish the flavor-dependent ratio:
\begin{equation}
\label{CeeCmumu}
c^{\phi}_{ee}:c^{\phi}_{\mu\mu}:c^{\phi}_{\tau\tau} = m_e : m_\mu : m_\tau,
\end{equation}
which we implement in Eq.~(\ref{GenMeDCoupl}) and maintain consistently throughout our analysis.

The dark matter sector consists of two Majorana fermion states $\chi_1$ and $\chi_2$, described by the Lagrangian:
\begin{equation}
\mathcal{L}^{\rm DM}_{\rm kin. term.} \supset  \sum_{i = 1,2} \left[ \frac{1}{2} \overline \chi_i \, i \gamma^\mu  \,\partial_\mu \chi_i
-\frac{1}{2} m_{\chi_i}\, \overline \chi_i \chi_i \right]\, ,
\label{KinTerMajoranaDM}
\end{equation}
where $m_{\chi_i}$ denotes the physical fermion masses, and we take $m_{\chi_1} \lesssim  m_{\chi_2}$ such that $\chi_1$ is the lightest stable DM candidate. 

We consider two  effective benchmark Lagrangians involving a spin-0 mediator that 
couples to a pair of Majorana fermions via either a CP-odd or a CP-even coupling, 
such that the typical interactions are given by~\cite{Dreiner:2008tw}:
\begin{align}
\mbox{Bench.~1:} \,\,       \mathcal{L}_{\rm  eff.}^{\rm (+)DM} 
& \supset 
 \text{Re}( \lambda^{\phi}_{\chi_1 \chi_2})  \overline{\chi}_1 \chi_2 \phi,   \label{eq:EffLagrangianScalarMEDMajoranaiCDM}
\\
  & \text{Im}( \lambda^{\phi}_{\chi_1 \chi_2}) = 0,  \nonumber
  \\
\mbox{Bench.~2:} \,\,    \mathcal{L}_{\rm eff}^{\rm (-) DM}
& \supset \label{eq:EffLagrangianScalarMEDMajoranaG5iCDM}
  - i \text{Im}     (\lambda^{\phi}_{\chi_1 \chi_2} )  \overline{\chi}_1  \gamma_5 \chi_2 \phi, 
  \\
  & \text{Re}( \lambda^{\phi}_{\chi_1 \chi_2}) = 0. \nonumber 
\end{align}
In our simplified scenario, we adopt the conservative assumption that diagonal couplings 
between the mediator and  Majorana DM are  suppressed, such that only off-diagonal terms contribute to  the effective interaction~\cite{DallaValleGarcia:2024zva}.

\section{Relic abundance of inelastic dark matter\label{sec:RelAbCalcEq}}

For the freeze-out mechanism considered in the present paper, we employ the standard Boltzmann equation technique  to compute thermal target curves~\cite{Berlin:2018bsc,Izaguirre:2017bqb,Krnjaic:2015mbs,Foguel:2024lca,Krnjaic:2025noj,Chen:2018vkr}, 
assuming a kinetic and chemical equilibrium between DM and the SM thermal bath 
at the early stages of the Universe expansion. 
The subsequent departure of 
DM from thermal equilibrium through the depletion 
processes  provides a relic abundance that can account 
for the observed DM density in the Universe.

Note that DM thermalization through the so-called secluded
$t$-channel reaction $\chi_{i} \chi_{i} \to \phi \phi$ (where $i=1,2$)
can provide a viable freeze-out mechanism~\cite{Krnjaic:2015mbs} for the DM relic abundance  as long as 
$m_{\chi_{i}}  \gtrsim  m_\phi$, for sufficiently large DM coupling to the mediator. 
However, a drawback of this framework is that the 
predicted relic density is largely insensitive to the coupling 
strength between the $\phi$ field and SM states, $c_{ll}^\phi$. This independence 
presents a challenge for experimental searches, such as those 
conducted at direct-detection facilities or accelerator-based experiments. 

In contrast to the well-established freeze-out scenario, the freeze-in 
mechanism~\cite{Belanger:2018ccd} provides an alternative framework to generating the 
observed relic abundance of dark matter. The assumption of 
freeze-in is that DM was never in thermal equilibrium with the SM plasma 
 in the early Universe. Instead, the initial abundance of 
DM is assumed to be negligible. The observed DM density is then 
accumulated gradually through the slow, continuous production of DM 
from interactions with particles in the hot  thermal bath~\cite{Krnjaic:2025zjl,Heeba:2023bik}. 

For  $2 m_l \gtrsim m_{\chi_1} + m_{\chi_2}$ and  
$m_\phi \gtrsim m_{\chi_1} + m_{\chi_2}$, the channels contributing to
freeze-in are the annihilation of SM fermions, 
$l^+ l^- \to \chi_1 \chi_2$,  or $\phi$ decays, $\phi \to \chi_1 \chi_2$.
The latter  channel dominates for sufficiently large DM coupling to the 
mediator~\cite{Belanger:2018ccd}.  
The drawback of the aforementioned freeze-in scenario via the fast decay $\phi \to \chi_1 \chi_2$ is analogous to that of freeze-out via the secluded channel $ \chi_{i} \chi_{i} \to \phi \phi$. Therefore, these scenarios are beyond the scope of the present paper.

\begin{figure}[tb!]
	\center{\includegraphics[scale=0.4]{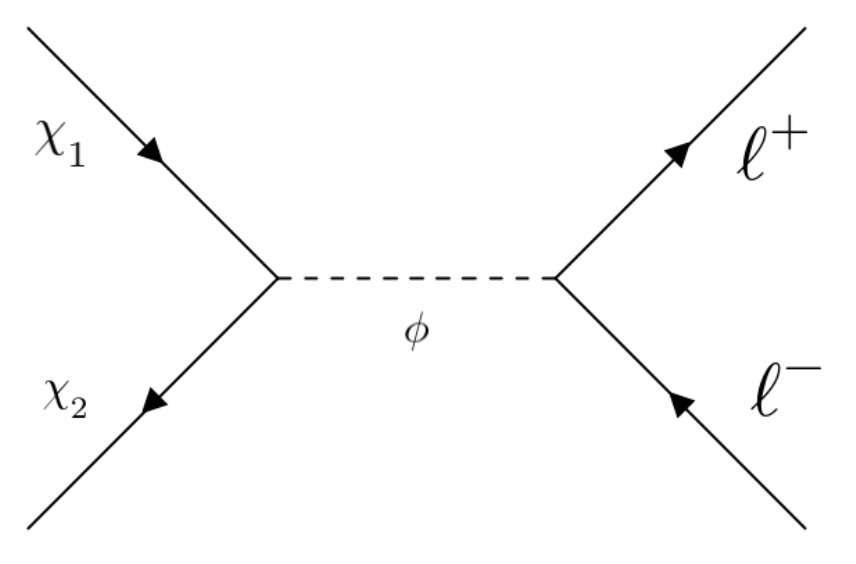}}
	\caption{ Feynman diagrams giving rise to $\chi_1 \chi_2 \to \phi \to l^+l^- $ annihilation in the early Universe. }
	\label{fig:chi1chi2Toll}
\end{figure}

Instead,  we focus on the freeze-out mechanism via the co-annihilation channel
$ \chi_1 \chi_2 \to l^+ l^-$ (see Fig.~\ref{fig:chi1chi2Toll}). Specifically, for  the parameter space of interest  $m_{\phi} \gtrsim m_{\chi_1} + m_{\chi_2}$, the on-shell inverse decay rate of the process  $\chi_1 \chi_2 \to \phi$ is  sharply suppressed relative to 
the $s$-channel process $\chi_1 \chi_2 \to l^+ l^-$~\cite{DAgnolo:2015ujb}, due to the thermal 
tail of DM velocity distribution. Thus, we do not include the inverse decay into the Boltzmann equation. 
As a result, the dark-matter co-annihilation  $\chi_1 \chi_2 \to l^+ l^-$ becomes  the dominant contribution to the DM depletion during the 
freeze-out mechanism~\cite{Fitzpatrick:2020vba}. 

The current value of the cold DM relic abundance obtained from 
the Planck 2018 combined analysis is~\cite{Planck:2018vyg,Husdal:2016haj}:
\[
    \Omega_c h^2 = 0.1200 \pm 0.0012.
\] 
As a result,  the  relic density is estimated to 
be~\cite{Kolb:1990vq}:
\begin{equation}\label{eq:RelDensDM1}
    \Omega_{c} h^2
 \propto  
    \left( \, \, \int\limits_{x_f}^{\infty} \frac{\Avrg{\sigma_{\rm eff} v}}{x^2} dx \, \right)^{-1}, 
\end{equation}
where $\Avrg{\sigma_{\rm eff} v}$ is a thermally averaged co-anihillation cross section and $x=m_{\chi_1}/T$ is a typical ratio of DM mass and temperature. 
We provide the details for the calculation of~(\ref{eq:RelDensDM1}) in Appendices~\ref{sec:RelicAbundance},~\ref{SectionCScalculation}, 
and~\ref{sectionAveraging}.

\section{Missing energy signatures
\label{sec:ExperimentSetup}}

\begin{figure}[tb!]
	\center{\includegraphics[scale=0.5]{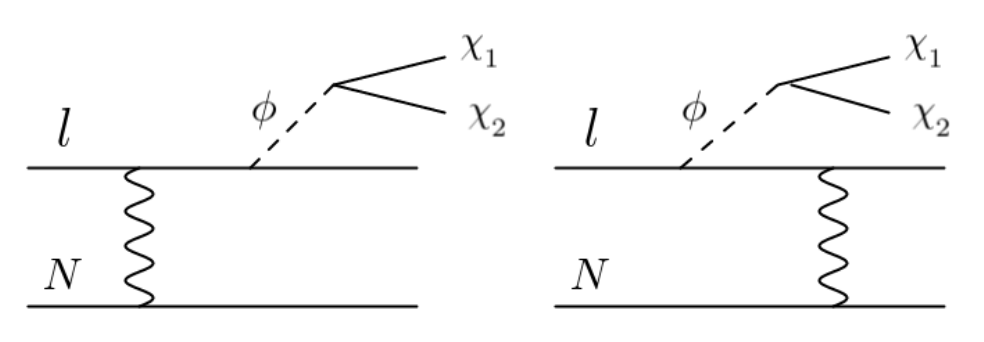}}
	\caption{Feynman diagrams for radiative scalar mediator production $l N \to l N \phi$ followed by invisible decay $\phi \to \chi_1 \chi_2$.}
	\label{fig:lNtolNchi1chi2}
\end{figure}

This section presents the experimental configurations of fixed-target facilities capable of probing the invisible decay channel $\phi\to \chi_1 \chi_2$ (see Fig.~\ref{fig:lNtolNchi1chi2}). We discuss two complementary experiments currently operating at the CERN SPS: 
NA64e~\cite{Gninenko:2016kpg,NA64:2016oww,NA64:2017vtt,Gninenko:2018ter,Banerjee:2019pds,Dusaev:2020gxi,Andreev:2021fzd,NA64:2022yly,NA64:2022rme,Arefyeva:2022eba,Zhevlakov:2022vio,NA64:2021acr,NA64:2021xzo} utilizes an electron beam to investigate the process $e N \to e N \phi (\to \chi_1 \chi_2)$;  
NA64$\mu$  
\cite{Gninenko:2014pea,Gninenko:2018tlp,Kirpichnikov:2021jev,Sieber:2021fue,NA64:2024nwj,NA64:2024klw} employs a muon beam to study 
$ \mu N \to \mu N \phi (\to \chi_1 \chi_2)$. For completeness, we also discuss the planned electron fixed-target experiment LDMX at SLAC~\cite{Berlin:2018bsc,Schuster:2021mlr,LDMX:2025bog}.
The differential cross section $d\sigma_{2\to 3}/dx $ of the relevant 
process is calculated in Ref.~\cite{Voronchikhin:2024vfu}, where  
$x=E_\phi/E_l$ is a missing-energy fraction,  $E_l$ is a lepton beam 
energy, $E_\phi$ is the energy of radiated scalar mediators,  and $l=(e,\mu)$ represents the incident lepton (electron or muon). 

In an electron fixed-target experiment, the initial energy  of the incident beam, $E_e$, decreases as it traverses the target material.  
The distribution of beam energies in the electromagnetic shower $E_s$ after propagation through a target thickness $t$  is~\cite{Tsai:1966js}:
\begin{equation}\label{eq:IeTsai}
	I_e(E_e, E_{s},t) 
= 
	\frac{1}{E_e} 
	\frac{ \left[ \ln(E_e/E_{s}) \right]^{\frac{4}{3}t-1} }
	     { \Gamma(\frac{4}{3}t)},
\end{equation}
where $t$ is expressed in units of radiation length  and $E_e$ denotes the initial beam energy at $t=0$.

We estimate the number of scalar mediators produced via bremsstrahlung at fixed-target facilities using $x \equiv E_\phi/E_e$ and $z \equiv E_\phi/E_s$ as follows~\cite{Tsai:1986tx,Andreas:2012mt}:
\begin{align}\label{eq:NradMEDwithIe}
	N^{\rm brem. }_\phi
= & 
	\;\mbox{EOT} \times 
	\frac{\rho N_{\rm A}  X_0 }{A} \times 
\\  & \nn 
\int\limits^{x_{\rm max}}_{x_{\rm cut}} dx	\int_{x}^{1} 
	\frac{E_e}{z}
	\frac{d\sigma}{dz} 
	dz
	\int_0^T  
	I_e(E_e, xE_e/z, t)
	dt,
\end{align}
where  $T$ is the thickness of the target in units of the radiation length $X_0$ (cm), $x_{\rm cut}$ and $x_{\rm max}$ are the minimum and maximum fractions of missing energy, respectively,
$\rho$ is the density of the target material, $N_{A}$ is Avogadro's number, $A$ is the atomic mass number, $Z$ is the atomic number, and $\mbox{EOT}$ denotes electrons accumulated on target. 
Explicit expressions for the differential cross sections and the corresponding discussion are given in Refs.~\cite{Liu:2016mqv,Liu:2017htz,Voronchikhin:2024vfu}.

In the approximation that mediator radiation occurs effectively in a thin layer of the target, i.e. for $T\ll 1$, the electron energy distribution $I_e(E_e,E_s,t)$ is~\cite{Bjorken:2009mm}:
\begin{equation}
	\lim\limits_{t \to 0} I_e(E_e, E_s, t) 
\to 
	\frac{x}{E_e}\delta(z - x).
\end{equation}
Thus, neglecting electron energy loss, the number of mediators produced via bremsstrahlung is:
 \begin{equation}\label{eq:NradGrav}
N^{\rm brem. }_\phi \simeq \mbox{EOT}\times \frac{\rho N_A}{A} L_T \int\limits^{x_{\rm max}}_{x_{\rm cut}}
dx \frac{d \sigma_{2\to3}(E_e)}{dx},
\end{equation}
where $L_T = T X_0$ the effective thickness in radiation lengths.
In our calculations, we employ Eq.~\eqref{eq:NradMEDwithIe} for NA64e since its target sufficiently thick, $T\gg 1$ and the Eq.~\eqref{eq:NradGrav}  for  LDMX due to thin target employed in the design of its experimental setup, $T \ll 1$. 
For NA64$\mu$ we also use thin target approach (\ref{eq:NradGrav}) implying the label replacement  $E_e \to E_\mu$ and $ \mbox{EOT} \to \mbox{MOT}$, where $\mbox{MOT}$ denotes the typical number of muons accumulated on target. Specifically,
 \begin{equation}\label{eq:MuonNradGrav}
N^{\rm brem. }_\phi \simeq \mbox{MOT}\times \frac{\rho N_A}{A} L_T \int\limits^{x_{\rm max}}_{x_{\rm cut}}
dx \frac{d \sigma_{2\to3}(E_\mu)}{dx}.
\end{equation}
In Tab.~\ref{tab:BenchLeptonFTexp} the parameters of the considered 
experiments are shown.

\subsection{NA64$e$
\label{NA64eSetupSect}}

The scalar mediator $\phi$  can be produced through the scattering of ultra-relativistic electrons with energies 
$E_e\simeq 100\, \mbox{GeV}$ on the nuclei of a target via the process $e N \to e N \phi$, followed by the prompt 
decay $\phi \to \chi_1 \chi_2$. 
A fraction of the initial electron energy, $E_{\rm miss}=x E_e$, may be carried away 
by the DM pair $\chi_1 \chi_2$, which traverses the NA64e detector without significant energy deposition in calorimeter modules. 
The  remaining energy fraction, $E_e^{\rm rec} \simeq (1-x) E_e$, can be measured in the electromagnetic calorimeter (ECAL) of the NA64e experiment through the detection of recoil electrons. Consequently, the production of the hidden scalar $\phi$ would manifest as an anomalous accumulation of events featuring a single electromagnetic 
shower  with energy $E^{\rm rec}_e$ in the signal box~\cite{NA64:2016oww}. In particular, candidate events are selected by requiring 
$E^{\rm rec}_e \lesssim 0.5 E_e \simeq 50\, \mbox{GeV}$, corresponding to an energy fraction threshold of 
$x_{\rm cut}\simeq 0.5$ in Eq.~(\ref{eq:NradGrav})  for the NA64e experiment.

Furthermore, it should be noted that the ECAL of the NA64e experiment serves 
as the active target for the incoming electron beam. The ECAL is composed of $6\times 6$ Shashlyk-type modules, 
each   constructed from alternating layers of plastic scintillator (Sc) and lead (Pb) plates. The typical 
parameters of  NA64e are summarized in Tab.~\ref{tab:BenchLeptonFTexp}.  
The NA64e aims   to collect approximately $1.5\times 10^{12}$ EOT at 
the $H4$ beam line after Long Shutdown~3.

To account for the up-stream acceptance effects of the NA64e detector, we rely on the analysis in Ref.~\cite{NA64:2023wbi}. 
The ECAL is positioned at a 20 mrad angle to the beam line, and the beam
electrons are deflected by the up-stream magnet to hit it.
This  beam line deflection was implemented to improve the high-energy electrons selection 
and suppress background  from the possible admixture of
low-energy electrons, as was suggested in Ref.~\cite{Gninenko:2013rka}. Specifically, the NA64e employs a tagging system
utilizing the synchrotron radiation from high-energy electrons~\cite{Dworkin:1986tk} in a dipole 
magnet installed up-stream of the detector, as illustrated schematically in
Fig. 1 of Ref.~\cite{NA64:2023wbi}.  Thus, we assume in our analysis that an electron hitting the NA64e active target is properly tagged up-stream and passes all event-selection criteria~\cite{NA64:2023wbi}.

We neglect the effect of signal acceptance related to the 
electron recoil angle in the active ECAL target, as the typical 
recoil angle of the outgoing electron in the reaction $e N \to e 
N \phi$ is estimated to be sufficiently small $m_\phi/E_e \simeq \mathcal{O}(10^{-3})$ (see 
e.~g.,~Ref.~\cite{Kirpichnikov:2021jev} and references therein). As a 
result, the dominant energy   of signal recoil electrons will be 
deposited  within the sufficiently thick and wide ECAL target, producing a single 
electromagnetic shower with energy in the signal region, as mentioned 
above. Detector efficiencies are incorporated into the effective number of electrons accumulated on the target.

The principal background processes in NA64e are~\cite{NA64:2023wbi}: (i) losses 
or decays of dimuons in the target; (ii) decays-in-flight along the beam line; 
(iii) insufficient calorimeter coverage; and (iv) particles flying through the 
calorimeters. 

We rely on the background analysis of Ref.~\cite{NA64:2023wbi}, which yields a conservative estimate of ${\rm N}_{\rm bckg.}\lesssim 0.75$ for the anticipated statistics of $\mbox{EOT}
\simeq 1.5\times 10^{12}$. However, we expect that upgraded detector electronics will 
improve background event rejection by a factor of $\mathcal{O}(1)$, which would 
lead to a sufficiently suppressed number,~${\rm N}_{\rm bckg.} \ll 1$, of such 
events.

\begin{table*}[tb!]
    \begin{tabular}{lccccc}
	\hline
 \hline
	& NA64e   & NA64$\mu$ & LDMX  \\ \hline
        target material & Pb  & Pb & Al  \\ \hline
	$Z$,~\mbox{atomic number} & $82$  & $82$ & 13   \\ \hline
	$A,~\mbox{g}\cdot\mbox{mole}^{-1}$  & $207$  & $207$ &  27   \\ \hline
	$L_T,~\mbox{cm}$ & $22.5$ & $22.5$  &  3.56   \\ \hline
	$x_{\rm cut}=E^{\rm cut}_\phi/E_l$ & $0.5$  & $0.5$ & 0.7   \\ \hline
	$l^\pm$,  primary beam & electron  & muon & electron   \\ \hline
        $E_l$,~GeV,~\mbox{beam energy} & $100$  & $160$ & 16  \\ 
        \hline 
        Expected statistics:  & EOT $=1.5\times 10^{12} $ & MOT $= 10^{11} $ & EOT $= 10^{15} $
        \\
        \hline
Expected background:  &  ${\rm N}_{\rm bckg.}\lesssim  0.75$~\cite{NA64:2023wbi}  & ${\rm N}_{\rm bckg.} \lesssim 0.35$~\cite{NA64:2024nwj} & ${\rm N}_{\rm bckg.} \ll 1$~\cite{Berlin:2018bsc}
        \\
        \hline 
        \hline
    \end{tabular} 
    \caption{
    The typical parameters governing the total production cross section 
    for scalar mediators via the process $l^\pm N \to l^\pm N + \phi $ 
    in  lepton missing-energy experiments are presented. The parameter 
    $E_\phi^{\rm cut}=x_{\rm cut} E_l$ represents the characteristic 
    minimum missing-energy threshold, which is determined by the 
    specific experimental configuration. The planned statistics of 
    $1.5~\times~10^{12}$~EOT for NA64e and $10^{11}$ MOT for NA64$\mu$ were 
    chosen in the analysis 
in order to suppress the expected background, $N_{\rm bckg.} \lesssim 1$. For the LDMX experiment we choose $10^{15}$~EOT.  
 We assume that upgrades of the experimental electronics and detectors will be implemented to further suppression of 
 the background by the final phase of running, achieving  $N_{\rm bckg.} \ll 1$.
    \label{tab:BenchLeptonFTexp}
	}
\end{table*}

\subsection{The LDMX experiment} 

The Light Dark Matter eXperiment (LDMX) is a planned fixed-target facility at SLAC designed to search for new dark-sector particles using the missing-momentum technique~\cite{Berlin:2018bsc,Schuster:2021mlr,Akesson:2022vza,LDMX:2025bog}. 
This experiment employs a high-intensity electron beam and a precise measurement of the missing momentum for each electron, providing sensitivity that is complementary to the missing-energy approach of NA64~\cite{NA64:2023wbi,NA64:2022yly}. 
The LDMX setup is designed to apply stringent requirements on the missing energy and to exploit  a comprehensive system of veto detectors, yielding an experimental configuration with  negligible background contamination~\cite{Berlin:2018bsc}.

A detailed discussion of the LDMX detector concept and its physics reach can be found in the Ref.~\cite{Berlin:2018bsc}. 
The LDMX detection system comprises a precision silicon tracking spectrometer with tracking stations located upstream and downstream of a thin aluminum target, followed by a high-granularity sampling electromagnetic calorimeter and a surrounding hadronic calorimeter~\cite{Berlin:2018bsc,Akesson:2022vza}. 
The selection criteria ensure that dark-sector particles typically carry away the majority of the beam energy, while background processes must produce additional particles that can be vetoed calorimetrically. 
In the analysis considered here, events are required to have reconstructed electron energy $E_{\rm e}^{\rm rec} \lesssim 0.3 E_{\rm e}$, which corresponds to a cut  $x_{\rm cut} = 0.7$, and the target thickness is 
chosen to be $L_T \simeq 0.4 X_0$. 
The benchmark parameters of the LDMX configuration used in this work are summarized in Tab.~\ref{tab:BenchLeptonFTexp}.  

According to current projections, residual backgrounds in LDMX arise from several distinct sources~\cite{LDMX:2025bog}: (i) unbiased electron interactions, (ii) photo-nuclear processes, (iii) electro-nuclear interactions, and (iv) muon conversion events. To preserve the intended  sensitivity at exposures up to $\mbox{EOT} \lesssim 10^{15}$, a dedicated program of  front-end electronics development is planned~\cite{LDMX:2025bog}. The goal of the detector upgrades is to enable an effectively background-free LDMX search in the region of the parameter space of light dark matter where the benchmark background is below the single-event level~\cite{Berlin:2018bsc,Akesson:2022vza}.

\subsection{NA64$\mu$}

NA64$\mu$  is a fixed-target facility at the CERN SPS that investigates dark sector particles through the missing-energy channel $\mu N \to \mu N \phi$, followed by the rapid 
decay $\phi \to \chi_1 \chi_2$.
This configuration serves as a muon-beam counterpart  of the electron-beam 
NA64e experiment, providing complementary sensitivity to the dark-sector 
parameter space. The typical scheme of the NA64$\mu$ facility can be found 
in  Ref.~\cite{NA64:2024nwj}. 

The NA64$\mu$ detection system employs two magnetic spectrometers to measure the energies of both incoming and 
outgoing muons. Our analysis implements a kinematic cut on the scattered muon, $E_\mu^{\rm rec} \lesssim 
0.5 E_\mu \simeq 80~\mbox{GeV}$, corresponding to $x_{\rm cut}=0.5$ 
in Eq.~(\ref{eq:NradGrav}).  The typical 
 parameters of  NA64$\mu$ are summarized in Tab.~\ref{tab:BenchLeptonFTexp}.  

For our sensitivity analysis of NA64$\mu$, 
we consider a muon beam energy of $E_\mu\simeq 160\, \mbox{GeV}$ and a anticipated muon-on-target (MOT) 
statistics of $\mbox{MOT}\simeq 10^{11}$.
The experimental setup utilizes a lead-based Shashlyk electromagnetic calorimeter, which 
serves simultaneously as the target with a effective thickness of $L_T\simeq 40 X_0 \simeq 22.5\, \mbox{cm}$.

We emphasize that for muons with energy $E_\mu \simeq 160~\mbox{GeV}$,
the energy loss in lead  is 
sufficiently small for traversal of a target medium of $40 X_0$ radiation lengths.   That  permits neglect of the average stopping power $ \langle d E_\mu /dx \rangle \simeq 12.7\times 10^{-3}~\mbox{GeV}/\mbox{cm}$  when numerically 
computing the $\phi$-boson  production yield at the NA64$\mu$ experiment. Specifically, by the end of the ECAL target the muon beam energy decreases to 
$$
    E^{\rm end}_\mu \simeq E_\mu - L_T \langle d E_\mu /dx \rangle \simeq 159.7~\mbox{GeV}.
$$
Consequently, this approximation 
validates the use of  Eq.~(\ref{eq:MuonNradGrav}) for estimating $N^{\rm brem.}_\phi$, with the target length 
parameter  $L_T\simeq 40 X_0$ appropriate for  the NA64$\mu$ experimental setup.  The typical precision of the muon momentum 
 reconstruction including multiple scattering in the target,  is at the level of $\simeq3\%$ 
 for the momentum $E_\mu^{\rm rec}\simeq 80~\mbox{GeV}$, and does not significantly affect the sensitivity of the 
 search~\cite{NA64:2018iqr}.   

We assume that there is no loss of efficiency in the energy measurement for $\phi$ particles produced near the end of the ECAL (i.e., for reconstructed outgoing muon energies $E_\mu^{\rm rec} \lesssim 0.5 E_\mu \simeq 80~\text{GeV}$). This assumption is justified as follows.

The final-state muons are identified by Micromegas trackers and two large hadronic 
calorimeters (HCAL) located downstream. Most muons that radiate a scalar mediator 
$\phi$ with masses $m_{\phi}\lesssim 1~\mbox{GeV}$
are deflected by a sufficiently small angle 
$\psi_{\mu'}^{\rm rec} \lesssim m_\phi/E_{\mu} \simeq 6.3 \times 10^{-3}$ (see, 
e.~g.~Ref.~\cite{Sieber:2023nkq}). Such collinear emission of $\phi$ implies that 
outgoing muons, whether deflected in the first layer or at the end of the ECAL, will 
be detected by the Micromegas trackers and the HCAL. The deflection angles remain 
within the hermeticity acceptance limits, which are estimated to be $\theta^{\rm in}_{\rm ECAL} \lesssim 0.105$ and $\theta^{\rm out}_{\rm ECAL} \lesssim 0.12$ 
for the first layer of ECAL and its end, respectively 
(see e.~g.~Refs.~\cite{Sieber:2021fue,NA64:2024nwj} and references therein). 
    
For NA64$\mu$, the main background sources are~\cite{NA64:2024nwj}: (i) momentum mis-reconstruction; (ii) hadron in-flight decays; (iii) single-hadron punch-through; (iv) dimuons production; and (v) detector non-hermeticity.

The conservative analysis based on~\cite{NA64:2024nwj} yields an upper limit of ${\rm N}_{\rm bckg.}\lesssim 
0.35$ for the anticipated $\mbox{MOT}\simeq 10^{11}$. 
Future mitigation of this background is foreseen through electronic upgrades, 
which are expected to improve event rejection by a factor of $\mathcal{O}(1)$ and 
consequently suppress the number of such events to ${\rm N}_{\rm bckg.} \ll 1$.

\subsection{Invisible signatures for fixed-target experiment}

The sufficiently large benchmark  coupling constant of mediator with DM,  
$|\lambda^{\phi}_{\chi_1 \chi_2}|^2  \gg (c^{\phi}_{ll})^2$,  implies that the total decay width of spin-0 boson is $\Gamma_\phi^{\rm tot} \simeq \Gamma_{\phi \to \chi_1 \chi_2}$ with visible decays being neglected 
$\Gamma_{\phi \to \chi_1 \chi_2} \gg   \Gamma_{\phi \to ll}$.
 We consider the following typical parameters:
\begin{equation}
    \alpha_{\rm iDM}~\simeq~0.5,
\,\,
    \Delta_2 \in (0.1,~0.5),  \,\,  m_{\LSDM}/m_\phi~\simeq~1/3,
    \label{BenchParCoupl}
\end{equation}
where $\alpha_{\rm iDM} = (\text{Re}[ \lambda^{\phi}_{\chi_1 \chi_2}])^2/(4\pi)$ or $ \alpha_{\rm iDM} = (\text{Im}[ \lambda^{\phi}_{\chi_1 \chi_2}])^2/(4\pi)$
depending on benchmark scenarios ~\eqref{eq:EffLagrangianScalarMEDMajoranaiCDM} 
and~\eqref{eq:EffLagrangianScalarMEDMajoranaG5iCDM}, respectively.  

To ensure that the invisible mode dominates over visible decay channel signatures of  the lepton-specific mediator, we require  that 
$$
    c_{\tau \tau}^{\phi} \ll \sqrt{4 \pi \alpha_{\text{iDM}}}
$$
which leads to
$$
    c_{ee}^{\phi} \ll (m_e/m_\tau) \sqrt{4 \pi \alpha_{\text{iDM}}}.
$$ 
In particular, one can find that $c_{ee}^{\phi}~\ll~7.2~\cdot~10^{-4}$ for $\alpha_{\text{iDM}}~=~0.5$. 
It is also worth noting that this value is comparable to the BaBar constraints.
We will show below that the benchmark choice (\ref{BenchParCoupl}) 
implies invisible decay signatures of mediator for both NA64e and NA64$\mu$.

\begin{figure}[tb!]
	\center{\includegraphics[scale=0.4]{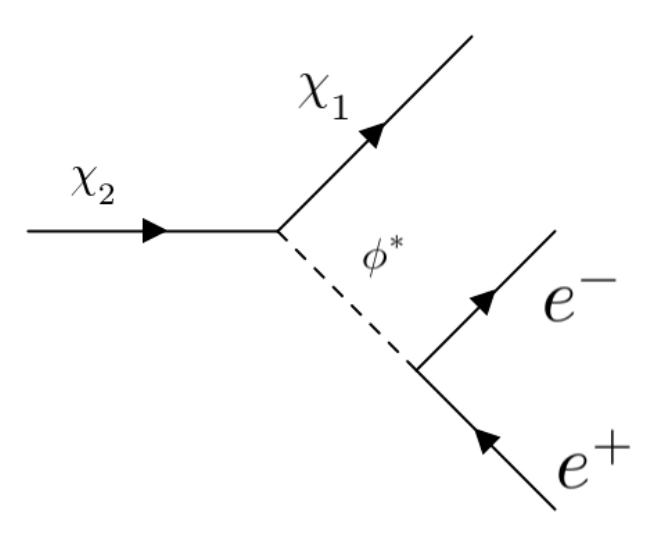}}
	\caption{ Feynman diagram describing visible decay of excited state $\chi_2$ to  DM particle $\chi_1$ and electron positron pair~$e^+ e^-$. }
	\label{fig:chi2Tochi1ll}
\end{figure}

\begin{figure}[tb!]
	\center{\includegraphics[scale=0.5]{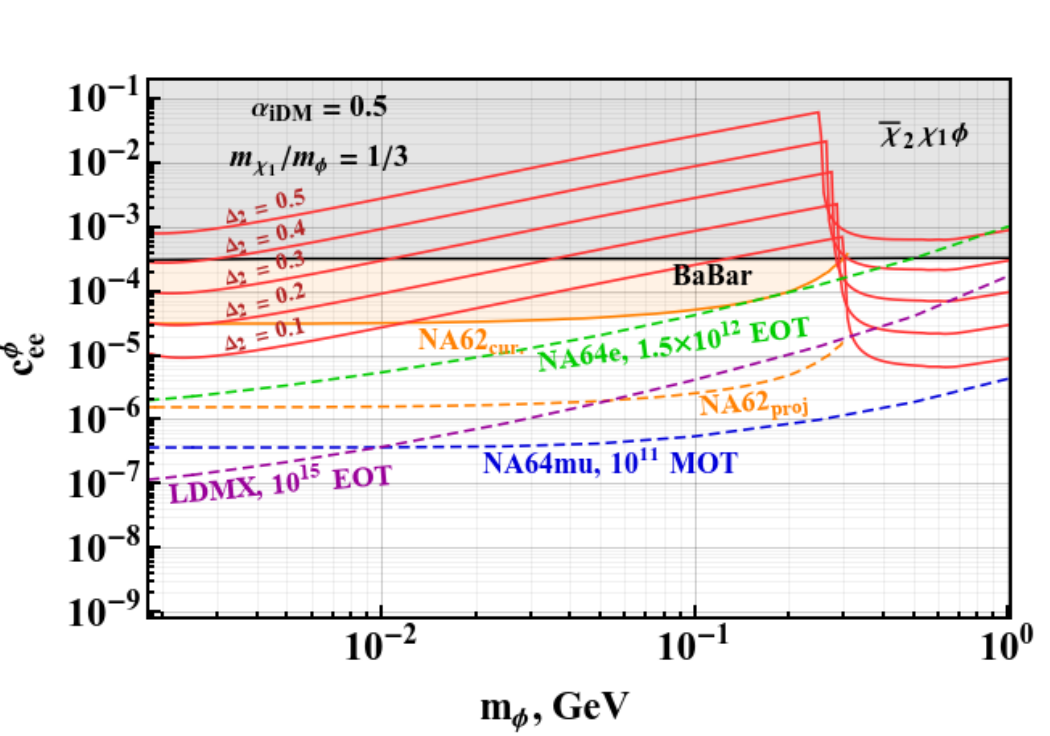}}
	\caption{ 
    The projected experimental reach at $90\, \%$ C.L. as a function of $m_\phi$ mass for the benchmark scenario~\eqref{eq:EffLagrangianScalarMEDMajoranaiCDM}.
    The grey shaded region corresponds to the existing BaBar~\cite{BaBar:2017tiz} monophoton limit from $e^+e^- \to \gamma \phi$. 
    The dashed green, blue, and purple lines show expected limits for
    NA64e, NA64$\mu$, and LDMX experiments, respectively.
    The current and projected limits of the NA62 experiment~\cite{NA62:2021bji,Krnjaic:2019rsv} are shown by solid  and dashed orange lines, respectively.  
    The red lines correspond to the typical thermal target curves  for a set of mass splittings $\Delta_2$. }
	\label{fig:LimitsFTEMajoranaiCDM}
\end{figure}

The choice of the benchmark parameter space (\ref{BenchParCoupl}) is motivated by the following considerations.
To maintain the thermally averaged cross section $\langle\sigma_{\rm eff} v \rangle$ at the value required by the 
relic-density constraint (\ref{eq:RelDensDM1}), a larger $\alpha_{\rm iDM}$ necessitates a smaller $c_{ll}^\phi$ on 
the thermal target curve. This dependence arises because 
$\langle\sigma_{\rm eff} v \rangle \propto \alpha_{\rm iDM} \times (c_{ll}^\phi)^2$. Thus, sufficiently small values of $\alpha_{\rm iDM}$ are ruled out by the BaBar experiment  (see Figs. ~\ref{fig:LimitsFTEMajoranaiCDM} and~\ref{fig:LimitsFTEMajoranaiG5CDM}).
As a result, we choose  $\alpha_{\rm iDM}$ as large as possible in order to ensure a perturbative regime,  $\alpha_{\rm iDM} \lesssim 1$ (see e.~g.~Ref.~\cite{Krnjaic:2015mbs,Chen:2018vkr,Davoudiasl:2015hxa} and references therein).   This implies that larger values of $ \alpha_{\rm iDM}$ shift the thermal target curves downward relative to the constraints, as depicted in Figs.~\ref{fig:LimitsFTEMajoranaiCDM} and~\ref{fig:LimitsFTEMajoranaiG5CDM}.  
Furthermore, we set the mass ratio $m_{\chi_1}/m_{\phi} \lesssim 1/2$ 
to avoid resonant enhancement~\cite{Foguel:2024lca}.

We refer the reader to Refs.~\cite{Giudice:2017zke,Foguel:2024lca} for the general 
expression for the 3-body decay width $\chi_2 \to \chi_1 l^+l^-$.
The relevant differential decay widths
for benchmarks~\eqref{eq:EffLagrangianScalarMEDMajoranaiCDM} 
and~\eqref{eq:EffLagrangianScalarMEDMajoranaG5iCDM} read, respectively:
\begin{multline}
    \frac{d \Gamma_{\chi_2 \to \chi_1 l^+l^-}}{d s_a} 
=
    \frac{(c_{ll}^{\phi})^2 (\text{Re}[ \lambda^{\phi}_{\chi_1 \chi_2}])^2}{(2 \pi)^3}
    \frac{1}{32 m_{\chi_2}^3}
    \frac{1}{2 m_{\phi}^4}
\\ \cdot
    \frac{4}{\sqrt{s_a}}
    (s_a - 4 m_l^2)^{3/2}
    ( ( m_{\chi_2} + m_{\chi_1} )^2  - s_a)
\\ \cdot
    \sqrt{\lambda(s_a, m_{\chi_1}^2, m_{\chi_2}^2)}
    \theta(( m_{\chi_2} - m_{\chi_1} )^2  - 4m_l^2),
\end{multline}
\begin{multline}
    \frac{d \Gamma_{\chi_2 \to \chi_1 l^+l^-}}{d s_a} 
=
    \frac{(c_{ll}^{\phi})^2 (\text{Im}[ \lambda^{\phi}_{\chi_1 \chi_2}])^2}{(2 \pi)^3}
    \frac{1}{32 m_{\chi_2}^3}
    \frac{1}{2 m_{\phi}^4}
\\ \cdot
    \frac{4}{\sqrt{s_a}}
    (s_a - 4 m_l^2)^{3/2}
    ( ( m_{\chi_2} - m_{\chi_1} )^2  - s_a)
\\ \cdot
    \sqrt{\lambda(s_a, m_{\chi_1}^2, m_{\chi_2}^2)}
    \theta(( m_{\chi_2} - m_{\chi_1} )^2  - 4m_l^2),
\end{multline}
where the variable~$s_a$ varies in the range from $s_a^-~=~4m_l^2$ to $s_a^+~=~( m_{\chi_2} - m_{\chi_1} )^2$. We also consider the approaches: 
\[
     s_a \ll m_{\phi}^2,
\quad
    \Gamma_{\rm tot} / m_{\phi} \ll 1,
\]
that leads to
\[
    (s_a - m_{\phi}^2)^2 + m_{\phi}^2 \Gamma_{\rm tot}^2 \simeq m_{\phi}^4.
\]
Additionally, by taking into account expressions for integration limits as
\[
     s_a \ll ( m_{\chi_2} + m_{\chi_1})^2,
\]
and the small mass splitting of inelastic dark matter 
\[
    m_l \ll ( m_{\chi_2} - m_{\chi_1}),
\quad
    \Delta_2 \ll 1,
\]
we get
\[
    \lambda(s_a, m_{\chi_1}^2, m_{\chi_2}^2) \simeq 2 m_{\chi_2}^2 \sqrt{( m_{\chi_2} - m_{\chi_1})^2 - s_a},
\]
\[
    ( m_{\chi_2} + m_{\chi_1})^2 - s_a \simeq 4 m_{\chi_2}^2,
\quad
    s_a - 4 m_l^2 \simeq s_a.
\]

Finally, one can obtain  the following expressions for 
the benchmarks~\eqref{eq:EffLagrangianScalarMEDMajoranaiCDM} and~\eqref{eq:EffLagrangianScalarMEDMajoranaG5iCDM}, 
respectively:
\begin{equation}\label{eq:WD1to3ScalarMEDMajoranaiCDM}
    \Gamma_{\chi_2 \to \chi_1 l^+l^-}
=
    \frac{(c_{ll}^{\phi})^2 (\text{Re}[ \lambda^{\phi}_{\chi_1 \chi_2}])^2}{60 \pi^3} 
    \frac{( m_{\chi_2} - m_{\chi_1})^5}{m_{\phi}^4},
\end{equation}
\begin{equation}\label{eq:WD1to3ScalarMEDMajoranaG5iCDM}
    \Gamma_{\chi_2 \to \chi_1 l^+l^-}
=
    \frac{(c_{ll}^{\phi})^2 (\text{Im}[ \lambda^{\phi}_{\chi_1 \chi_2}])^2}{560 \pi^3} 
    \frac{( m_{\chi_2} - m_{\chi_1})^7}{m_{\chi_2}^2 m_{\phi}^4}.
\end{equation}
The relevant decays  imply the kinematic threshold~$m_{\LSDM}\Delta_2 > 2 m_l$.
Thus, one can see that the muon channel $\chi_2 \to \chi_1 \mu^+\mu^-$ is allowed for sufficiently large mass 
splitting~$\Delta_2~\gtrsim~ 6 m_\mu/m_{\phi} \simeq  0.61$ and 
the masses of interest~$r=m_{\LSDM} / m_\phi = 1/3$ with $m_\phi \lesssim 1~\mbox{GeV}$. 
As a result, for $\Delta_2 \lesssim 0.5$ the dominant decay channel of excited state $\chi_2$ is associated with three-body decay $\chi_2 \to \chi_1 e^+e^-$ (see Fig.~\ref{fig:chi2Tochi1ll}). 

The decay length of the heavier inelastic dark matter state in the laboratory frame is:
\begin{equation}\label{eq:LenDecay}
    l_{\chi_2} = \frac{E_{\chi_2}}{m_{\chi_2}} \frac{1}{\Gamma^{\rm tot}_{\chi_2}},
\end{equation}
where we assume that~$\Gamma^{\rm tot}_{\chi_2}~\simeq~\Gamma_{\chi_2 \to \chi_1 e^+ e^-}$ for ~$\Delta_2~<~0.5$. Using the approximation~$E_{\chi_2} \simeq E_{\phi}/2 \simeq E_{l}/2$,
 the decay lengths for~\eqref{eq:EffLagrangianScalarMEDMajoranaiCDM} and~\eqref{eq:EffLagrangianScalarMEDMajoranaG5iCDM} read, respectively:
\begin{multline}\label{eq:LenDecayApproxMajorana}
    l_{\chi_2}
\simeq
    8.76 \times 10^{6}~\mbox{cm}
\times
    \left( \frac{0.3}{\Delta_2} \right)^{5}
    \left( \frac{1}{1 + \Delta_2} \right)
    \left( \frac{1/3}{r} \right)^{6}
\\ \times
    \left( \frac{E_{l}}{10~\mbox{GeV}} \right)
    \left( \frac{0.1~\mbox{GeV}}{m_{\phi}} \right)^{2}
    \left( \frac{10^{-5}}{c_{\rm ee}^{\phi}} \right)^{2}
    \left( \frac{0.5}{\alpha_{\rm iDM}} \right),
\end{multline}
\begin{multline}\label{eq:LenDecayApproxMajoranaG5}
    l_{\chi_2}
\simeq
    9.09 \times 10^{8}~\mbox{cm}
\times
    \left( \frac{0.3}{\Delta_2} \right)^{7}
    \left( \frac{1 + \Delta_2}{1} \right)
    \left( \frac{1/3}{r} \right)^{6}
\\ \times
    \left( \frac{E_{l}}{10~\mbox{GeV}} \right)
    \left( \frac{0.1~\mbox{GeV}}{m_{\phi}} \right)^{2}
    \left( \frac{10^{-5}}{c_{\rm ee}^{\phi}} \right)^{2}
    \left( \frac{0.5}{\alpha_{\rm iDM}} \right).
\end{multline}

Thus, within the relevant parameter space, a rough estimate gives $l_{\chi_2} \gtrsim \mathcal{O}(1)~\mbox{km}$.
The typical number of $\chi_2$ decays at the fixed-target facilities can be approximated as
\begin{equation}
N_{\rm sign} \simeq N^{\rm brems.}_{\phi} \exp(-L_{\rm det.}/l_{\chi_2}),
\end{equation}
where $L_{\rm det.}$ is the typical length of the experimental facility, that is 
estimated to be~$\lesssim (1 - 10) \mbox{m}$ for considered fixed-target experiments,  this 
however implies~$L_{\rm det.} \ll l_{\chi_2}$ and $N_{\rm sign}\simeq N^{\rm brems.}_{\phi}$.
This means that the visible decay of $\chi_2$ occurs outside LDMX, NA64e and NA64$\mu$ 
detectors for the parameter space of interest (\ref{BenchParCoupl}).
However,
that results in the invisible signature at the considered fixed-target facilities, since the decays $\chi_2 \to \chi_1 e^+ e^-$ evade 
detection in their calorimetric system.

\section{The experimental reach \label{sec:ExpectedReach}}

In this section, we discuss the expected experimental reach of the fixed-target facilities such as NA64e, LDMX, and NA64$\mu$ assuming that the scalar mediator decays into the invisible  mode. 

We set the $90\% \mbox{~C.~L.}$ exclusion limit on the coupling constant between the electron and the scalar mediator, using the typical number of signal events $N_{\rm sign.} \gtrsim 2.3$.
Also, we imply the suppressed background, $ {\rm N}_{\rm bckg.} \lesssim 1$,   and the null results of the observed missing-energy events for experiments, ${\rm N}_{\rm obs.}=0$. 
For the evaluation of experimental limits, we employ
the projected statistics of fixed-target facilities that are presented 
in~Tab.~\ref{tab:BenchLeptonFTexp}.

\begin{figure}[tb!]
	\center{\includegraphics[scale=0.5]{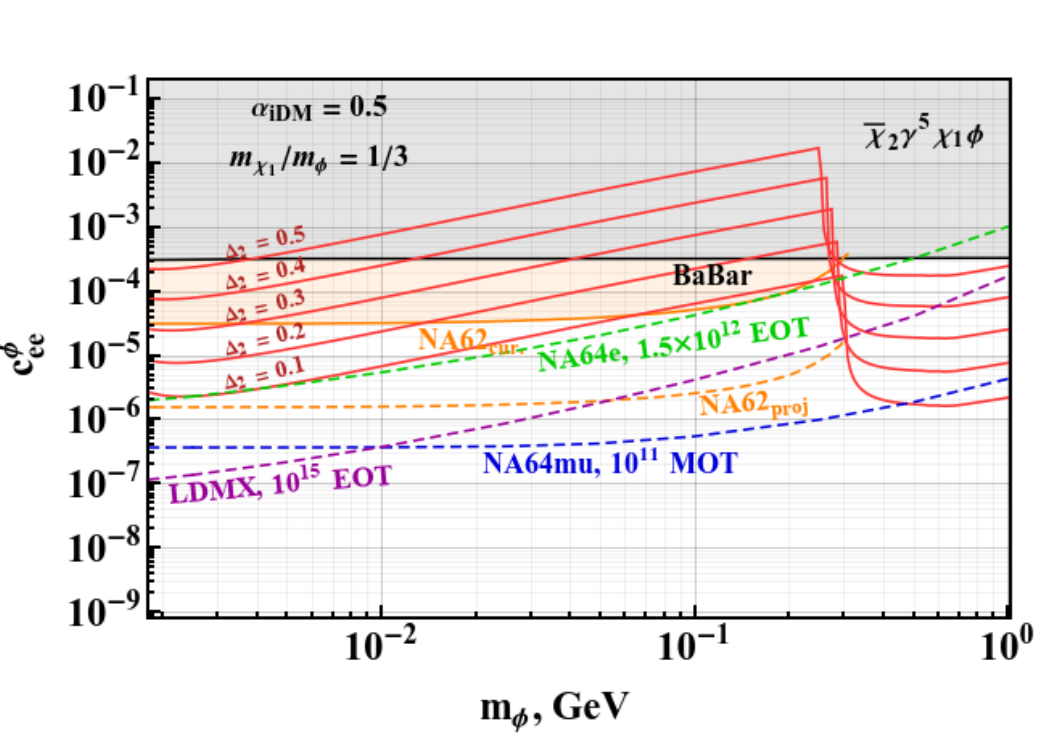}}
	\caption{ 
    The same as in Fig.~\ref{fig:LimitsFTEMajoranaiCDM} but for
    the benchmark  model~\eqref{eq:EffLagrangianScalarMEDMajoranaG5iCDM}.  
    }
	\label{fig:LimitsFTEMajoranaiG5CDM}
\end{figure}

As discussed above, 
the decay length of the heavier dark-sector state $\chi_2$ is much 
larger than the detector base of the considered experiments.
Thus, we focus on the invisible mode at the lepton fixed-target facilities, 
$ l N \to l N \phi (\to \chi_1 \chi_2)$.  

In Figs.~\ref{fig:LimitsFTEMajoranaiCDM} 
and~\ref{fig:LimitsFTEMajoranaiG5CDM}, we present the typical thermal target 
curves for the benchmark couplings~\eqref{eq:EffLagrangianScalarMEDMajoranaiCDM} 
and~\eqref{eq:EffLagrangianScalarMEDMajoranaG5iCDM}, respectively. We also display the expected sensitivities of NA64e and NA64$\mu$ lepton missing-energy 
experiments for the anticipated statistics of $\mbox{EOT }=1.5\times 10^{12} $ 
and $\mbox{MOT} = 10^{11} $, respectively. We adopt the benchmark 
relation between couplings of charged leptons and scalar mediator 
as (\ref{CeeCmumu}). 

We emphasize that the heavier dark-sector state $\chi_2$ predominantly decays into the 
lightest  state and an electron-positron pair  $\chi_2 \to \chi_1 e^+ e^-$ for 
the considered parameter space (\ref{BenchParCoupl})  of  the fixed-target 
experiments.
The muon decay channel $\chi_2 \to \chi_1 \mu^+ \mu^-$ requires a sufficiently 
large mass splitting $\Delta_2~\gtrsim~0.61$ which was excluded by the Babar 
experiment.

The coupling of a scalar mediator to leptons induces an additional tree-level 
decay of the charged kaon, $K^{+}\to\mu^{+}\nu_{\mu}\phi$~\cite{Blinov:2024gcw}. 
In this process, the scalar mediator can be produced invisibly, carrying energy away from the detector.  In Ref.~\cite{Krnjaic:2019rsv}, the projected reach of the NA62 experiment
for the muon-philic scalar  coupling $\mathcal{L} \supset c_{\mu \mu }^\phi \phi \bar{\mu} \mu$ was obtained. This leads to the typical bound on the coupling  at the level of  $c_{ee}^{\phi}\lesssim\mathcal{O}(10^{-6})$ for mediator masses $m_{\phi} \lesssim 300~\mathrm{MeV}$. 

However, these projected bounds can be recast using the current NA62 data~\cite{NA62:2021bji} on the invisible branching fraction of charged kaons at the 90 \% C.L., yielding:
$$\mathcal{O}(10^{-7}) \lesssim {\rm Br}(K^+\to \mu^+ \nu_\mu \phi)_{\rm data} \lesssim \mathcal{O}(10^{-5}).$$
The data from Ref.~\cite{NA62:2021bji} imply that for $m_\phi \simeq 50~\mbox{MeV}$
$$
{\rm Br}(K^+\to \mu^+ \nu_\mu \phi)_{\rm data} \simeq 2\times10^{-6}.
$$
On the other hand, the authors of Ref.~\cite{Krnjaic:2019rsv} have  shown that for $m_\phi\simeq 50~\mbox{MeV}$ the  
projected luminosity data of  NA62 can  yield typical bound on 
branching fraction
$$
{\rm Br}(K^+\to \mu^+ \nu_\mu \phi)_{\rm proj.} \simeq 5\times10^{-9}.
$$
That can be translated to the current limits on $(c_{ee}^\phi)_{\rm data}$ in the following form
$$
(c_{ee}^\phi)_{\rm data} \simeq (c_{ee}^\phi)_{\rm proj.} \left( \frac{{\rm Br}(K^+\to \mu^+ \nu_\mu \phi)_{\rm data}}{
{\rm Br}(K^+\to \mu^+ \nu_\mu \phi)_{\rm proj.}
} \right)^{1/2},
$$
where we use the relation (\ref{CeeCmumu}) between muon and electron coupling to the mediator, this yields the recasting coefficient $(c_{ee}^\phi)_{\rm data} \simeq 20 \times (c_{ee}^\phi)_{\rm proj.}$. 
Consequently, the NA62 limits 
rule out several of the splittings for the thermal Majorana  inelastic dark matter below $m_{\phi}\lesssim 300 ~\mathrm{MeV}$.

The  NA64$\mu$ experiment can be sensitive to couplings at the 
level of~$c_{ee}^{\phi} \gtrsim 5\times 10^{-7}$, 
also the NA64e experiment can 
rule out the coupling  in the range~$c_{ee}^{\phi} \gtrsim 10^{-6}$ for the small 
mass region $m_\phi \lesssim 10~\mbox{MeV}$. Moreover, the anticipated number 
of electrons collected on target~$\mbox{EOT}\simeq 1.5\times 10^{12}$
for NA64e will allow us to constrain the inelastic DM model with $ \Delta_2 \gtrsim 0.1$, 
$\alpha_{\rm iDM}=0.5$ and $m_{\LSDM}/m_\phi =1/3$ for the mediator in the mass range 
$m_\phi \lesssim 300~\mbox{MeV}$.  
Furthermore, the NA64$\mu$ with 
$\mbox{MOT}\simeq 10^{11}$ can rule out the relevant parameter space in the wider 
mass range of the mediator,  $2~\mbox{MeV} \lesssim m_\phi \lesssim 1~\mbox{GeV}$ for the scalar coupling between iDM and mediator. 
 
In addition, the LDMX experiment can be sensitive to couplings at the 
level of~$c_{ee}^{\phi} \gtrsim 10^{-7}$ and will
provide new limits on the interaction coupling for $\mMED \lesssim 10~\mbox{MeV}$.

To address the limits from the production of heavy-flavor leptons and their decays - 
specifically, $\tau^- \to \mu^- \bar{\nu_\mu} \nu_\tau \phi $, $\mu^- \to e^- \bar{\nu_e} \nu_\mu \phi $, and $\tau^- \to e^- \bar{\nu_e} \nu_\tau \phi$—we refer the reader 
to Ref.~\cite{Chen:2018vkr}. Its authors demonstrate that the corresponding bounds are 
weaker than those from current accelerator-based experiments~\cite{BaBar:2017tiz}. This 
implies a limit of $ c_{ee}^\phi \simeq 5\times 10^{-4}$, which has been ruled out by BaBar.

For completeness, we refer the reader to Ref.~\cite{Boos:2022gtt}, which addresses the 
projected bounds from the Charm-Tau factory. This facility implies the potential for 
heavy lepton production associated with the invisible decay of a scalar mediator, such 
as in the process $e^+ e^- \to \tau^+\tau^- \phi$. The relevant limits could be as small 
as $c_{ee}^\phi \lesssim 10^{-8}$ for a projected luminosity of 
$\mathcal{L}\simeq 3~\mbox{ab}^{-1}$ with collider energy at $\sqrt{s}\simeq 4.2~\mbox{GeV}$. 
As a result, a dominant portion of the benchmark iDM parameter space, $ 1~\mbox{MeV} \lesssim m_\phi \lesssim 1~\mbox{GeV}$,  can be ruled out by Charm-Tau factory.

Let us discuss the impact of the mass splitting on the thermal target curves for inelastic dark matter in Figs.~\ref{fig:LimitsFTEMajoranaiCDM} and~\ref{fig:LimitsFTEMajoranaiG5CDM}.
Increasing the mass splitting value shifts the relic curves upward relative to the constraints. That can be explained as follows. 
By taking into account the expression for the effective thermally averaged cross 
section~\eqref{eq:DefEffThermalCS} and approximations for the fraction of heavy inelastic 
dark matter~\eqref{eq:RatPartDensDMapproxCDM}, we can estimate the typical scaling for the 
coupling~$c_{ee}^{\phi}~\propto~e^{\Delta_2 x_f}$. Thus, the larger $\Delta_2$, the larger 
$ c_{ee}^\phi$ for the typical thermal target curve. On the other hand, smaller values 
$\Delta_2 x_f \ll 1$ would not impact on the DM relic abundance coupling $c_{ee}^{\phi}$.
A  sizable  contribution of the  mass splitting to the thermal target curves occurs for ~$\Delta_2~\gtrsim~1/x_f~\simeq~0.04$.

Regions beneath the red contours are excluded as they correspond to a 
cosmological over-abundance of  DM, $\Omega_c h^2 \gtrsim 0.12$, for the specified mass 
ratio, $m_\phi/m_{\chi_1}=3$. The discontinuities observed near $m_\phi \simeq 2 m_\mu$ 
arise from the kinematic opening of the new co-annihilation channel 
 $ \chi_1 \chi_2 \to \mu^+ \mu^-$ in addition to the $e^+e^-$ pair production. 

In order to conclude this section we note that one-loop elastic  direct detection signature $\chi_1 e^- \to \chi_1 e^-$ provides the bound that has been already ruled out by BaBar (see Appendix~\ref{sec:DD}).

\section{Conclusion
\label{sec:Conclusion}}

In this work, we explore the consequences of sub-GeV inelastic dark matter scenarios 
mediated by a lepton-specific scalar portal, by solving the Boltzmann equations, we derive thermal targets for a set of DM mass splittings, $0.1 \lesssim \Delta_2 \lesssim 0.5$, focusing on the sub-GeV regime for the scalar mediator mass. Considering a minimal spin-0 extension of SM lepton sector, we evaluate the sensitivity of lepton fixed-target experiments such as NA64e, LDMX, and NA64$\mu$.
In particular, we use missing-energy signature to obtain projected constraint on coupling of scalar 
mediator and electron.  That channel can be represented as $\phi$-strahlung processes with invisible 
decay~$\phi~\to~\chi_2 \chi_1$.
We focus on a scalar mediator with the masses from~$\mathcal{O}(1)$~MeV to~$\mathcal{O}(1)$~GeV.
For mass splitting greater than $0.5$, the corresponding models are excluded from the Babar experiment. 
For the coupling constant relation,~$c_{ee}^{\phi}:c_{\mu\mu}^{\phi}=m_{e}:m_{\mu}$, the 
NA64$\mu$ experiment can be sensitive 
to~$c_{ee}^{\phi} \gtrsim  5\times 10^{-7}$ for mass window sub-GeV mediator mass window, 
$m_\phi \lesssim 1~\mbox{GeV}$. In addition, the NA64e experiment can rule out the typical 
couplings~$c_{ee}^{\phi} \gtrsim 2\times 10^{-6} - 10^{-4}$ for mediator masses in the range 
$m_\phi \lesssim 300~\mbox{MeV}$. The LDMX electron fixed target facility will probe couplings~$c_{ee}^{\phi} \gtrsim  10^{-7} - 10^{-4}$ for sub-GeV mediator masses. 
We  show that current data of the NA62 
experiment~\cite{NA62:2021bji} rule out relatively large splitting $\Delta_2 \gtrsim 0.2 - 0.3$ for the  
masses below $m_\phi \lesssim 300~\mbox{MeV}$. 
We argue that one-loop induced direct detection signature $\chi_1 e^- \to \chi_1 e^-$
constraints from XENON1T (2019) data
have been ruled out by BaBar experiment.

\begin{acknowledgments} 
 The work of IV and DK on calculation of the DM thermal target curves and estimation of NA64e and NA64$\mu$ sensitivities   was supported 
 by the Foundation for the Advancement of Theoretical Physics 
 and Mathematics BASIS (Project No.~\text{24-1-2-11-2} and 
 No.~\text{24-1-2-11-1}).
The work of DK on evaluation of the current constraints from the NA62 experiment and the projected limits from the LDMX experiment was supported by RSF Grant No. 24-72-10110.
\end{acknowledgments}

\appendix

\section{Relic density of inelastic dark matter}\label{sec:RelicAbundance}

In this section, we discuss the form of the Boltzmann equation in the case of inelastic dark matter.
The relic abundance of the lightest state of dark matter in the freeze-out mechanism can be 
described by the sum of the densities of all particles of the hidden sector~\cite{Griest:1990kh,Edsjo:1997bg}.
Next, summing the Boltzmann equations for particle densities of k-th type of iDM,~$n_k$, one can get the following expression~\cite{Gondolo:1990dk,Griest:1990kh,Edsjo:1997bg}:
\begin{align}\label{eq:BoltzmannEqForTotPartDens}
    \dot{n} 
=  
    & -   3 H n
- \nn \\ & -
        \sum\limits_{f} \!
        \sum\limits_{ i,j = 1 }^{2} 
        \Avrg{\sigma_{ij} v_{ij}}
        \left( n_i n_j - n_i^{\rm eq} n_j^{\rm eq} \right),
\end{align}
where~$\sum_{f}$ is the sum over final state of SM particles, $n~=~\sum_{l=1}^{2} n_l$~is a total particle density of DM and $n_{\rm eq}~=~\sum_{k=1}^{2} n_k^{\rm eq}$ is~total particle density of DM in equilibrium.
The thermally averaged cross section reads~\cite{Gondolo:1990dk}:
\begin{equation}\label{eq:ThermalAveregedCS}
    \Avrg{\sigma_{ij} v_{ij}}
=
    \frac{g_i g_j}{n_{i}^{\rm eq} n_{j}^{\rm eq}}
    \int \sigma_{ij} v_{ij} f_i f_j 
    \frac{ d \vect{p}_i}{(2\pi)^3} \frac{ d \vect{p}_j}{(2\pi)^3},
\end{equation}
where $f_i~=~e^{-E_i/T}$ is equilibrium distribution function in the Maxwell-Boltzmann approximation 
for the typical temperature.
The Moller velocity has the following form:
\begin{equation}\label{eq:MollerVelocityGeneral}
    v_{lk} =  I_{lk} / \left( E_l E_k \right),
\end{equation}
where Moller invariant is:
\[
    I_{lk} 
= 
    \sqrt{(p_l, p_k)^2 - m_{\chi_l}^2 m_{\chi_k}^2}
=
    (1/2)\sqrt{ \lambda(s, m_{\chi_l}^2,m_{\chi_k}^2)},
\]
$\lambda(s, m_{\chi_l}^2,m_{\chi_k}^2) = (s - ( m_{\chi_l} +  m_{\chi_k})^2)(s - ( m_{\chi_l} -  m_{\chi_k})^2)$ is triangular function and~$s~=~(p_i~+~p_j)^2$ is the Mandelstam variable.

Taking into account~$n_i/n~\simeq~n_i^{\rm eq}/n^{\rm eq}$, one can get the Boltzmann equation for total particle density of DM in the standard form as:
\begin{equation}\label{eq:BoltzmannEqForTotPartDensEffThermalCS}
    \dot{n} 
= 
    - 3 H n
    -    \Avrg{\sigma_{\rm eff} v}
        \left( n^2 - n_{\rm eq}^2 \right),
\end{equation}
where~effective thermally averaged cross section is:
\begin{equation}\label{eq:DefEffThermalCS}
    \Avrg{\sigma_{\rm eff} v}
=   
    \sum\limits_{f} \!
    \sum\limits_{ i,j = 1 }^{2} 
    \Avrg{\sigma_{ij} v_{ij}}
    \frac{n_i^{\rm eq} n_j^{\rm eq}}{n_{\rm eq}^2}.
\end{equation}
The explicit form of the effective thermally averaged cross section is shown in~\eqref{eq:NonRelAvEffCS} for the non-relativistic case.

Let us consider the dependence of the critical dark matter density  on the thermally averaged cross section.
Taking into account the Friedmann Universe, the Hubble parameter is defined as \cite{Kolb:1990vq}:
\[
    H(x) 
 \! \simeq  \! 
    \frac{H(m_{\LSDM})}{x^2},
\quad
    H(m_{\LSDM}) 
 \! \simeq  \!
    0.331 \;
    g_{*\varepsilon}^{1/2}(m_{\LSDM})
    \frac{m_{\LSDM}^2}{M_{Pl}},
\]
where $M_{Pl} \simeq  2.4 \cdot 10^{18} \;\mbox{GeV}$ is the reduced Planck 
mass, $g_{*\varepsilon}(T)$ is the  effective degree of 
freedom~\cite{Husdal:2016haj}, and $m_{\LSDM}$ is the lightest state mass of DM, the variable $x$ is the typical ratio,  defined as $x = m_{\LSDM} / T$. 
Assuming that the effective number of entropy degrees of freedom,~$g_{*s}(T)$, depends weakly on temperature,
one can obtain expression for Hubble parameter from the conservation of comoving entropy,~$d(a^3(t)s)/dt~=~0$, as:
\[
    H(t)~=~(-1/T)dT/dt,  
\]
where~$a(t)$ is the scale factor and the total entropy density takes the following forms:
\[
    s(T) 
= 
    \frac{2 \pi^2}{45}
    g_{*s} \left(T\right) 
    T^3,
\;
    s(x) 
= 
    \frac{2 \pi^2}{45}
    g_{*s} \left(T = \frac{m_{\LSDM}}{x}\right) 
    \frac{m_{\LSDM}^3}{x^3}.
\]
Thus, time is related to the parameter~$x$ as~$dt~=~x dx/H(m_{\LSDM})$.

The Boltzmann equation~\eqref{eq:BoltzmannEqForTotPartDensEffThermalCS} can be written by defining the term  $Y(x) = n(x)/s(x)$ as a ratio of DM particle density $n(x)$ over the total SM entropy density $s(x)$ in the following form:
\begin{equation}\label{eq:BoltzmannEqInYvarible}
    \frac{dY}{dx}  
\! = \!
    - \frac{s(x)}{x H(x)} \Avrg{\sigma_{\rm eff} v} 
    \left( Y^2(x) \! - \! Y^2_{\rm eq}(x)\right),
\end{equation}
where~$Y_{\rm eq}(x)~=~n_{\rm eq}(x) / s(x)$.

The decoupling of the considered type of particles is achieved if the following condition is 
fulfilled~\cite{Kolb:1990vq}:
\begin{equation}\label{DecouplingCondition}
    \left. n_{\rm eq} \Avrg{\sigma_{\rm eff} v} \right|_{x = x_f}
\simeq   
    H(x_f), 
\end{equation}
where $T_f$ is the critical temperature  and $x_f = m_{\LSDM} / T_f$ is the parameter of freeze-out, that 
is estimated to be~$x_f \simeq 25$. 
 
The current value of variable $Y$ tends to the relic value $Y(\infty)$, that can be estimated on the interval $(x_f, \infty)$ from the Boltzmann equation~\eqref{eq:BoltzmannEqInYvarible} as:
\begin{equation}\label{eq:exprYinfty}
    Y^{-1}\!(\infty) 
\! = \!
    \frac{\; s(T = m_{\LSDM}) }{H(m_{\LSDM})} 
    \! J(x_f),
\;\;
    J(x_f)
\! = \!
    \! \int\limits_{x_f}^{\infty} \!
    \frac{\Avrg{\sigma_{\rm eff} v}}{x^2} dx,
\end{equation}
where we take into account that $Y(x)~\gg~Y_{\rm eq}(x)$ for $x~>~x_f$. 
In case of s-wave annihilation one can exploit $J(x_f) = \TAvrgCS / x_f$.

The critical density of cold dark matter is:
\begin{equation}
    \Omega_{c}
=
    \frac{\varepsilon_{\DM}}{\varepsilon_{\rm cr}}
=
    \frac{s_0\sum_{k=1}^{2} m_{\chi_k} Y_k}{\varepsilon_{\rm cr}}
\simeq
    \frac{m_{\LSDM}s_0 Y_1}{\varepsilon_{\rm cr}},
\end{equation}
where~$Y_1~\simeq~Y(\infty)$, $Y_k~=~n_k(x)/s(x)$, $\varepsilon_{\DM}~=~\sum_{k=1}^{2} m_{\chi_k} n_k$~is total energy density of cold dark matter and~$\varepsilon_{\rm cr}~=~3H_0^{2}M_{\rm Pl}^2$ is critical density.
We assume for mass of inelastic dark matter that $m_{\chi_i}~=~m_{\LSDM}(1~+~\Delta_i)~\simeq~m_{\LSDM}$.
Thus, one can get:
\begin{equation}
    \Omega_{c} 
\simeq 
    \frac{ m_{\LSDM} s_0}{3 M_{Pl}^2 H_0^2}
    \frac{H(m_{\LSDM})}{\; s(T = m_{\LSDM})}
    \left(\, \int\limits_{x_f}^{\infty} \frac{\Avrg{\sigma_{\rm eff} v}}{x^2} dx \right)^{-1} \!\!, 
\end{equation}
where~$s_0 = s(T_0)$ is the current total entropy density and current temperature of the Universe is:
\[
    T_0 = 2.726 \;\mbox{K} = 2.35 \times 10^{-13} \;\mbox{GeV}.
\]
The current values of the Hubble parameter,~$H_0$, and the dimensionless constant,~$h$, read, respectively~\cite{Planck:2018vyg}:
\[
    H_0~=~2.13~h~\times~10^{-42}~\mbox{GeV},
\quad
    h \simeq 0.674 \pm 0.005.
\]
In addition, 
we take into account that the current value of relic abundance of cold DM obtained from 
the Planck 2018 combined analysis is~\cite{Planck:2018vyg}:
\[
    \Omega_c h^2 = 0.1200 \pm 0.0012.
\] 
As a result,  the expression of relic density takes the following form \cite{Kolb:1990vq}:
\begin{equation}\label{eq:RelDensDM}
    \Omega_{c} h^2
\!  \simeq  \! 
    \frac{0.85  \! \cdot \! 10^{-10}}{\mbox{GeV}^{2}}
    g^{-1/2}_{*s}(m_{\LSDM})
    \left(\, \int\limits_{x_f}^{\infty} \frac{\Avrg{\sigma_{\rm eff} v}}{x^2} dx \right)^{-1},
\end{equation}
where $g_{*\varepsilon}(T)$,~$g_{*s}(T)$ is the energy and entropy effective degrees of freedom~\cite{Husdal:2016haj}, respectively, and it is taken into account that $g_{*s}(T_0)~\approx~3.91$ and for the temperature $T~\gtrsim~1~\mbox{MeV}$ we can set $g_{*\varepsilon}(T)~\simeq~g_{*s}(T)$.

\section{Annihilation cross sections of inelastic dark matter
\label{SectionCScalculation}}

In this section, we  provide the cross sections of inelastic DM annihilation into SM particles in the case of 
a lepton-specific scalar mediator. 
For the calculation of both the decay width and cross section, we employ the state-of-the-art FeynCalc  
package~\cite{Shtabovenko:2020gxv,Shtabovenko:2016sxi} for the Wolfram Mathematica routine~\cite{Mathematica}. 

The decay width of scalar mediator in cases of lepton and considered benchmarks are, respectively:
\begin{align}
    \Gamma_{\phi \to l^+l^-}(s)
& =
    \frac{(c_{ll}^{\phi})^2}{8 \pi} 
    \sqrt{s} 
    \beta^3(4 m_l^2, s), 
\\
    \Gamma_{\phi \to \chi_2 \chi_1}(s)
& =
    \frac{(\text{Re}[ \lambda^{\phi}_{\chi_1 \chi_2}])^2}{8 \pi} 
    \sqrt{s} 
    \beta^3(s_0, s) 
    \beta(\delta_0, s), 
\\
    \Gamma_{\phi \to \chi_2 \chi_1}(s)
& =
    \frac{(\text{Im}[ \lambda^{\phi}_{\chi_1 \chi_2}])^2}{8 \pi} 
    \sqrt{s} 
    \beta(s_0, s)  
    \beta^3(\delta_0, s), 
\end{align}
where~$s_0~=~(m_{\chi_2} + m_{\chi_1})^2$ and~$\delta_0~=~(m_{\chi_2} - m_{\chi_1})^2$, also we denote~$\beta(a, b) = \left( 1 - a/b \right)^{1/2}$.
Moreover, the decay widths reduce to well-known expressions~\cite{Liu:2016mqv,Liu:2017htz} for the chosen Lagrangian densities in the case of equal  masses of dark matter.

In the case of two-body final-state process, the resonant total cross section for $\chi_i \chi_j \to~\phi\to~l^+l^-$ can be approximated by the Breit-Wigner (BW) resonant 
formula~\cite{Foguel:2024lca}:
\begin{multline}\label{eq:BWformiula}
    \sigma_{\chi_i \chi_j \to  \phi\to l^+l^-}
\! = \!
    4 \pi N_{\sigma}^{\rm BW} 
    \frac{s^2}{I_{ij}^2} 
    \frac{ \Gamma_{ \phi~\to~\chi_i \chi_j}(s)
           \Gamma_{ \phi~\to~l^+l^-}(s)
         }{\DBWformula^2(s)}, 
\end{multline}
where~$N_{\sigma}^{\rm BW}~=~S_{\rm BW} (2 J \! + \! 1) (2 S_i \!+ \! 1)^{-1} (2 S_j \!+ \! 1)^{-1}$,
$\Gamma_{\rm tot}$ is the total decay width of resonance, $J$ is a spin of resonance, 
$S_{\rm BW} = 1$ for different particles and $S_{\rm BW} = 2$ for identical particles in the initial state, $S_i$ and $S_j$ are spins of initial particles. In Eq.~(\ref{eq:BWformiula}) we denote:
\begin{equation}\label{eq:defNoteD}
    \DBWformula^2(s)  
    = 
        (s - m_{\phi}^2)^2  
        +   
        m_{\phi}^2 \left(\Gamma_\phi^{\rm tot}(m_{\phi}^2)\right)^2.
\end{equation}
The total cross sections for the considered benchmarks are:
\begin{multline}
    \sigma_{  \chi_1 \chi_2 \to \phi \to l^+l^-}(s)
=
    \frac{(\text{Re}[ \lambda^{\phi}_{\chi_1 \chi_2}])^2 (c_{ll}^{\phi})^2}{16 \pi \DBWformula^2(s) } 
    s 
\\ \cdot
    \beta^3(4 m_l^2, s)
    \beta(s_0, s) 
    \beta^{-1}(\delta_0, s), 
\end{multline}
\begin{multline}
    \sigma_{  \chi_1 \chi_2 \to \phi \to l^+l^-}(s)
=
    \frac{(\text{Im}[ \lambda^{\phi}_{\chi_1 \chi_2}])^2 (c_{ll}^{\phi})^2}{16 \pi \DBWformula^2(s) } 
    \sqrt{s}
\\ \cdot
    \beta^3(4 m_l^2, s)
    \beta^{-1}(s_0, s) 
    \beta^{-1}(\delta_0, s), 
\end{multline}
In particular, non-relativistic leading terms of total cross sections are:
\begin{multline}
    \sigma_{  \chi_1 \chi_2 \to \phi \to l^+l^-}^{\rm low~vel.}(v_{-})
= 
    \frac{(\text{Re}[ \lambda^{\phi}_{\chi_1 \chi_2}])^2 (c_{ll}^{\phi})^2}{32 \pi \DBWformula^2(s_0) } 
\\ \cdot
    s_0 \beta^3(4 m_l^2, s_0)  v_{-},
\end{multline}
\begin{multline}
    \sigma_{  \chi_1 \chi_2 \to \phi \to l^+l^- }^{\rm low~vel.}(v_{-})
= 
    \frac{(\text{Im}[ \lambda^{\phi}_{\chi_1 \chi_2}])^2 (c_{ll}^{\phi})^2}{8 \pi \DBWformula^2(s_0) }
\\ \cdot
     s_0  \beta^3(4 m_l^2, s_0) v_{-}^{-1}.
\end{multline}
This means that co-annihilation for the benchmark scenarios (\ref{eq:EffLagrangianScalarMEDMajoranaiCDM}) and (\ref{eq:EffLagrangianScalarMEDMajoranaG5iCDM}) implies a p-wave and s-wave channel, respectively~\cite{Hooper:2018kfv}.

\section{Thermally averaged cross section in low-velocity approach
\label{sectionAveraging}}

In this section, we  discuss the thermally averaged cross section in the non-relativistic approach. 
In particular,  the expression of the thermally averaged cross section~\eqref{eq:ThermalAveregedCS} in non-relativistic limit and the center-of-mass system is~\cite{Choi:2017mkk}:
\begin{equation}\label{eq:ThermalAveregedCSforDefNonRel}
    \Avrg{\sigma_{ij} v_{ij}}_{\rm non. rel.}
=
    \frac{\int \sigma_{ij} v_{ij} e^{-\sqrt{s}/T}  \delta(\vect{V}) d\vect{v}_i d \vect{v}_j}
         {\int e^{-\sqrt{s}/T} \delta(\vect{V}) d \vect{v}_i d\vect{v}_j},
\end{equation}
where~$\vect{V} = ( m_{\chi_i} \vect{v}_i +  m_{\chi_j} \vect{v}_j) / ( m_{\chi_i} +  m_{\chi_j})$ is the center-of-mass velocity. 
In general, the Mandelstam variable,~$s~=~(p_i~+~p_j)^2$, expanded in terms of the relative velocity,~$\vect{v}_{-}~=~\vect{v}_i~-~\vect{v}_j$, reads as:
\begin{equation}\label{eq:sVar}
    s  \simeq s_0 +  m_{\chi_i}  m_{\chi_j} v_{-}^2,
\quad
    \sqrt{s}  \simeq \sqrt{s_0} + (\mu_{ij} / 2) v_{-}^2,
\end{equation}
where~$s_0~=~( m_{\chi_i} +  m_{\chi_j})^2$, $v_{-}~=~|\vect{v}_i - \vect{v}_j|$ and~$\mu_{ij}~=~ m_{\chi_i}  m_{\chi_j} / \sqrt{s_0}$ is the reduced mass.
Accounting~$v_{ij}~\approx~v_{-}$ and~$d \vect{v}_i d\vect{v}_j~=~ d \vect{V} d \vect{v}_{-}$, the low-velocity limit of the thermally averaged cross section takes the following form:
\begin{multline}\label{eq:ThermalAveregedCSNonRel}
    \Avrg{\sigma_{ij} v_{ij}}_{\rm non. rel.}
=
    \frac{x^{3/2}}{2 \sqrt{\pi}}
    \left( \frac{2 \mu_{ij} }{m_{\LSDM}} \right)^{3/2}
\\ \cdot
    \int\limits_{0}^{\infty} \sigma(v_{-}) v_{-}^3 \exp\left[- \frac{\mu_{ij} }{2 m_{\LSDM}} x v_{-}^2 \right]
    d v_{-}.
\end{multline}
Thus, in order to obtain a non-relativistic expansion of the total cross section for the corresponding process, one can use the substitution~$(s~-~s_0)/( m_{\chi_i}  m_{\chi_j})~\to~v_{-}^2$.

By using the non-relativistic expansion for the effective thermally averaged o
cross section~\eqref{eq:ThermalAveregedCSNonRel} in the  low-velocity approach as~$\sigma(v_{-}) v_{-}~=~\sum_{k=0}^{\infty} a_k v_{-}^{2k}/k!$, one can get:
\begin{multline}\label{eq:LowVelocityExpan}
    \Avrg{\sigma_{ij} v_{ij}}_{\rm non. rel.}
= 
    \frac{4}{2 \sqrt{\pi}}
    \sum\limits_{k=0}^{\infty}
    b^{k}
    \Gamma(k + 3/2)
\;
    \frac{a_k}{k!}
    x^{-k}
\approx \\  \approx
    a_0 + (3/2) b a_1 x^{-1} + (15/8) b^2 a_2 x^{-2},
\end{multline}
where $b = \left(2 m_{\LSDM} / \mu_{ij}   \right)$, $a_k$ are the expansion coefficients of cross section for the low-velocity series for each channel and $\Gamma(k)$ is the Gamma function. 

Moreover, for $m_{\chi_1}=m_{\chi_2}$ equation~\eqref{eq:LowVelocityExpan} reduces to the well-known expansion of the thermally averaged cross section~\cite{Gondolo:1990dk,Wells:1994qy,Choi:2017mkk}.
It is worth mentioning that $n = 0$, $n = 1$ and $n = 2$  correspond to the  s-wave, p-wave and 
d-wave annihilations, respectively~\cite{Kolb:1990vq}.
Thus, explicit analytical integrated expressions for the relic density can be obtained by considering a first non-zero term in the low-velocity approach~\eqref{eq:LowVelocityExpan} as $\Avrg{\sigma_{ij} v_{ij}}_{\rm non. rel.} = \sigma_{ij}^0 x^{-n}$.

In the case of~$T~\lesssim~m_{\chi_k}$  the particle net density in equilibrium reads~\cite{Ellis:1999mm}:
\begin{equation}\label{eq:PartDensDMapproxCDM}
    n_{k}^{\rm eq}
\approx
    g_k \left(\frac{m_{\chi_k} T}{2 \pi} \right)^{3/2}
    e^{-m_{\chi_k}/T}
    \left( 1 + \frac{15 T}{8 m_{\chi_k}} + \dots \right).
\end{equation}
Thus, one can estimate~\cite{Griest:1990kh}:
\begin{equation}\label{eq:RatPartDensDMapproxCDM}
    \frac{n_i^{\rm eq}}{n_{\rm eq}}
=
    \frac{ g_i (1 + \Delta_i)^{3/2}
           e^{- x \Delta_i }}
         { \sum_{k=1}^{2} g_k (1 + \Delta_k)^{3/2} e^{- x \Delta_k } }
\sim
    e^{- x \Delta_i },
\end{equation}
where~$\Delta_i~=~(m_{\chi_i} - m_{\LSDM})/m_{\LSDM}$. 

As result, in the approach of the approximation for density of DM in equilibrium~\eqref{eq:PartDensDMapproxCDM} and the low-velocity approach~\eqref{eq:LowVelocityExpan}  one can see that effective thermally averaged cross section~\eqref{eq:DefEffThermalCS} takes the following form~\cite{Griest:1990kh}:
 
\begin{multline}\label{eq:NonRelAvEffCS}
    \Avrg{\sigma_{\rm eff} v}
=   
    \sum\limits_{f} \!
    \sum\limits_{ i,j = 1 }^{2} 
    \sigma_{ij}^0 x^{-n}
    \frac{n_i^{\rm eq}n_j^{\rm eq}}{n_{\rm eq}^2}
=   
    \sum\limits_{f} \!
    \sum\limits_{ i,j = 1 }^{2} 
     \sigma_{ij}^0  x^{-n} 
\\ \cdot
        \frac{ 
           g_i g_j (1 + \Delta_i)^{3/2} (1 + \Delta_j)^{3/2}
           e^{ - x (\Delta_i + \Delta_j) }}
         { \left( \sum_{l=1}^{2} g_l (1 + \Delta_l)^{3/2} e^{- x \Delta_l } \right)^2 }.
\end{multline}

It is also worth noticing that heavier component 
the hidden sector can decay into the lightest mass-state particle. The contribution 
of these decays to the DM relic density is expected to be 
negligible~\cite{Griest:1990kh}. This also holds for the up-scattering 
cross section $\chi_i l \to \chi_j l$ 
(see e.~g.~Ref.~\cite{Griest:1990kh} and references therein). It means that  
co-anihillation channel $\chi_1 \chi_2 \to l^+ l^- $ provides a dominant 
contribution  to the  observed DM net density.

\begin{figure}[!tbh]
\centering
\begin{tikzpicture}[scale=1.299]
  \begin{feynhand}
    \vertex (p') at (2.5, -2) {};
    \vertex (p) at (-1, -2) {};
    \vertex (k') at (2.5, 2) {};
    \vertex (k) at (-1, 2) {};
    \vertex  [dot] (Chi1Chi2Phi) at (0, 0.75) {};
    \vertex  [dot] (Chi2Chi1Phi) at (1.5, 0.75) {};
    \vertex  [dot] (p1p2Phi) at (0, -0.75) {};
    \vertex  [dot] (p2p1Phi) at (1.5, -0.75) {};
    \propag[fer] (p) to [edge label =$e^-$] (p1p2Phi);
    \propag[fer] (p1p2Phi) to [edge label' =$ e^- $] (p2p1Phi);
    \propag[fer] (p2p1Phi) to [edge label =$e^-$] (p');
    \propag[sca] (p1p2Phi) to [edge label =$\phi$] (Chi1Chi2Phi);
    \propag[sca] (p2p1Phi) to [edge label' =$\phi$] (Chi2Chi1Phi);
    \propag[fer] (k) to [edge label' =$\chi_1$] (Chi1Chi2Phi);
    \propag[fer] (Chi1Chi2Phi) to [edge label =$ \chi_2 $] (Chi2Chi1Phi);
    \propag[fer] (Chi2Chi1Phi) to [edge label' =$\chi_1$] (k');
  \end{feynhand}
\end{tikzpicture}
\caption{A Feynman diagram responsible for the one-loop coupling of DM to electrons. We don't show crossed diagram.
\label{Fig1loopDD}}
\end{figure}

 \section{Direct Detection
 \label{sec:DD}}

    For the model under consideration, the inelastic scattering of the lightest state off electrons, 
    $\chi_1 e \to \chi_2 e$, is kinematically suppressed for relatively large mass splittings, 
    $m_{\chi_1}\Delta \gtrsim 1 - 100~\mathrm{keV}$, leading to a weak signal in direct-
    detection experiments (see e.g. Refs.~\cite{Harigaya:2020ckz,Wang:2025uwh} and references 
    therein). On the other hand, the elastic scattering $\chi_1 e^- \to \chi_1 e^-$ of DM
     can be induced at the one-loop level via the scalar exchange 
    depicted in Fig.~\ref{Fig1loopDD}.

The one-loop contribution to the elastic scattering $\chi_1 f \to \chi_1 f$ of a SM fermion $f$, 
mediated by a light hidden vector, is discussed in 
Ref.~\cite{Weiner:2012cb,Batell:2009vb,Berlin:2018jbm}. Unlike the inelastic channel, these 
processes are not kinematically suppressed for large mass splittings, but they are suppressed 
due to the  leading non-vanishing quantum correction.

    Specifically, the effective low-energy Lagrangian in this case is
    $\mathcal{L}_{\rm EFT} \supset C^{(0)}_e \overline{\chi}_1 \chi_1 m_e \overline{e} e$.
    The spin-independent scattering cross section is then given by
    \begin{equation}
    \sigma_{\rm SI}^{\rm el} = \frac{4}{\pi} (\mu_{e \chi_1})^2 |\mathcal{M}_{e \chi_1}|^2,
    \end{equation}
    where $\mu_{e \chi_1} = m_e m_{\chi_1}/(m_e+m_{\chi_1})$ is the DM-electron reduced mass 
    and $\mathcal{M}_{e \chi_1}$ is the spin-independent amplitude for electron scattering. 
    This amplitude reads
    \begin{equation}
    \mathcal{M}_{e \chi_1} = m_e C^{(0)}_e.
    \end{equation}
    We estimate the direct detection sensitivity for the one-loop scalar exchange by adapting 
    the results of Ref.~\cite{Berlin:2018jbm} for a vector mediator. For a sufficiently heavy 
    mediator, $m_\phi \gg m_{\chi_1} \gg m_e$, the Wilson coefficient $C^{(0)}_e$ typically 
    scales with the model parameters as~\cite{Berlin:2018jbm}
    \begin{equation}
    C^{(0)}_e \propto \frac{(c_{ee}^\phi )^2 \, \alpha_{\rm iDM} \, m_{\chi_1} }{4\pi (m_\phi)^4}.
    \label{WilsonCoeff1}
    \end{equation}
    Although a detailed calculation of this Wilson coefficient is beyond the scope of this 
    work, we do not expect these estimates to change our final scalar results by more than an order      of magnitude.

    As a result, the DM scattering cross section is estimated to be
    \begin{equation}
    \sigma_{\rm SI}^{\rm el} \propto \frac{1}{4 \pi^3} (c_{ee}^\phi)^4 (\alpha_{\rm iDM})^2 \frac{m_e^4 m_{\chi_1}^2}{(m_{\phi})^8}.
    \label{SI_CS_DD_electron}
    \end{equation}
    We compare the predicted DM-electron elastic scattering cross section from 
    Eq.~(\ref{SI_CS_DD_electron}) to the limit from XENON1T (2019) data, 
    $\sigma_{\rm SI}^{\rm el} \lesssim 10^{-40}~\mbox{cm}^2$ for 
    $m_{\chi} \simeq 100~\mbox{MeV}$ (see, e.g., Fig.~5.12 of Ref.~\cite{Cirelli:2024ssz}).
    This comparison yields a typical lower limit of $c_{ee}^\phi \gtrsim 1.9$, which is 
    already ruled out by the BaBar experiment. Hence, current direct detection limits are not 
    competitive with those from electron-positron colliders for the sub-GeV mass range.

\bibliography{bibl}

\begin{thebibliography}{126}%
\makeatletter
\providecommand \@ifxundefined [1]{%
 \@ifx{#1\undefined}
}%
\providecommand \@ifnum [1]{%
 \ifnum #1\expandafter \@firstoftwo
 \else \expandafter \@secondoftwo
 \fi
}%
\providecommand \@ifx [1]{%
 \ifx #1\expandafter \@firstoftwo
 \else \expandafter \@secondoftwo
 \fi
}%
\providecommand \natexlab [1]{#1}%
\providecommand \enquote  [1]{``#1''}%
\providecommand \bibnamefont  [1]{#1}%
\providecommand \bibfnamefont [1]{#1}%
\providecommand \citenamefont [1]{#1}%
\providecommand \href@noop [0]{\@secondoftwo}%
\providecommand \href [0]{\begingroup \@sanitize@url \@href}%
\providecommand \@href[1]{\@@startlink{#1}\@@href}%
\providecommand \@@href[1]{\endgroup#1\@@endlink}%
\providecommand \@sanitize@url [0]{\catcode `\\12\catcode `\$12\catcode `\&12\catcode `\#12\catcode `\^12\catcode `\_12\catcode `\%12\relax}%
\providecommand \@@startlink[1]{}%
\providecommand \@@endlink[0]{}%
\providecommand \url  [0]{\begingroup\@sanitize@url \@url }%
\providecommand \@url [1]{\endgroup\@href {#1}{\urlprefix }}%
\providecommand \urlprefix  [0]{URL }%
\providecommand \Eprint [0]{\href }%
\providecommand \doibase [0]{http://dx.doi.org/}%
\providecommand \selectlanguage [0]{\@gobble}%
\providecommand \bibinfo  [0]{\@secondoftwo}%
\providecommand \bibfield  [0]{\@secondoftwo}%
\providecommand \translation [1]{[#1]}%
\providecommand \BibitemOpen [0]{}%
\providecommand \bibitemStop [0]{}%
\providecommand \bibitemNoStop [0]{.\EOS\space}%
\providecommand \EOS [0]{\spacefactor3000\relax}%
\providecommand \BibitemShut  [1]{\csname bibitem#1\endcsname}%
\let\auto@bib@innerbib\@empty
\bibitem [{\citenamefont {Bergstrom}(2012)}]{Bergstrom:2012fi}%
  \BibitemOpen
  \bibfield  {author} {\bibinfo {author} {\bibfnamefont {Lars}\ \bibnamefont {Bergstrom}},\ }\bibfield  {title} {\enquote {\bibinfo {title} {{Dark Matter Evidence, Particle Physics Candidates and Detection Methods}},}\ }\href {\doibase 10.1002/andp.201200116} {\bibfield  {journal} {\bibinfo  {journal} {Annalen Phys.}\ }\textbf {\bibinfo {volume} {524}},\ \bibinfo {pages} {479--496} (\bibinfo {year} {2012})},\ \Eprint {http://arxiv.org/abs/1205.4882} {arXiv:1205.4882 [astro-ph.HE]} \BibitemShut {NoStop}%
\bibitem [{\citenamefont {Bertone}\ and\ \citenamefont {Hooper}(2018)}]{Bertone:2016nfn}%
  \BibitemOpen
  \bibfield  {author} {\bibinfo {author} {\bibfnamefont {Gianfranco}\ \bibnamefont {Bertone}}\ and\ \bibinfo {author} {\bibfnamefont {Dan}\ \bibnamefont {Hooper}},\ }\bibfield  {title} {\enquote {\bibinfo {title} {{History of dark matter}},}\ }\href {\doibase 10.1103/RevModPhys.90.045002} {\bibfield  {journal} {\bibinfo  {journal} {Rev. Mod. Phys.}\ }\textbf {\bibinfo {volume} {90}},\ \bibinfo {pages} {045002} (\bibinfo {year} {2018})},\ \Eprint {http://arxiv.org/abs/1605.04909} {arXiv:1605.04909 [astro-ph.CO]} \BibitemShut {NoStop}%
\bibitem [{\citenamefont {Cirelli}\ \emph {et~al.}(2024)\citenamefont {Cirelli}, \citenamefont {Strumia},\ and\ \citenamefont {Zupan}}]{Cirelli:2024ssz}%
  \BibitemOpen
  \bibfield  {author} {\bibinfo {author} {\bibfnamefont {Marco}\ \bibnamefont {Cirelli}}, \bibinfo {author} {\bibfnamefont {Alessandro}\ \bibnamefont {Strumia}}, \ and\ \bibinfo {author} {\bibfnamefont {Jure}\ \bibnamefont {Zupan}},\ }\bibfield  {title} {\enquote {\bibinfo {title} {{Dark Matter}},}\ }\href@noop {} {\  (\bibinfo {year} {2024})},\ \Eprint {http://arxiv.org/abs/2406.01705} {arXiv:2406.01705 [hep-ph]} \BibitemShut {NoStop}%
\bibitem [{\citenamefont {Bertone}\ \emph {et~al.}(2005)\citenamefont {Bertone}, \citenamefont {Hooper},\ and\ \citenamefont {Silk}}]{Bertone:2004pz}%
  \BibitemOpen
  \bibfield  {author} {\bibinfo {author} {\bibfnamefont {Gianfranco}\ \bibnamefont {Bertone}}, \bibinfo {author} {\bibfnamefont {Dan}\ \bibnamefont {Hooper}}, \ and\ \bibinfo {author} {\bibfnamefont {Joseph}\ \bibnamefont {Silk}},\ }\bibfield  {title} {\enquote {\bibinfo {title} {{Particle dark matter: Evidence, candidates and constraints}},}\ }\href {\doibase 10.1016/j.physrep.2004.08.031} {\bibfield  {journal} {\bibinfo  {journal} {Phys. Rept.}\ }\textbf {\bibinfo {volume} {405}},\ \bibinfo {pages} {279--390} (\bibinfo {year} {2005})},\ \Eprint {http://arxiv.org/abs/hep-ph/0404175} {arXiv:hep-ph/0404175} \BibitemShut {NoStop}%
\bibitem [{\citenamefont {Gelmini}(2015)}]{Gelmini:2015zpa}%
  \BibitemOpen
  \bibfield  {author} {\bibinfo {author} {\bibfnamefont {Graciela~B.}\ \bibnamefont {Gelmini}},\ }\bibfield  {title} {\enquote {\bibinfo {title} {{The hunt for dark matter.}}}\ }in\ \href {\doibase 10.1142/9789814678766_0012} {\emph {\bibinfo {booktitle} {{Theoretical Advanced Study Institute in Elementary Particle Physics}: {Journeys Through the Precision Frontier: Amplitudes for Colliders}}}}\ (\bibinfo {year} {2015})\ pp.\ \bibinfo {pages} {559--616},\ \Eprint {http://arxiv.org/abs/1502.01320} {arXiv:1502.01320 [hep-ph]} \BibitemShut {NoStop}%
\bibitem [{\citenamefont {Ade}\ \emph {et~al.}(2016)\citenamefont {Ade} \emph {et~al.}}]{Planck:2015fie}%
  \BibitemOpen
  \bibfield  {author} {\bibinfo {author} {\bibfnamefont {P.~A.~R.}\ \bibnamefont {Ade}} \emph {et~al.} (\bibinfo {collaboration} {Planck}),\ }\bibfield  {title} {\enquote {\bibinfo {title} {{Planck 2015 results. XIII. Cosmological parameters}},}\ }\href {\doibase 10.1051/0004-6361/201525830} {\bibfield  {journal} {\bibinfo  {journal} {Astron. Astrophys.}\ }\textbf {\bibinfo {volume} {594}},\ \bibinfo {pages} {A13} (\bibinfo {year} {2016})},\ \Eprint {http://arxiv.org/abs/1502.01589} {arXiv:1502.01589 [astro-ph.CO]} \BibitemShut {NoStop}%
\bibitem [{\citenamefont {Aghanim}\ \emph {et~al.}(2020)\citenamefont {Aghanim} \emph {et~al.}}]{Planck:2018vyg}%
  \BibitemOpen
  \bibfield  {author} {\bibinfo {author} {\bibfnamefont {N.}~\bibnamefont {Aghanim}} \emph {et~al.} (\bibinfo {collaboration} {Planck}),\ }\bibfield  {title} {\enquote {\bibinfo {title} {{Planck 2018 results. VI. Cosmological parameters}},}\ }\href {\doibase 10.1051/0004-6361/201833910} {\bibfield  {journal} {\bibinfo  {journal} {Astron. Astrophys.}\ }\textbf {\bibinfo {volume} {641}},\ \bibinfo {pages} {A6} (\bibinfo {year} {2020})},\ \bibinfo {note} {[Erratum: Astron.Astrophys. 652, C4 (2021)]},\ \Eprint {http://arxiv.org/abs/1807.06209} {arXiv:1807.06209 [astro-ph.CO]} \BibitemShut {NoStop}%
\bibitem [{\citenamefont {Ponten}\ \emph {et~al.}(2024)\citenamefont {Ponten}, \citenamefont {Sieber}, \citenamefont {Oberhauser}, \citenamefont {Crivelli}, \citenamefont {Kirpichnikov}, \citenamefont {Gninenko}, \citenamefont {H\"osgen}, \citenamefont {Bueno}, \citenamefont {Mongillo},\ and\ \citenamefont {Zhevlakov}}]{Ponten:2024grp}%
  \BibitemOpen
  \bibfield  {author} {\bibinfo {author} {\bibfnamefont {A.}~\bibnamefont {Ponten}}, \bibinfo {author} {\bibfnamefont {H.}~\bibnamefont {Sieber}}, \bibinfo {author} {\bibfnamefont {B.~Banto}\ \bibnamefont {Oberhauser}}, \bibinfo {author} {\bibfnamefont {P.}~\bibnamefont {Crivelli}}, \bibinfo {author} {\bibfnamefont {D.}~\bibnamefont {Kirpichnikov}}, \bibinfo {author} {\bibfnamefont {S.~N.}\ \bibnamefont {Gninenko}}, \bibinfo {author} {\bibfnamefont {M.}~\bibnamefont {H\"osgen}}, \bibinfo {author} {\bibfnamefont {L.~Molina}\ \bibnamefont {Bueno}}, \bibinfo {author} {\bibfnamefont {M.}~\bibnamefont {Mongillo}}, \ and\ \bibinfo {author} {\bibfnamefont {A.}~\bibnamefont {Zhevlakov}},\ }\bibfield  {title} {\enquote {\bibinfo {title} {{Probing hidden leptonic scalar portals using the NA64 experiment at CERN}},}\ }\href {\doibase 10.1140/epjc/s10052-024-13421-1} {\bibfield  {journal} {\bibinfo  {journal} {Eur. Phys. J. C}\ }\textbf {\bibinfo {volume} {84}},\ \bibinfo {pages} {1035} (\bibinfo {year} {2024})},\
  \Eprint {http://arxiv.org/abs/2404.15931} {arXiv:2404.15931 [hep-ph]} \BibitemShut {NoStop}%
\bibitem [{\citenamefont {Kolay}\ and\ \citenamefont {Nandi}(2025)}]{Kolay:2025jip}%
  \BibitemOpen
  \bibfield  {author} {\bibinfo {author} {\bibfnamefont {Lipika}\ \bibnamefont {Kolay}}\ and\ \bibinfo {author} {\bibfnamefont {Soumitra}\ \bibnamefont {Nandi}},\ }\bibfield  {title} {\enquote {\bibinfo {title} {{Flavour and Electroweak Precision Constraints on a Simplified Dark Matter Model with a Light Spin-0 Mediator}},}\ }\href@noop {} {\  (\bibinfo {year} {2025})},\ \Eprint {http://arxiv.org/abs/2503.15609} {arXiv:2503.15609 [hep-ph]} \BibitemShut {NoStop}%
\bibitem [{\citenamefont {Kolay}\ and\ \citenamefont {Nandi}(2024)}]{Kolay:2024wns}%
  \BibitemOpen
  \bibfield  {author} {\bibinfo {author} {\bibfnamefont {Lipika}\ \bibnamefont {Kolay}}\ and\ \bibinfo {author} {\bibfnamefont {Soumitra}\ \bibnamefont {Nandi}},\ }\bibfield  {title} {\enquote {\bibinfo {title} {{Exploring constraints on Simplified Dark Matter model through flavour and electroweak observables}},}\ }\href {\doibase 10.1007/JHEP10(2024)008} {\bibfield  {journal} {\bibinfo  {journal} {JHEP}\ }\textbf {\bibinfo {volume} {10}},\ \bibinfo {pages} {008} (\bibinfo {year} {2024})},\ \Eprint {http://arxiv.org/abs/2403.20303} {arXiv:2403.20303 [hep-ph]} \BibitemShut {NoStop}%
\bibitem [{\citenamefont {McDonald}(1994)}]{McDonald:1993ex}%
  \BibitemOpen
  \bibfield  {author} {\bibinfo {author} {\bibfnamefont {John}\ \bibnamefont {McDonald}},\ }\bibfield  {title} {\enquote {\bibinfo {title} {{Gauge singlet scalars as cold dark matter}},}\ }\href {\doibase 10.1103/PhysRevD.50.3637} {\bibfield  {journal} {\bibinfo  {journal} {Phys. Rev. D}\ }\textbf {\bibinfo {volume} {50}},\ \bibinfo {pages} {3637--3649} (\bibinfo {year} {1994})},\ \Eprint {http://arxiv.org/abs/hep-ph/0702143} {arXiv:hep-ph/0702143} \BibitemShut {NoStop}%
\bibitem [{\citenamefont {Burgess}\ \emph {et~al.}(2001)\citenamefont {Burgess}, \citenamefont {Pospelov},\ and\ \citenamefont {ter Veldhuis}}]{Burgess:2000yq}%
  \BibitemOpen
  \bibfield  {author} {\bibinfo {author} {\bibfnamefont {C.~P.}\ \bibnamefont {Burgess}}, \bibinfo {author} {\bibfnamefont {Maxim}\ \bibnamefont {Pospelov}}, \ and\ \bibinfo {author} {\bibfnamefont {Tonnis}\ \bibnamefont {ter Veldhuis}},\ }\bibfield  {title} {\enquote {\bibinfo {title} {{The Minimal model of nonbaryonic dark matter: A Singlet scalar}},}\ }\href {\doibase 10.1016/S0550-3213(01)00513-2} {\bibfield  {journal} {\bibinfo  {journal} {Nucl. Phys. B}\ }\textbf {\bibinfo {volume} {619}},\ \bibinfo {pages} {709--728} (\bibinfo {year} {2001})},\ \Eprint {http://arxiv.org/abs/hep-ph/0011335} {arXiv:hep-ph/0011335} \BibitemShut {NoStop}%
\bibitem [{\citenamefont {Wells}(2008)}]{Wells:2008xg}%
  \BibitemOpen
  \bibfield  {author} {\bibinfo {author} {\bibfnamefont {James~D.}\ \bibnamefont {Wells}},\ }\bibfield  {title} {\enquote {\bibinfo {title} {{How to Find a Hidden World at the Large Hadron Collider}},}\ }\href@noop {} {\bibfield  {journal} {\bibinfo  {journal} {Perspectives on LHC Physics}\ ,\ \bibinfo {pages} {283--298}} (\bibinfo {year} {2008})},\ \Eprint {http://arxiv.org/abs/0803.1243} {arXiv:0803.1243 [hep-ph]} \BibitemShut {NoStop}%
\bibitem [{\citenamefont {Schabinger}\ and\ \citenamefont {Wells}(2005)}]{Schabinger:2005ei}%
  \BibitemOpen
  \bibfield  {author} {\bibinfo {author} {\bibfnamefont {Robert~M.}\ \bibnamefont {Schabinger}}\ and\ \bibinfo {author} {\bibfnamefont {James~D.}\ \bibnamefont {Wells}},\ }\bibfield  {title} {\enquote {\bibinfo {title} {{A Minimal spontaneously broken hidden sector and its impact on Higgs boson physics at the large hadron collider}},}\ }\href {\doibase 10.1103/PhysRevD.72.093007} {\bibfield  {journal} {\bibinfo  {journal} {Phys. Rev. D}\ }\textbf {\bibinfo {volume} {72}},\ \bibinfo {pages} {093007} (\bibinfo {year} {2005})},\ \Eprint {http://arxiv.org/abs/hep-ph/0509209} {arXiv:hep-ph/0509209} \BibitemShut {NoStop}%
\bibitem [{\citenamefont {Bickendorf}\ and\ \citenamefont {Drees}(2022)}]{Bickendorf:2022buy}%
  \BibitemOpen
  \bibfield  {author} {\bibinfo {author} {\bibfnamefont {Gerrit}\ \bibnamefont {Bickendorf}}\ and\ \bibinfo {author} {\bibfnamefont {Manuel}\ \bibnamefont {Drees}},\ }\bibfield  {title} {\enquote {\bibinfo {title} {{Constraints on light leptophilic dark matter mediators from decay experiments}},}\ }\href {\doibase 10.1140/epjc/s10052-022-11128-9} {\bibfield  {journal} {\bibinfo  {journal} {Eur. Phys. J. C}\ }\textbf {\bibinfo {volume} {82}},\ \bibinfo {pages} {1163} (\bibinfo {year} {2022})},\ \Eprint {http://arxiv.org/abs/2206.05038} {arXiv:2206.05038 [hep-ph]} \BibitemShut {NoStop}%
\bibitem [{\citenamefont {Boos}\ \emph {et~al.}(2023)\citenamefont {Boos}, \citenamefont {Bunichev},\ and\ \citenamefont {Trykov}}]{Boos:2022gtt}%
  \BibitemOpen
  \bibfield  {author} {\bibinfo {author} {\bibfnamefont {E.~E.}\ \bibnamefont {Boos}}, \bibinfo {author} {\bibfnamefont {V.~E.}\ \bibnamefont {Bunichev}}, \ and\ \bibinfo {author} {\bibfnamefont {S.~S.}\ \bibnamefont {Trykov}},\ }\bibfield  {title} {\enquote {\bibinfo {title} {{Prospects for dark matter search at a super c-tau factory}},}\ }\href {\doibase 10.1103/PhysRevD.107.075021} {\bibfield  {journal} {\bibinfo  {journal} {Phys. Rev. D}\ }\textbf {\bibinfo {volume} {107}},\ \bibinfo {pages} {075021} (\bibinfo {year} {2023})},\ \Eprint {http://arxiv.org/abs/2205.07364} {arXiv:2205.07364 [hep-ph]} \BibitemShut {NoStop}%
\bibitem [{\citenamefont {Sieber}\ \emph {et~al.}(2023)\citenamefont {Sieber}, \citenamefont {Kirpichnikov}, \citenamefont {Voronchikhin}, \citenamefont {Crivelli}, \citenamefont {Gninenko}, \citenamefont {Kirsanov}, \citenamefont {Krasnikov}, \citenamefont {Molina-Bueno},\ and\ \citenamefont {Sekatskii}}]{Sieber:2023nkq}%
  \BibitemOpen
  \bibfield  {author} {\bibinfo {author} {\bibfnamefont {H.}~\bibnamefont {Sieber}}, \bibinfo {author} {\bibfnamefont {D.~V.}\ \bibnamefont {Kirpichnikov}}, \bibinfo {author} {\bibfnamefont {I.~V.}\ \bibnamefont {Voronchikhin}}, \bibinfo {author} {\bibfnamefont {P.}~\bibnamefont {Crivelli}}, \bibinfo {author} {\bibfnamefont {S.~N.}\ \bibnamefont {Gninenko}}, \bibinfo {author} {\bibfnamefont {M.~M.}\ \bibnamefont {Kirsanov}}, \bibinfo {author} {\bibfnamefont {N.~V.}\ \bibnamefont {Krasnikov}}, \bibinfo {author} {\bibfnamefont {L.}~\bibnamefont {Molina-Bueno}}, \ and\ \bibinfo {author} {\bibfnamefont {S.~K.}\ \bibnamefont {Sekatskii}},\ }\bibfield  {title} {\enquote {\bibinfo {title} {{Probing hidden sectors with a muon beam: Implication of spin-0 dark matter mediators for the muon (g-2) anomaly and the validity of the Weisz\"acker-Williams approach}},}\ }\href {\doibase 10.1103/PhysRevD.108.056018} {\bibfield  {journal} {\bibinfo  {journal} {Phys. Rev. D}\ }\textbf {\bibinfo {volume} {108}},\ \bibinfo {pages}
  {056018} (\bibinfo {year} {2023})},\ \Eprint {http://arxiv.org/abs/2305.09015} {arXiv:2305.09015 [hep-ph]} \BibitemShut {NoStop}%
\bibitem [{\citenamefont {Guo}\ \emph {et~al.}(2025)\citenamefont {Guo}, \citenamefont {Liu}, \citenamefont {Peng},\ and\ \citenamefont {Wang}}]{Guo:2025qes}%
  \BibitemOpen
  \bibfield  {author} {\bibinfo {author} {\bibfnamefont {Jinhui}\ \bibnamefont {Guo}}, \bibinfo {author} {\bibfnamefont {Jia}\ \bibnamefont {Liu}}, \bibinfo {author} {\bibfnamefont {Chenhao}\ \bibnamefont {Peng}}, \ and\ \bibinfo {author} {\bibfnamefont {Xiao-Ping}\ \bibnamefont {Wang}},\ }\bibfield  {title} {\enquote {\bibinfo {title} {{Probing Purely Inelastic Scalar Dark Matter Across Colliders and Gravitational Wave Observatories}},}\ }\href@noop {} {\  (\bibinfo {year} {2025})},\ \Eprint {http://arxiv.org/abs/2508.13276} {arXiv:2508.13276 [hep-ph]} \BibitemShut {NoStop}%
\bibitem [{\citenamefont {Voronchikhin}\ and\ \citenamefont {Kirpichnikov}(2024{\natexlab{a}})}]{Voronchikhin:2023qig}%
  \BibitemOpen
  \bibfield  {author} {\bibinfo {author} {\bibfnamefont {I.~V.}\ \bibnamefont {Voronchikhin}}\ and\ \bibinfo {author} {\bibfnamefont {D.~V.}\ \bibnamefont {Kirpichnikov}},\ }\bibfield  {title} {\enquote {\bibinfo {title} {{Probing scalar, Dirac, Majorana, and vector dark matter through a spin-0 electron-specific mediator at electron fixed-target experiments}},}\ }\href {\doibase 10.1103/PhysRevD.109.075012} {\bibfield  {journal} {\bibinfo  {journal} {Phys. Rev. D}\ }\textbf {\bibinfo {volume} {109}},\ \bibinfo {pages} {075012} (\bibinfo {year} {2024}{\natexlab{a}})},\ \Eprint {http://arxiv.org/abs/2312.15697} {arXiv:2312.15697 [hep-ph]} \BibitemShut {NoStop}%
\bibitem [{\citenamefont {Catena}\ and\ \citenamefont {Gray}(2023)}]{Catena:2023use}%
  \BibitemOpen
  \bibfield  {author} {\bibinfo {author} {\bibfnamefont {Riccardo}\ \bibnamefont {Catena}}\ and\ \bibinfo {author} {\bibfnamefont {Taylor~R.}\ \bibnamefont {Gray}},\ }\bibfield  {title} {\enquote {\bibinfo {title} {{Spin-1 thermal targets for dark matter searches at beam dump and fixed target experiments}},}\ }\href {\doibase 10.1088/1475-7516/2023/11/058} {\bibfield  {journal} {\bibinfo  {journal} {JCAP}\ }\textbf {\bibinfo {volume} {11}},\ \bibinfo {pages} {058} (\bibinfo {year} {2023})},\ \Eprint {http://arxiv.org/abs/2307.02207} {arXiv:2307.02207 [hep-ph]} \BibitemShut {NoStop}%
\bibitem [{\citenamefont {Holdom}(1986)}]{Holdom:1985ag}%
  \BibitemOpen
  \bibfield  {author} {\bibinfo {author} {\bibfnamefont {Bob}\ \bibnamefont {Holdom}},\ }\bibfield  {title} {\enquote {\bibinfo {title} {{Two U(1)'s and Epsilon Charge Shifts}},}\ }\href {\doibase 10.1016/0370-2693(86)91377-8} {\bibfield  {journal} {\bibinfo  {journal} {Phys. Lett. B}\ }\textbf {\bibinfo {volume} {166}},\ \bibinfo {pages} {196--198} (\bibinfo {year} {1986})}\BibitemShut {NoStop}%
\bibitem [{\citenamefont {Izaguirre}\ \emph {et~al.}(2015)\citenamefont {Izaguirre}, \citenamefont {Krnjaic}, \citenamefont {Schuster},\ and\ \citenamefont {Toro}}]{Izaguirre:2015yja}%
  \BibitemOpen
  \bibfield  {author} {\bibinfo {author} {\bibfnamefont {Eder}\ \bibnamefont {Izaguirre}}, \bibinfo {author} {\bibfnamefont {Gordan}\ \bibnamefont {Krnjaic}}, \bibinfo {author} {\bibfnamefont {Philip}\ \bibnamefont {Schuster}}, \ and\ \bibinfo {author} {\bibfnamefont {Natalia}\ \bibnamefont {Toro}},\ }\bibfield  {title} {\enquote {\bibinfo {title} {{Analyzing the Discovery Potential for Light Dark Matter}},}\ }\href {\doibase 10.1103/PhysRevLett.115.251301} {\bibfield  {journal} {\bibinfo  {journal} {Phys. Rev. Lett.}\ }\textbf {\bibinfo {volume} {115}},\ \bibinfo {pages} {251301} (\bibinfo {year} {2015})},\ \Eprint {http://arxiv.org/abs/1505.00011} {arXiv:1505.00011 [hep-ph]} \BibitemShut {NoStop}%
\bibitem [{\citenamefont {Essig}\ \emph {et~al.}(2011)\citenamefont {Essig}, \citenamefont {Schuster}, \citenamefont {Toro},\ and\ \citenamefont {Wojtsekhowski}}]{Essig:2010xa}%
  \BibitemOpen
  \bibfield  {author} {\bibinfo {author} {\bibfnamefont {Rouven}\ \bibnamefont {Essig}}, \bibinfo {author} {\bibfnamefont {Philip}\ \bibnamefont {Schuster}}, \bibinfo {author} {\bibfnamefont {Natalia}\ \bibnamefont {Toro}}, \ and\ \bibinfo {author} {\bibfnamefont {Bogdan}\ \bibnamefont {Wojtsekhowski}},\ }\bibfield  {title} {\enquote {\bibinfo {title} {{An Electron Fixed Target Experiment to Search for a New Vector Boson A' Decaying to e+e-}},}\ }\href {\doibase 10.1007/JHEP02(2011)009} {\bibfield  {journal} {\bibinfo  {journal} {JHEP}\ }\textbf {\bibinfo {volume} {02}},\ \bibinfo {pages} {009} (\bibinfo {year} {2011})},\ \Eprint {http://arxiv.org/abs/1001.2557} {arXiv:1001.2557 [hep-ph]} \BibitemShut {NoStop}%
\bibitem [{\citenamefont {Kahn}\ \emph {et~al.}(2015)\citenamefont {Kahn}, \citenamefont {Krnjaic}, \citenamefont {Thaler},\ and\ \citenamefont {Toups}}]{Kahn:2014sra}%
  \BibitemOpen
  \bibfield  {author} {\bibinfo {author} {\bibfnamefont {Yonatan}\ \bibnamefont {Kahn}}, \bibinfo {author} {\bibfnamefont {Gordan}\ \bibnamefont {Krnjaic}}, \bibinfo {author} {\bibfnamefont {Jesse}\ \bibnamefont {Thaler}}, \ and\ \bibinfo {author} {\bibfnamefont {Matthew}\ \bibnamefont {Toups}},\ }\bibfield  {title} {\enquote {\bibinfo {title} {{DAE\ensuremath{\delta}ALUS and dark matter detection}},}\ }\href {\doibase 10.1103/PhysRevD.91.055006} {\bibfield  {journal} {\bibinfo  {journal} {Phys. Rev. D}\ }\textbf {\bibinfo {volume} {91}},\ \bibinfo {pages} {055006} (\bibinfo {year} {2015})},\ \Eprint {http://arxiv.org/abs/1411.1055} {arXiv:1411.1055 [hep-ph]} \BibitemShut {NoStop}%
\bibitem [{\citenamefont {Batell}\ \emph {et~al.}(2014)\citenamefont {Batell}, \citenamefont {Essig},\ and\ \citenamefont {Surujon}}]{Batell:2014mga}%
  \BibitemOpen
  \bibfield  {author} {\bibinfo {author} {\bibfnamefont {Brian}\ \bibnamefont {Batell}}, \bibinfo {author} {\bibfnamefont {Rouven}\ \bibnamefont {Essig}}, \ and\ \bibinfo {author} {\bibfnamefont {Ze'ev}\ \bibnamefont {Surujon}},\ }\bibfield  {title} {\enquote {\bibinfo {title} {{Strong Constraints on Sub-GeV Dark Sectors from SLAC Beam Dump E137}},}\ }\href {\doibase 10.1103/PhysRevLett.113.171802} {\bibfield  {journal} {\bibinfo  {journal} {Phys. Rev. Lett.}\ }\textbf {\bibinfo {volume} {113}},\ \bibinfo {pages} {171802} (\bibinfo {year} {2014})},\ \Eprint {http://arxiv.org/abs/1406.2698} {arXiv:1406.2698 [hep-ph]} \BibitemShut {NoStop}%
\bibitem [{\citenamefont {Izaguirre}\ \emph {et~al.}(2013)\citenamefont {Izaguirre}, \citenamefont {Krnjaic}, \citenamefont {Schuster},\ and\ \citenamefont {Toro}}]{Izaguirre:2013uxa}%
  \BibitemOpen
  \bibfield  {author} {\bibinfo {author} {\bibfnamefont {Eder}\ \bibnamefont {Izaguirre}}, \bibinfo {author} {\bibfnamefont {Gordan}\ \bibnamefont {Krnjaic}}, \bibinfo {author} {\bibfnamefont {Philip}\ \bibnamefont {Schuster}}, \ and\ \bibinfo {author} {\bibfnamefont {Natalia}\ \bibnamefont {Toro}},\ }\bibfield  {title} {\enquote {\bibinfo {title} {{New Electron Beam-Dump Experiments to Search for MeV to few-GeV Dark Matter}},}\ }\href {\doibase 10.1103/PhysRevD.88.114015} {\bibfield  {journal} {\bibinfo  {journal} {Phys. Rev. D}\ }\textbf {\bibinfo {volume} {88}},\ \bibinfo {pages} {114015} (\bibinfo {year} {2013})},\ \Eprint {http://arxiv.org/abs/1307.6554} {arXiv:1307.6554 [hep-ph]} \BibitemShut {NoStop}%
\bibitem [{\citenamefont {Kachanovich}\ \emph {et~al.}(2022)\citenamefont {Kachanovich}, \citenamefont {Kovalenko}, \citenamefont {Kuleshov}, \citenamefont {Lyubovitskij},\ and\ \citenamefont {Zhevlakov}}]{Kachanovich:2021eqa}%
  \BibitemOpen
  \bibfield  {author} {\bibinfo {author} {\bibfnamefont {Aliaksei}\ \bibnamefont {Kachanovich}}, \bibinfo {author} {\bibfnamefont {Sergey}\ \bibnamefont {Kovalenko}}, \bibinfo {author} {\bibfnamefont {Serguei}\ \bibnamefont {Kuleshov}}, \bibinfo {author} {\bibfnamefont {Valery~E.}\ \bibnamefont {Lyubovitskij}}, \ and\ \bibinfo {author} {\bibfnamefont {Alexey~S.}\ \bibnamefont {Zhevlakov}},\ }\bibfield  {title} {\enquote {\bibinfo {title} {{Lepton phenomenology of Stueckelberg portal to dark sector}},}\ }\href {\doibase 10.1103/PhysRevD.105.075004} {\bibfield  {journal} {\bibinfo  {journal} {Phys. Rev. D}\ }\textbf {\bibinfo {volume} {105}},\ \bibinfo {pages} {075004} (\bibinfo {year} {2022})},\ \Eprint {http://arxiv.org/abs/2111.12522} {arXiv:2111.12522 [hep-ph]} \BibitemShut {NoStop}%
\bibitem [{\citenamefont {Lyubovitskij}\ \emph {et~al.}(2023)\citenamefont {Lyubovitskij}, \citenamefont {Zhevlakov}, \citenamefont {Kachanovich},\ and\ \citenamefont {Kuleshov}}]{Lyubovitskij:2022hna}%
  \BibitemOpen
  \bibfield  {author} {\bibinfo {author} {\bibfnamefont {Valery~E.}\ \bibnamefont {Lyubovitskij}}, \bibinfo {author} {\bibfnamefont {Alexey~S.}\ \bibnamefont {Zhevlakov}}, \bibinfo {author} {\bibfnamefont {Aliaksei}\ \bibnamefont {Kachanovich}}, \ and\ \bibinfo {author} {\bibfnamefont {Serguei}\ \bibnamefont {Kuleshov}},\ }\bibfield  {title} {\enquote {\bibinfo {title} {{Dark $SU(2)$ Stueckelberg portal}},}\ }\href {\doibase 10.1103/PhysRevD.107.055006} {\bibfield  {journal} {\bibinfo  {journal} {Phys. Rev. D}\ }\textbf {\bibinfo {volume} {107}},\ \bibinfo {pages} {055006} (\bibinfo {year} {2023})},\ \Eprint {http://arxiv.org/abs/2210.05555} {arXiv:2210.05555 [hep-ph]} \BibitemShut {NoStop}%
\bibitem [{\citenamefont {Gorbunov}\ and\ \citenamefont {Kalashnikov}(2023)}]{Gorbunov:2022dgw}%
  \BibitemOpen
  \bibfield  {author} {\bibinfo {author} {\bibfnamefont {Dmitry}\ \bibnamefont {Gorbunov}}\ and\ \bibinfo {author} {\bibfnamefont {Dmitry}\ \bibnamefont {Kalashnikov}},\ }\bibfield  {title} {\enquote {\bibinfo {title} {{Probing light exotics from a hidden sector at c-\ensuremath{\tau} factories with polarized electron beams}},}\ }\href {\doibase 10.1103/PhysRevD.107.015014} {\bibfield  {journal} {\bibinfo  {journal} {Phys. Rev. D}\ }\textbf {\bibinfo {volume} {107}},\ \bibinfo {pages} {015014} (\bibinfo {year} {2023})},\ \Eprint {http://arxiv.org/abs/2211.06270} {arXiv:2211.06270 [hep-ph]} \BibitemShut {NoStop}%
\bibitem [{\citenamefont {Claude}\ \emph {et~al.}(2023)\citenamefont {Claude}, \citenamefont {Dutra},\ and\ \citenamefont {Godfrey}}]{Claude:2022rho}%
  \BibitemOpen
  \bibfield  {author} {\bibinfo {author} {\bibfnamefont {J\'er\^ome}\ \bibnamefont {Claude}}, \bibinfo {author} {\bibfnamefont {Ma\'\i{}ra}\ \bibnamefont {Dutra}}, \ and\ \bibinfo {author} {\bibfnamefont {Stephen}\ \bibnamefont {Godfrey}},\ }\bibfield  {title} {\enquote {\bibinfo {title} {{Probing feebly interacting dark matter with monojet searches}},}\ }\href {\doibase 10.1103/PhysRevD.107.075006} {\bibfield  {journal} {\bibinfo  {journal} {Phys. Rev. D}\ }\textbf {\bibinfo {volume} {107}},\ \bibinfo {pages} {075006} (\bibinfo {year} {2023})},\ \Eprint {http://arxiv.org/abs/2208.09422} {arXiv:2208.09422 [hep-ph]} \BibitemShut {NoStop}%
\bibitem [{\citenamefont {Wang}\ \emph {et~al.}(2023)\citenamefont {Wang}, \citenamefont {Xu}, \citenamefont {Yang},\ and\ \citenamefont {Zhu}}]{Wang:2023wrx}%
  \BibitemOpen
  \bibfield  {author} {\bibinfo {author} {\bibfnamefont {Wenyu}\ \bibnamefont {Wang}}, \bibinfo {author} {\bibfnamefont {Wu-Long}\ \bibnamefont {Xu}}, \bibinfo {author} {\bibfnamefont {Jin~Min}\ \bibnamefont {Yang}}, \ and\ \bibinfo {author} {\bibfnamefont {Rui}\ \bibnamefont {Zhu}},\ }\bibfield  {title} {\enquote {\bibinfo {title} {{Direct detection of cosmic ray-boosted puffy dark matter}},}\ }\href@noop {} {\  (\bibinfo {year} {2023})},\ \Eprint {http://arxiv.org/abs/2305.12668} {arXiv:2305.12668 [hep-ph]} \BibitemShut {NoStop}%
\bibitem [{\citenamefont {Lee}\ \emph {et~al.}(2014)\citenamefont {Lee}, \citenamefont {Park},\ and\ \citenamefont {Sanz}}]{Lee:2013bua}%
  \BibitemOpen
  \bibfield  {author} {\bibinfo {author} {\bibfnamefont {Hyun~Min}\ \bibnamefont {Lee}}, \bibinfo {author} {\bibfnamefont {Myeonghun}\ \bibnamefont {Park}}, \ and\ \bibinfo {author} {\bibfnamefont {Veronica}\ \bibnamefont {Sanz}},\ }\bibfield  {title} {\enquote {\bibinfo {title} {{Gravity-mediated (or Composite) Dark Matter}},}\ }\href {\doibase 10.1140/epjc/s10052-014-2715-8} {\bibfield  {journal} {\bibinfo  {journal} {Eur. Phys. J. C}\ }\textbf {\bibinfo {volume} {74}},\ \bibinfo {pages} {2715} (\bibinfo {year} {2014})},\ \Eprint {http://arxiv.org/abs/1306.4107} {arXiv:1306.4107 [hep-ph]} \BibitemShut {NoStop}%
\bibitem [{\citenamefont {Kang}\ and\ \citenamefont {Lee}(2020)}]{Kang:2020huh}%
  \BibitemOpen
  \bibfield  {author} {\bibinfo {author} {\bibfnamefont {Yoo-Jin}\ \bibnamefont {Kang}}\ and\ \bibinfo {author} {\bibfnamefont {Hyun~Min}\ \bibnamefont {Lee}},\ }\bibfield  {title} {\enquote {\bibinfo {title} {{Lightening Gravity-Mediated Dark Matter}},}\ }\href {\doibase 10.1140/epjc/s10052-020-8153-x} {\bibfield  {journal} {\bibinfo  {journal} {Eur. Phys. J. C}\ }\textbf {\bibinfo {volume} {80}},\ \bibinfo {pages} {602} (\bibinfo {year} {2020})},\ \Eprint {http://arxiv.org/abs/2001.04868} {arXiv:2001.04868 [hep-ph]} \BibitemShut {NoStop}%
\bibitem [{\citenamefont {Bernal}\ \emph {et~al.}(2018)\citenamefont {Bernal}, \citenamefont {Dutra}, \citenamefont {Mambrini}, \citenamefont {Olive}, \citenamefont {Peloso},\ and\ \citenamefont {Pierre}}]{Bernal:2018qlk}%
  \BibitemOpen
  \bibfield  {author} {\bibinfo {author} {\bibfnamefont {Nicol\'as}\ \bibnamefont {Bernal}}, \bibinfo {author} {\bibfnamefont {Ma\'\i{}ra}\ \bibnamefont {Dutra}}, \bibinfo {author} {\bibfnamefont {Yann}\ \bibnamefont {Mambrini}}, \bibinfo {author} {\bibfnamefont {Keith}\ \bibnamefont {Olive}}, \bibinfo {author} {\bibfnamefont {Marco}\ \bibnamefont {Peloso}}, \ and\ \bibinfo {author} {\bibfnamefont {Mathias}\ \bibnamefont {Pierre}},\ }\bibfield  {title} {\enquote {\bibinfo {title} {{Spin-2 Portal Dark Matter}},}\ }\href {\doibase 10.1103/PhysRevD.97.115020} {\bibfield  {journal} {\bibinfo  {journal} {Phys. Rev. D}\ }\textbf {\bibinfo {volume} {97}},\ \bibinfo {pages} {115020} (\bibinfo {year} {2018})},\ \Eprint {http://arxiv.org/abs/1803.01866} {arXiv:1803.01866 [hep-ph]} \BibitemShut {NoStop}%
\bibitem [{\citenamefont {Folgado}\ \emph {et~al.}(2020)\citenamefont {Folgado}, \citenamefont {Donini},\ and\ \citenamefont {Rius}}]{Folgado:2019gie}%
  \BibitemOpen
  \bibfield  {author} {\bibinfo {author} {\bibfnamefont {Miguel~G.}\ \bibnamefont {Folgado}}, \bibinfo {author} {\bibfnamefont {Andrea}\ \bibnamefont {Donini}}, \ and\ \bibinfo {author} {\bibfnamefont {Nuria}\ \bibnamefont {Rius}},\ }\bibfield  {title} {\enquote {\bibinfo {title} {{Gravity-mediated Dark Matter in Clockwork/Linear Dilaton Extra-Dimensions}},}\ }\href {\doibase 10.1007/JHEP04(2020)036} {\bibfield  {journal} {\bibinfo  {journal} {JHEP}\ }\textbf {\bibinfo {volume} {04}},\ \bibinfo {pages} {036} (\bibinfo {year} {2020})},\ \Eprint {http://arxiv.org/abs/1912.02689} {arXiv:1912.02689 [hep-ph]} \BibitemShut {NoStop}%
\bibitem [{\citenamefont {Kang}\ and\ \citenamefont {Lee}(2021)}]{Kang:2020yul}%
  \BibitemOpen
  \bibfield  {author} {\bibinfo {author} {\bibfnamefont {Yoo-Jin}\ \bibnamefont {Kang}}\ and\ \bibinfo {author} {\bibfnamefont {Hyun~Min}\ \bibnamefont {Lee}},\ }\bibfield  {title} {\enquote {\bibinfo {title} {{Dark matter self-interactions from spin-2 mediators}},}\ }\href {\doibase 10.1140/epjc/s10052-021-09610-x} {\bibfield  {journal} {\bibinfo  {journal} {Eur. Phys. J. C}\ }\textbf {\bibinfo {volume} {81}},\ \bibinfo {pages} {868} (\bibinfo {year} {2021})},\ \Eprint {http://arxiv.org/abs/2002.12779} {arXiv:2002.12779 [hep-ph]} \BibitemShut {NoStop}%
\bibitem [{\citenamefont {Dutra}(2019)}]{Dutra:2019xet}%
  \BibitemOpen
  \bibfield  {author} {\bibinfo {author} {\bibfnamefont {Ma\'\i{}ra}\ \bibnamefont {Dutra}},\ }\bibfield  {title} {\enquote {\bibinfo {title} {{Freeze-in production of dark matter through spin-1 and spin-2 portals}},}\ }\href {\doibase 10.22323/1.367.0076} {\bibfield  {journal} {\bibinfo  {journal} {PoS}\ }\textbf {\bibinfo {volume} {LeptonPhoton2019}},\ \bibinfo {pages} {076} (\bibinfo {year} {2019})},\ \Eprint {http://arxiv.org/abs/1911.11844} {arXiv:1911.11844 [hep-ph]} \BibitemShut {NoStop}%
\bibitem [{\citenamefont {Clery}\ \emph {et~al.}(2022)\citenamefont {Clery}, \citenamefont {Mambrini}, \citenamefont {Olive}, \citenamefont {Shkerin},\ and\ \citenamefont {Verner}}]{Clery:2022wib}%
  \BibitemOpen
  \bibfield  {author} {\bibinfo {author} {\bibfnamefont {Simon}\ \bibnamefont {Clery}}, \bibinfo {author} {\bibfnamefont {Yann}\ \bibnamefont {Mambrini}}, \bibinfo {author} {\bibfnamefont {Keith~A.}\ \bibnamefont {Olive}}, \bibinfo {author} {\bibfnamefont {Andrey}\ \bibnamefont {Shkerin}}, \ and\ \bibinfo {author} {\bibfnamefont {Sarunas}\ \bibnamefont {Verner}},\ }\bibfield  {title} {\enquote {\bibinfo {title} {{Gravitational portals with nonminimal couplings}},}\ }\href {\doibase 10.1103/PhysRevD.105.095042} {\bibfield  {journal} {\bibinfo  {journal} {Phys. Rev. D}\ }\textbf {\bibinfo {volume} {105}},\ \bibinfo {pages} {095042} (\bibinfo {year} {2022})},\ \Eprint {http://arxiv.org/abs/2203.02004} {arXiv:2203.02004 [hep-ph]} \BibitemShut {NoStop}%
\bibitem [{\citenamefont {Gill}\ \emph {et~al.}(2023)\citenamefont {Gill}, \citenamefont {Sengupta}, \citenamefont {G},\ and\ \citenamefont {Williams}}]{Gill:2023kyz}%
  \BibitemOpen
  \bibfield  {author} {\bibinfo {author} {\bibfnamefont {Joshua~A.}\ \bibnamefont {Gill}}, \bibinfo {author} {\bibfnamefont {Dipan}\ \bibnamefont {Sengupta}}, \bibinfo {author} {\bibfnamefont {Anthony}\ \bibnamefont {G}}, \ and\ \bibinfo {author} {\bibnamefont {Williams}},\ }\bibfield  {title} {\enquote {\bibinfo {title} {{Graviton-photon production with a massive spin-2 particle}},}\ }\href@noop {} {\  (\bibinfo {year} {2023})},\ \Eprint {http://arxiv.org/abs/2303.04329} {arXiv:2303.04329 [hep-ph]} \BibitemShut {NoStop}%
\bibitem [{\citenamefont {Wang}\ \emph {et~al.}(2020)\citenamefont {Wang}, \citenamefont {Wu}, \citenamefont {Yang}, \citenamefont {Zhou},\ and\ \citenamefont {Zhu}}]{Wang:2019jtk}%
  \BibitemOpen
  \bibfield  {author} {\bibinfo {author} {\bibfnamefont {Wenyu}\ \bibnamefont {Wang}}, \bibinfo {author} {\bibfnamefont {Lei}\ \bibnamefont {Wu}}, \bibinfo {author} {\bibfnamefont {Jin~Min}\ \bibnamefont {Yang}}, \bibinfo {author} {\bibfnamefont {Hang}\ \bibnamefont {Zhou}}, \ and\ \bibinfo {author} {\bibfnamefont {Bin}\ \bibnamefont {Zhu}},\ }\bibfield  {title} {\enquote {\bibinfo {title} {{Cosmic ray boosted sub-GeV gravitationally interacting dark matter in direct detection}},}\ }\href {\doibase 10.1007/JHEP12(2020)072} {\bibfield  {journal} {\bibinfo  {journal} {JHEP}\ }\textbf {\bibinfo {volume} {12}},\ \bibinfo {pages} {072} (\bibinfo {year} {2020})},\ \bibinfo {note} {[Erratum: JHEP 02, 052 (2021)]},\ \Eprint {http://arxiv.org/abs/1912.09904} {arXiv:1912.09904 [hep-ph]} \BibitemShut {NoStop}%
\bibitem [{\citenamefont {de~Giorgi}\ and\ \citenamefont {Vogl}(2021)}]{deGiorgi:2021xvm}%
  \BibitemOpen
  \bibfield  {author} {\bibinfo {author} {\bibfnamefont {Arturo}\ \bibnamefont {de~Giorgi}}\ and\ \bibinfo {author} {\bibfnamefont {Stefan}\ \bibnamefont {Vogl}},\ }\bibfield  {title} {\enquote {\bibinfo {title} {{Dark matter interacting via a massive spin-2 mediator in warped extra-dimensions}},}\ }\href {\doibase 10.1007/JHEP11(2021)036} {\bibfield  {journal} {\bibinfo  {journal} {JHEP}\ }\textbf {\bibinfo {volume} {11}},\ \bibinfo {pages} {036} (\bibinfo {year} {2021})},\ \Eprint {http://arxiv.org/abs/2105.06794} {arXiv:2105.06794 [hep-ph]} \BibitemShut {NoStop}%
\bibitem [{\citenamefont {de~Giorgi}\ and\ \citenamefont {Vogl}(2023)}]{deGiorgi:2022yha}%
  \BibitemOpen
  \bibfield  {author} {\bibinfo {author} {\bibfnamefont {Arturo}\ \bibnamefont {de~Giorgi}}\ and\ \bibinfo {author} {\bibfnamefont {Stefan}\ \bibnamefont {Vogl}},\ }\bibfield  {title} {\enquote {\bibinfo {title} {{Warm dark matter from a gravitational freeze-in in extra dimensions}},}\ }\href {\doibase 10.1007/JHEP04(2023)032} {\bibfield  {journal} {\bibinfo  {journal} {JHEP}\ }\textbf {\bibinfo {volume} {04}},\ \bibinfo {pages} {032} (\bibinfo {year} {2023})},\ \Eprint {http://arxiv.org/abs/2208.03153} {arXiv:2208.03153 [hep-ph]} \BibitemShut {NoStop}%
\bibitem [{\citenamefont {Jod\l{}owski}(2023)}]{Jodlowski:2023yne}%
  \BibitemOpen
  \bibfield  {author} {\bibinfo {author} {\bibfnamefont {Krzysztof}\ \bibnamefont {Jod\l{}owski}},\ }\bibfield  {title} {\enquote {\bibinfo {title} {{Looking forward to photon-coupled long-lived particles I: massive spin-2 portal}},}\ }\href@noop {} {\  (\bibinfo {year} {2023})},\ \Eprint {http://arxiv.org/abs/2305.05710} {arXiv:2305.05710 [hep-ph]} \BibitemShut {NoStop}%
\bibitem [{\citenamefont {Voronchikhin}\ and\ \citenamefont {Kirpichnikov}(2024{\natexlab{b}})}]{Voronchikhin:2024ygo}%
  \BibitemOpen
  \bibfield  {author} {\bibinfo {author} {\bibfnamefont {I.~V.}\ \bibnamefont {Voronchikhin}}\ and\ \bibinfo {author} {\bibfnamefont {D.~V.}\ \bibnamefont {Kirpichnikov}},\ }\bibfield  {title} {\enquote {\bibinfo {title} {{The bremsstrahlung-like production of the massive spin-2 dark matter mediator}},}\ }\href@noop {} {\  (\bibinfo {year} {2024}{\natexlab{b}})},\ \Eprint {http://arxiv.org/abs/2412.10150} {arXiv:2412.10150 [hep-ph]} \BibitemShut {NoStop}%
\bibitem [{\citenamefont {Berlin}\ and\ \citenamefont {Kling}(2019)}]{Berlin:2018jbm}%
  \BibitemOpen
  \bibfield  {author} {\bibinfo {author} {\bibfnamefont {Asher}\ \bibnamefont {Berlin}}\ and\ \bibinfo {author} {\bibfnamefont {Felix}\ \bibnamefont {Kling}},\ }\bibfield  {title} {\enquote {\bibinfo {title} {{Inelastic Dark Matter at the LHC Lifetime Frontier: ATLAS, CMS, LHCb, CODEX-b, FASER, and MATHUSLA}},}\ }\href {\doibase 10.1103/PhysRevD.99.015021} {\bibfield  {journal} {\bibinfo  {journal} {Phys. Rev. D}\ }\textbf {\bibinfo {volume} {99}},\ \bibinfo {pages} {015021} (\bibinfo {year} {2019})},\ \Eprint {http://arxiv.org/abs/1810.01879} {arXiv:1810.01879 [hep-ph]} \BibitemShut {NoStop}%
\bibitem [{\citenamefont {Dienes}\ \emph {et~al.}(2023)\citenamefont {Dienes}, \citenamefont {Feng}, \citenamefont {Fieg}, \citenamefont {Huang}, \citenamefont {Lee},\ and\ \citenamefont {Thomas}}]{Dienes:2023uve}%
  \BibitemOpen
  \bibfield  {author} {\bibinfo {author} {\bibfnamefont {Keith~R.}\ \bibnamefont {Dienes}}, \bibinfo {author} {\bibfnamefont {Jonathan~L.}\ \bibnamefont {Feng}}, \bibinfo {author} {\bibfnamefont {Max}\ \bibnamefont {Fieg}}, \bibinfo {author} {\bibfnamefont {Fei}\ \bibnamefont {Huang}}, \bibinfo {author} {\bibfnamefont {Seung~J.}\ \bibnamefont {Lee}}, \ and\ \bibinfo {author} {\bibfnamefont {Brooks}\ \bibnamefont {Thomas}},\ }\bibfield  {title} {\enquote {\bibinfo {title} {{Extending the discovery potential for inelastic-dipole dark matter with FASER}},}\ }\href {\doibase 10.1103/PhysRevD.107.115006} {\bibfield  {journal} {\bibinfo  {journal} {Phys. Rev. D}\ }\textbf {\bibinfo {volume} {107}},\ \bibinfo {pages} {115006} (\bibinfo {year} {2023})},\ \Eprint {http://arxiv.org/abs/2301.05252} {arXiv:2301.05252 [hep-ph]} \BibitemShut {NoStop}%
\bibitem [{\citenamefont {Mongillo}\ \emph {et~al.}(2023)\citenamefont {Mongillo}, \citenamefont {Abdullahi}, \citenamefont {Oberhauser}, \citenamefont {Crivelli}, \citenamefont {Hostert}, \citenamefont {Massaro}, \citenamefont {Bueno},\ and\ \citenamefont {Pascoli}}]{Mongillo:2023hbs}%
  \BibitemOpen
  \bibfield  {author} {\bibinfo {author} {\bibfnamefont {Martina}\ \bibnamefont {Mongillo}}, \bibinfo {author} {\bibfnamefont {Asli}\ \bibnamefont {Abdullahi}}, \bibinfo {author} {\bibfnamefont {Benjamin~Banto}\ \bibnamefont {Oberhauser}}, \bibinfo {author} {\bibfnamefont {Paolo}\ \bibnamefont {Crivelli}}, \bibinfo {author} {\bibfnamefont {Matheus}\ \bibnamefont {Hostert}}, \bibinfo {author} {\bibfnamefont {Daniele}\ \bibnamefont {Massaro}}, \bibinfo {author} {\bibfnamefont {Laura~Molina}\ \bibnamefont {Bueno}}, \ and\ \bibinfo {author} {\bibfnamefont {Silvia}\ \bibnamefont {Pascoli}},\ }\bibfield  {title} {\enquote {\bibinfo {title} {{Constraining light thermal inelastic dark matter with NA64}},}\ }\href {\doibase 10.1140/epjc/s10052-023-11536-5} {\bibfield  {journal} {\bibinfo  {journal} {Eur. Phys. J. C}\ }\textbf {\bibinfo {volume} {83}},\ \bibinfo {pages} {391} (\bibinfo {year} {2023})},\ \Eprint {http://arxiv.org/abs/2302.05414} {arXiv:2302.05414 [hep-ph]} \BibitemShut {NoStop}%
\bibitem [{\citenamefont {Dalla Valle~Garcia}(2025{\natexlab{a}})}]{DallaValleGarcia:2025giv}%
  \BibitemOpen
  \bibfield  {author} {\bibinfo {author} {\bibfnamefont {Giovani}\ \bibnamefont {Dalla Valle~Garcia}},\ }\emph {\bibinfo {title} {{New directions in inelastic Dark Matter - The Role of Parity in Dark Sectors}}},\ \href {\doibase 10.5445/IR/1000186680} {Ph.D. thesis},\ \bibinfo  {school} {Karlsruhe Institute of Technology} (\bibinfo {year} {2025}{\natexlab{a}})\BibitemShut {NoStop}%
\bibitem [{\citenamefont {Jod{\l}owski}\ \emph {et~al.}(2020)\citenamefont {Jod{\l}owski}, \citenamefont {Kling}, \citenamefont {Roszkowski},\ and\ \citenamefont {Trojanowski}}]{Jodlowski:2019ycu}%
  \BibitemOpen
  \bibfield  {author} {\bibinfo {author} {\bibfnamefont {Krzysztof}\ \bibnamefont {Jod{\l}owski}}, \bibinfo {author} {\bibfnamefont {Felix}\ \bibnamefont {Kling}}, \bibinfo {author} {\bibfnamefont {Leszek}\ \bibnamefont {Roszkowski}}, \ and\ \bibinfo {author} {\bibfnamefont {Sebastian}\ \bibnamefont {Trojanowski}},\ }\bibfield  {title} {\enquote {\bibinfo {title} {{Extending the reach of FASER, MATHUSLA, and SHiP towards smaller lifetimes using secondary particle production}},}\ }\href {\doibase 10.1103/PhysRevD.101.095020} {\bibfield  {journal} {\bibinfo  {journal} {Phys. Rev. D}\ }\textbf {\bibinfo {volume} {101}},\ \bibinfo {pages} {095020} (\bibinfo {year} {2020})},\ \Eprint {http://arxiv.org/abs/1911.11346} {arXiv:1911.11346 [hep-ph]} \BibitemShut {NoStop}%
\bibitem [{\citenamefont {Izaguirre}\ \emph {et~al.}(2016)\citenamefont {Izaguirre}, \citenamefont {Krnjaic},\ and\ \citenamefont {Shuve}}]{Izaguirre:2015zva}%
  \BibitemOpen
  \bibfield  {author} {\bibinfo {author} {\bibfnamefont {Eder}\ \bibnamefont {Izaguirre}}, \bibinfo {author} {\bibfnamefont {Gordan}\ \bibnamefont {Krnjaic}}, \ and\ \bibinfo {author} {\bibfnamefont {Brian}\ \bibnamefont {Shuve}},\ }\bibfield  {title} {\enquote {\bibinfo {title} {{Discovering Inelastic Thermal-Relic Dark Matter at Colliders}},}\ }\href {\doibase 10.1103/PhysRevD.93.063523} {\bibfield  {journal} {\bibinfo  {journal} {Phys. Rev. D}\ }\textbf {\bibinfo {volume} {93}},\ \bibinfo {pages} {063523} (\bibinfo {year} {2016})},\ \Eprint {http://arxiv.org/abs/1508.03050} {arXiv:1508.03050 [hep-ph]} \BibitemShut {NoStop}%
\bibitem [{\citenamefont {Gninenko}\ \emph {et~al.}(2016)\citenamefont {Gninenko}, \citenamefont {Krasnikov}, \citenamefont {Kirsanov},\ and\ \citenamefont {Kirpichnikov}}]{Gninenko:2016kpg}%
  \BibitemOpen
  \bibfield  {author} {\bibinfo {author} {\bibfnamefont {S.~N.}\ \bibnamefont {Gninenko}}, \bibinfo {author} {\bibfnamefont {N.~V.}\ \bibnamefont {Krasnikov}}, \bibinfo {author} {\bibfnamefont {M.~M.}\ \bibnamefont {Kirsanov}}, \ and\ \bibinfo {author} {\bibfnamefont {D.~V.}\ \bibnamefont {Kirpichnikov}},\ }\bibfield  {title} {\enquote {\bibinfo {title} {{Missing energy signature from invisible decays of dark photons at the CERN SPS}},}\ }\href {\doibase 10.1103/PhysRevD.94.095025} {\bibfield  {journal} {\bibinfo  {journal} {Phys. Rev. D}\ }\textbf {\bibinfo {volume} {94}},\ \bibinfo {pages} {095025} (\bibinfo {year} {2016})},\ \Eprint {http://arxiv.org/abs/1604.08432} {arXiv:1604.08432 [hep-ph]} \BibitemShut {NoStop}%
\bibitem [{\citenamefont {Banerjee}\ \emph {et~al.}(2017)\citenamefont {Banerjee} \emph {et~al.}}]{NA64:2016oww}%
  \BibitemOpen
  \bibfield  {author} {\bibinfo {author} {\bibfnamefont {D.}~\bibnamefont {Banerjee}} \emph {et~al.} (\bibinfo {collaboration} {NA64}),\ }\bibfield  {title} {\enquote {\bibinfo {title} {{Search for invisible decays of sub-GeV dark photons in missing-energy events at the CERN SPS}},}\ }\href {\doibase 10.1103/PhysRevLett.118.011802} {\bibfield  {journal} {\bibinfo  {journal} {Phys. Rev. Lett.}\ }\textbf {\bibinfo {volume} {118}},\ \bibinfo {pages} {011802} (\bibinfo {year} {2017})},\ \Eprint {http://arxiv.org/abs/1610.02988} {arXiv:1610.02988 [hep-ex]} \BibitemShut {NoStop}%
\bibitem [{\citenamefont {Banerjee}\ \emph {et~al.}(2018)\citenamefont {Banerjee} \emph {et~al.}}]{NA64:2017vtt}%
  \BibitemOpen
  \bibfield  {author} {\bibinfo {author} {\bibfnamefont {D.}~\bibnamefont {Banerjee}} \emph {et~al.} (\bibinfo {collaboration} {NA64}),\ }\bibfield  {title} {\enquote {\bibinfo {title} {{Search for vector mediator of Dark Matter production in invisible decay mode}},}\ }\href {\doibase 10.1103/PhysRevD.97.072002} {\bibfield  {journal} {\bibinfo  {journal} {Phys. Rev. D}\ }\textbf {\bibinfo {volume} {97}},\ \bibinfo {pages} {072002} (\bibinfo {year} {2018})},\ \Eprint {http://arxiv.org/abs/1710.00971} {arXiv:1710.00971 [hep-ex]} \BibitemShut {NoStop}%
\bibitem [{\citenamefont {Gninenko}\ \emph {et~al.}(2019)\citenamefont {Gninenko}, \citenamefont {Kirpichnikov},\ and\ \citenamefont {Krasnikov}}]{Gninenko:2018ter}%
  \BibitemOpen
  \bibfield  {author} {\bibinfo {author} {\bibfnamefont {S.~N.}\ \bibnamefont {Gninenko}}, \bibinfo {author} {\bibfnamefont {D.~V.}\ \bibnamefont {Kirpichnikov}}, \ and\ \bibinfo {author} {\bibfnamefont {N.~V.}\ \bibnamefont {Krasnikov}},\ }\bibfield  {title} {\enquote {\bibinfo {title} {{Probing millicharged particles with NA64 experiment at CERN}},}\ }\href {\doibase 10.1103/PhysRevD.100.035003} {\bibfield  {journal} {\bibinfo  {journal} {Phys. Rev. D}\ }\textbf {\bibinfo {volume} {100}},\ \bibinfo {pages} {035003} (\bibinfo {year} {2019})},\ \Eprint {http://arxiv.org/abs/1810.06856} {arXiv:1810.06856 [hep-ph]} \BibitemShut {NoStop}%
\bibitem [{\citenamefont {Banerjee}\ \emph {et~al.}(2019)\citenamefont {Banerjee} \emph {et~al.}}]{Banerjee:2019pds}%
  \BibitemOpen
  \bibfield  {author} {\bibinfo {author} {\bibfnamefont {D.}~\bibnamefont {Banerjee}} \emph {et~al.},\ }\bibfield  {title} {\enquote {\bibinfo {title} {{Dark matter search in missing energy events with NA64}},}\ }\href {\doibase 10.1103/PhysRevLett.123.121801} {\bibfield  {journal} {\bibinfo  {journal} {Phys. Rev. Lett.}\ }\textbf {\bibinfo {volume} {123}},\ \bibinfo {pages} {121801} (\bibinfo {year} {2019})},\ \Eprint {http://arxiv.org/abs/1906.00176} {arXiv:1906.00176 [hep-ex]} \BibitemShut {NoStop}%
\bibitem [{\citenamefont {Dusaev}\ \emph {et~al.}(2020)\citenamefont {Dusaev}, \citenamefont {Kirpichnikov},\ and\ \citenamefont {Kirsanov}}]{Dusaev:2020gxi}%
  \BibitemOpen
  \bibfield  {author} {\bibinfo {author} {\bibfnamefont {R.~R.}\ \bibnamefont {Dusaev}}, \bibinfo {author} {\bibfnamefont {D.~V.}\ \bibnamefont {Kirpichnikov}}, \ and\ \bibinfo {author} {\bibfnamefont {M.~M.}\ \bibnamefont {Kirsanov}},\ }\bibfield  {title} {\enquote {\bibinfo {title} {{Photoproduction of axionlike particles in the NA64 experiment}},}\ }\href {\doibase 10.1103/PhysRevD.102.055018} {\bibfield  {journal} {\bibinfo  {journal} {Phys. Rev. D}\ }\textbf {\bibinfo {volume} {102}},\ \bibinfo {pages} {055018} (\bibinfo {year} {2020})},\ \Eprint {http://arxiv.org/abs/2004.04469} {arXiv:2004.04469 [hep-ph]} \BibitemShut {NoStop}%
\bibitem [{\citenamefont {Andreev}\ \emph {et~al.}(2021{\natexlab{a}})\citenamefont {Andreev} \emph {et~al.}}]{Andreev:2021fzd}%
  \BibitemOpen
  \bibfield  {author} {\bibinfo {author} {\bibfnamefont {Yu.~M.}\ \bibnamefont {Andreev}} \emph {et~al.},\ }\bibfield  {title} {\enquote {\bibinfo {title} {{Improved exclusion limit for light dark matter from e+e- annihilation in NA64}},}\ }\href {\doibase 10.1103/PhysRevD.104.L091701} {\bibfield  {journal} {\bibinfo  {journal} {Phys. Rev. D}\ }\textbf {\bibinfo {volume} {104}},\ \bibinfo {pages} {L091701} (\bibinfo {year} {2021}{\natexlab{a}})},\ \Eprint {http://arxiv.org/abs/2108.04195} {arXiv:2108.04195 [hep-ex]} \BibitemShut {NoStop}%
\bibitem [{\citenamefont {Andreev}\ \emph {et~al.}(2022{\natexlab{a}})\citenamefont {Andreev} \emph {et~al.}}]{NA64:2022yly}%
  \BibitemOpen
  \bibfield  {author} {\bibinfo {author} {\bibfnamefont {Yu.~M.}\ \bibnamefont {Andreev}} \emph {et~al.} (\bibinfo {collaboration} {NA64}),\ }\bibfield  {title} {\enquote {\bibinfo {title} {{Search for a New B-L Z' Gauge Boson with the NA64 Experiment at CERN}},}\ }\href {\doibase 10.1103/PhysRevLett.129.161801} {\bibfield  {journal} {\bibinfo  {journal} {Phys. Rev. Lett.}\ }\textbf {\bibinfo {volume} {129}},\ \bibinfo {pages} {161801} (\bibinfo {year} {2022}{\natexlab{a}})},\ \Eprint {http://arxiv.org/abs/2207.09979} {arXiv:2207.09979 [hep-ex]} \BibitemShut {NoStop}%
\bibitem [{\citenamefont {Andreev}\ \emph {et~al.}(2022{\natexlab{b}})\citenamefont {Andreev} \emph {et~al.}}]{NA64:2022rme}%
  \BibitemOpen
  \bibfield  {author} {\bibinfo {author} {\bibfnamefont {Yu.~M.}\ \bibnamefont {Andreev}} \emph {et~al.} (\bibinfo {collaboration} {NA64}),\ }\bibfield  {title} {\enquote {\bibinfo {title} {{Search for a light Z' in the L\ensuremath{\mu}-L\ensuremath{\tau} scenario with the NA64-e experiment at CERN}},}\ }\href {\doibase 10.1103/PhysRevD.106.032015} {\bibfield  {journal} {\bibinfo  {journal} {Phys. Rev. D}\ }\textbf {\bibinfo {volume} {106}},\ \bibinfo {pages} {032015} (\bibinfo {year} {2022}{\natexlab{b}})},\ \Eprint {http://arxiv.org/abs/2206.03101} {arXiv:2206.03101 [hep-ex]} \BibitemShut {NoStop}%
\bibitem [{\citenamefont {Arefyeva}\ \emph {et~al.}(2022)\citenamefont {Arefyeva}, \citenamefont {Gninenko}, \citenamefont {Gorbunov},\ and\ \citenamefont {Kirpichnikov}}]{Arefyeva:2022eba}%
  \BibitemOpen
  \bibfield  {author} {\bibinfo {author} {\bibfnamefont {Nataliya}\ \bibnamefont {Arefyeva}}, \bibinfo {author} {\bibfnamefont {Sergei}\ \bibnamefont {Gninenko}}, \bibinfo {author} {\bibfnamefont {Dmitry}\ \bibnamefont {Gorbunov}}, \ and\ \bibinfo {author} {\bibfnamefont {Dmitry}\ \bibnamefont {Kirpichnikov}},\ }\bibfield  {title} {\enquote {\bibinfo {title} {{Passage of millicharged particles in the electron beam-dump: Refining constraints from SLACmQ and estimating sensitivity of NA64e}},}\ }\href {\doibase 10.1103/PhysRevD.106.035029} {\bibfield  {journal} {\bibinfo  {journal} {Phys. Rev. D}\ }\textbf {\bibinfo {volume} {106}},\ \bibinfo {pages} {035029} (\bibinfo {year} {2022})},\ \Eprint {http://arxiv.org/abs/2204.03984} {arXiv:2204.03984 [hep-ph]} \BibitemShut {NoStop}%
\bibitem [{\citenamefont {Zhevlakov}\ \emph {et~al.}(2022)\citenamefont {Zhevlakov}, \citenamefont {Kirpichnikov},\ and\ \citenamefont {Lyubovitskij}}]{Zhevlakov:2022vio}%
  \BibitemOpen
  \bibfield  {author} {\bibinfo {author} {\bibfnamefont {Alexey~S.}\ \bibnamefont {Zhevlakov}}, \bibinfo {author} {\bibfnamefont {Dmitry~V.}\ \bibnamefont {Kirpichnikov}}, \ and\ \bibinfo {author} {\bibfnamefont {Valery~E.}\ \bibnamefont {Lyubovitskij}},\ }\bibfield  {title} {\enquote {\bibinfo {title} {{Implication of the dark axion portal for the EDM of fermions and dark matter probing with NA64e, NA64\ensuremath{\mu}, LDMX, M3, and BaBar}},}\ }\href {\doibase 10.1103/PhysRevD.106.035018} {\bibfield  {journal} {\bibinfo  {journal} {Phys. Rev. D}\ }\textbf {\bibinfo {volume} {106}},\ \bibinfo {pages} {035018} (\bibinfo {year} {2022})},\ \Eprint {http://arxiv.org/abs/2204.09978} {arXiv:2204.09978 [hep-ph]} \BibitemShut {NoStop}%
\bibitem [{\citenamefont {Cazzaniga}\ \emph {et~al.}(2021)\citenamefont {Cazzaniga} \emph {et~al.}}]{NA64:2021acr}%
  \BibitemOpen
  \bibfield  {author} {\bibinfo {author} {\bibfnamefont {C.}~\bibnamefont {Cazzaniga}} \emph {et~al.} (\bibinfo {collaboration} {NA64}),\ }\bibfield  {title} {\enquote {\bibinfo {title} {{Probing the explanation of the muon (g-2) anomaly and thermal light dark matter with the semi-visible dark photon channel}},}\ }\href {\doibase 10.1140/epjc/s10052-021-09705-5} {\bibfield  {journal} {\bibinfo  {journal} {Eur. Phys. J. C}\ }\textbf {\bibinfo {volume} {81}},\ \bibinfo {pages} {959} (\bibinfo {year} {2021})},\ \Eprint {http://arxiv.org/abs/2107.02021} {arXiv:2107.02021 [hep-ex]} \BibitemShut {NoStop}%
\bibitem [{\citenamefont {Andreev}\ \emph {et~al.}(2021{\natexlab{b}})\citenamefont {Andreev} \emph {et~al.}}]{NA64:2021xzo}%
  \BibitemOpen
  \bibfield  {author} {\bibinfo {author} {\bibfnamefont {Yu.~M.}\ \bibnamefont {Andreev}} \emph {et~al.} (\bibinfo {collaboration} {NA64}),\ }\bibfield  {title} {\enquote {\bibinfo {title} {{Constraints on New Physics in Electron $g-2$ from a Search for Invisible Decays of a Scalar, Pseudoscalar, Vector, and Axial Vector}},}\ }\href {\doibase 10.1103/PhysRevLett.126.211802} {\bibfield  {journal} {\bibinfo  {journal} {Phys. Rev. Lett.}\ }\textbf {\bibinfo {volume} {126}},\ \bibinfo {pages} {211802} (\bibinfo {year} {2021}{\natexlab{b}})},\ \Eprint {http://arxiv.org/abs/2102.01885} {arXiv:2102.01885 [hep-ex]} \BibitemShut {NoStop}%
\bibitem [{\citenamefont {Andreev}\ \emph {et~al.}(2024{\natexlab{a}})\citenamefont {Andreev} \emph {et~al.}}]{NA64:2024klw}%
  \BibitemOpen
  \bibfield  {author} {\bibinfo {author} {\bibfnamefont {Yu.~M.}\ \bibnamefont {Andreev}} \emph {et~al.} (\bibinfo {collaboration} {NA64}),\ }\bibfield  {title} {\enquote {\bibinfo {title} {{First Results in the Search for Dark Sectors at NA64 with the CERN SPS High Energy Muon Beam}},}\ }\href {\doibase 10.1103/PhysRevLett.132.211803} {\bibfield  {journal} {\bibinfo  {journal} {Phys. Rev. Lett.}\ }\textbf {\bibinfo {volume} {132}},\ \bibinfo {pages} {211803} (\bibinfo {year} {2024}{\natexlab{a}})},\ \Eprint {http://arxiv.org/abs/2401.01708} {arXiv:2401.01708 [hep-ex]} \BibitemShut {NoStop}%
\bibitem [{\citenamefont {Sieber}\ \emph {et~al.}(2022)\citenamefont {Sieber}, \citenamefont {Banerjee}, \citenamefont {Crivelli}, \citenamefont {Depero}, \citenamefont {Gninenko}, \citenamefont {Kirpichnikov}, \citenamefont {Kirsanov}, \citenamefont {Poliakov},\ and\ \citenamefont {Molina~Bueno}}]{Sieber:2021fue}%
  \BibitemOpen
  \bibfield  {author} {\bibinfo {author} {\bibfnamefont {H.}~\bibnamefont {Sieber}}, \bibinfo {author} {\bibfnamefont {D.}~\bibnamefont {Banerjee}}, \bibinfo {author} {\bibfnamefont {P.}~\bibnamefont {Crivelli}}, \bibinfo {author} {\bibfnamefont {E.}~\bibnamefont {Depero}}, \bibinfo {author} {\bibfnamefont {S.~N.}\ \bibnamefont {Gninenko}}, \bibinfo {author} {\bibfnamefont {D.~V.}\ \bibnamefont {Kirpichnikov}}, \bibinfo {author} {\bibfnamefont {M.~M.}\ \bibnamefont {Kirsanov}}, \bibinfo {author} {\bibfnamefont {V.}~\bibnamefont {Poliakov}}, \ and\ \bibinfo {author} {\bibfnamefont {L.}~\bibnamefont {Molina~Bueno}},\ }\bibfield  {title} {\enquote {\bibinfo {title} {{Prospects in the search for a new light Z' boson with the NA64\ensuremath{\mu} experiment at the CERN SPS}},}\ }\href {\doibase 10.1103/PhysRevD.105.052006} {\bibfield  {journal} {\bibinfo  {journal} {Phys. Rev. D}\ }\textbf {\bibinfo {volume} {105}},\ \bibinfo {pages} {052006} (\bibinfo {year} {2022})},\ \Eprint {http://arxiv.org/abs/2110.15111}
  {arXiv:2110.15111 [hep-ex]} \BibitemShut {NoStop}%
\bibitem [{\citenamefont {Kirpichnikov}\ \emph {et~al.}(2021)\citenamefont {Kirpichnikov}, \citenamefont {Sieber}, \citenamefont {Bueno}, \citenamefont {Crivelli},\ and\ \citenamefont {Kirsanov}}]{Kirpichnikov:2021jev}%
  \BibitemOpen
  \bibfield  {author} {\bibinfo {author} {\bibfnamefont {D.~V.}\ \bibnamefont {Kirpichnikov}}, \bibinfo {author} {\bibfnamefont {H.}~\bibnamefont {Sieber}}, \bibinfo {author} {\bibfnamefont {L.~Molina}\ \bibnamefont {Bueno}}, \bibinfo {author} {\bibfnamefont {P.}~\bibnamefont {Crivelli}}, \ and\ \bibinfo {author} {\bibfnamefont {M.~M.}\ \bibnamefont {Kirsanov}},\ }\bibfield  {title} {\enquote {\bibinfo {title} {{Probing hidden sectors with a muon beam: Total and differential cross sections for vector boson production in muon bremsstrahlung}},}\ }\href {\doibase 10.1103/PhysRevD.104.076012} {\bibfield  {journal} {\bibinfo  {journal} {Phys. Rev. D}\ }\textbf {\bibinfo {volume} {104}},\ \bibinfo {pages} {076012} (\bibinfo {year} {2021})},\ \Eprint {http://arxiv.org/abs/2107.13297} {arXiv:2107.13297 [hep-ph]} \BibitemShut {NoStop}%
\bibitem [{\citenamefont {Berlin}\ \emph {et~al.}(2019)\citenamefont {Berlin} \emph {et~al.}}]{Berlin:2018bsc}%
  \BibitemOpen
  \bibfield  {author} {\bibinfo {author} {\bibfnamefont {Asher}\ \bibnamefont {Berlin}} \emph {et~al.},\ }\bibfield  {title} {\enquote {\bibinfo {title} {{Dark Matter, Millicharges, Axion and Scalar Particles, Gauge Bosons, and Other New Physics with LDMX}},}\ }\href {\doibase 10.1103/PhysRevD.99.075001} {\bibfield  {journal} {\bibinfo  {journal} {Phys. Rev. D}\ }\textbf {\bibinfo {volume} {99}},\ \bibinfo {pages} {075001} (\bibinfo {year} {2019})},\ \Eprint {http://arxiv.org/abs/1807.01730} {arXiv:1807.01730 [hep-ph]} \BibitemShut {NoStop}%
\bibitem [{\citenamefont {Schuster}\ \emph {et~al.}(2022)\citenamefont {Schuster}, \citenamefont {Toro},\ and\ \citenamefont {Zhou}}]{Schuster:2021mlr}%
  \BibitemOpen
  \bibfield  {author} {\bibinfo {author} {\bibfnamefont {Philip}\ \bibnamefont {Schuster}}, \bibinfo {author} {\bibfnamefont {Natalia}\ \bibnamefont {Toro}}, \ and\ \bibinfo {author} {\bibfnamefont {Kevin}\ \bibnamefont {Zhou}},\ }\bibfield  {title} {\enquote {\bibinfo {title} {{Probing invisible vector meson decays with the NA64 and LDMX experiments}},}\ }\href {\doibase 10.1103/PhysRevD.105.035036} {\bibfield  {journal} {\bibinfo  {journal} {Phys. Rev. D}\ }\textbf {\bibinfo {volume} {105}},\ \bibinfo {pages} {035036} (\bibinfo {year} {2022})},\ \Eprint {http://arxiv.org/abs/2112.02104} {arXiv:2112.02104 [hep-ph]} \BibitemShut {NoStop}%
\bibitem [{\citenamefont {{\r{A}}kesson}\ \emph {et~al.}(2022)\citenamefont {{\r{A}}kesson} \emph {et~al.}}]{Akesson:2022vza}%
  \BibitemOpen
  \bibfield  {author} {\bibinfo {author} {\bibfnamefont {Torsten}\ \bibnamefont {{\r{A}}kesson}} \emph {et~al.},\ }\bibfield  {title} {\enquote {\bibinfo {title} {{Current Status and Future Prospects for the Light Dark Matter eXperiment}},}\ }in\ \href@noop {} {\emph {\bibinfo {booktitle} {{Snowmass 2021}}}}\ (\bibinfo {year} {2022})\ \Eprint {http://arxiv.org/abs/2203.08192} {arXiv:2203.08192 [hep-ex]} \BibitemShut {NoStop}%
\bibitem [{\citenamefont {Akesson}\ \emph {et~al.}(2025)\citenamefont {Akesson} \emph {et~al.}}]{LDMX:2025bog}%
  \BibitemOpen
  \bibfield  {author} {\bibinfo {author} {\bibfnamefont {Torsten}\ \bibnamefont {Akesson}} \emph {et~al.} (\bibinfo {collaboration} {LDMX}),\ }\bibfield  {title} {\enquote {\bibinfo {title} {{LDMX - The Light Dark Matter eXperiment}},}\ }\href@noop {} {\  (\bibinfo {year} {2025})},\ \Eprint {http://arxiv.org/abs/2508.11833} {arXiv:2508.11833 [hep-ex]} \BibitemShut {NoStop}%
\bibitem [{\citenamefont {Tucker-Smith}\ and\ \citenamefont {Weiner}(2001)}]{Tucker-Smith:2001myb}%
  \BibitemOpen
  \bibfield  {author} {\bibinfo {author} {\bibfnamefont {David}\ \bibnamefont {Tucker-Smith}}\ and\ \bibinfo {author} {\bibfnamefont {Neal}\ \bibnamefont {Weiner}},\ }\bibfield  {title} {\enquote {\bibinfo {title} {{Inelastic dark matter}},}\ }\href {\doibase 10.1103/PhysRevD.64.043502} {\bibfield  {journal} {\bibinfo  {journal} {Phys. Rev. D}\ }\textbf {\bibinfo {volume} {64}},\ \bibinfo {pages} {043502} (\bibinfo {year} {2001})},\ \Eprint {http://arxiv.org/abs/hep-ph/0101138} {arXiv:hep-ph/0101138} \BibitemShut {NoStop}%
\bibitem [{\citenamefont {Tucker-Smith}\ and\ \citenamefont {Weiner}(2005)}]{Tucker-Smith:2004mxa}%
  \BibitemOpen
  \bibfield  {author} {\bibinfo {author} {\bibfnamefont {David}\ \bibnamefont {Tucker-Smith}}\ and\ \bibinfo {author} {\bibfnamefont {Neal}\ \bibnamefont {Weiner}},\ }\bibfield  {title} {\enquote {\bibinfo {title} {{The Status of inelastic dark matter}},}\ }\href {\doibase 10.1103/PhysRevD.72.063509} {\bibfield  {journal} {\bibinfo  {journal} {Phys. Rev. D}\ }\textbf {\bibinfo {volume} {72}},\ \bibinfo {pages} {063509} (\bibinfo {year} {2005})},\ \Eprint {http://arxiv.org/abs/hep-ph/0402065} {arXiv:hep-ph/0402065} \BibitemShut {NoStop}%
\bibitem [{\citenamefont {Chang}\ \emph {et~al.}(2009)\citenamefont {Chang}, \citenamefont {Kribs}, \citenamefont {Tucker-Smith},\ and\ \citenamefont {Weiner}}]{Chang:2008gd}%
  \BibitemOpen
  \bibfield  {author} {\bibinfo {author} {\bibfnamefont {Spencer}\ \bibnamefont {Chang}}, \bibinfo {author} {\bibfnamefont {Graham~D.}\ \bibnamefont {Kribs}}, \bibinfo {author} {\bibfnamefont {David}\ \bibnamefont {Tucker-Smith}}, \ and\ \bibinfo {author} {\bibfnamefont {Neal}\ \bibnamefont {Weiner}},\ }\bibfield  {title} {\enquote {\bibinfo {title} {{Inelastic Dark Matter in Light of DAMA/LIBRA}},}\ }\href {\doibase 10.1103/PhysRevD.79.043513} {\bibfield  {journal} {\bibinfo  {journal} {Phys. Rev. D}\ }\textbf {\bibinfo {volume} {79}},\ \bibinfo {pages} {043513} (\bibinfo {year} {2009})},\ \Eprint {http://arxiv.org/abs/0807.2250} {arXiv:0807.2250 [hep-ph]} \BibitemShut {NoStop}%
\bibitem [{\citenamefont {Bernabei}\ \emph {et~al.}(2013)\citenamefont {Bernabei} \emph {et~al.}}]{Bernabei:2013xsa}%
  \BibitemOpen
  \bibfield  {author} {\bibinfo {author} {\bibfnamefont {R.}~\bibnamefont {Bernabei}} \emph {et~al.},\ }\bibfield  {title} {\enquote {\bibinfo {title} {{Final model independent result of DAMA/LIBRA-phase1}},}\ }\href {\doibase 10.1140/epjc/s10052-013-2648-7} {\bibfield  {journal} {\bibinfo  {journal} {Eur. Phys. J. C}\ }\textbf {\bibinfo {volume} {73}},\ \bibinfo {pages} {2648} (\bibinfo {year} {2013})},\ \Eprint {http://arxiv.org/abs/1308.5109} {arXiv:1308.5109 [astro-ph.GA]} \BibitemShut {NoStop}%
\bibitem [{\citenamefont {Jordan}\ \emph {et~al.}(2018)\citenamefont {Jordan}, \citenamefont {Kahn}, \citenamefont {Krnjaic}, \citenamefont {Moschella},\ and\ \citenamefont {Spitz}}]{Jordan:2018gcd}%
  \BibitemOpen
  \bibfield  {author} {\bibinfo {author} {\bibfnamefont {Johnathon~R.}\ \bibnamefont {Jordan}}, \bibinfo {author} {\bibfnamefont {Yonatan}\ \bibnamefont {Kahn}}, \bibinfo {author} {\bibfnamefont {Gordan}\ \bibnamefont {Krnjaic}}, \bibinfo {author} {\bibfnamefont {Matthew}\ \bibnamefont {Moschella}}, \ and\ \bibinfo {author} {\bibfnamefont {Joshua}\ \bibnamefont {Spitz}},\ }\bibfield  {title} {\enquote {\bibinfo {title} {{Signatures of Pseudo-Dirac Dark Matter at High-Intensity Neutrino Experiments}},}\ }\href {\doibase 10.1103/PhysRevD.98.075020} {\bibfield  {journal} {\bibinfo  {journal} {Phys. Rev. D}\ }\textbf {\bibinfo {volume} {98}},\ \bibinfo {pages} {075020} (\bibinfo {year} {2018})},\ \Eprint {http://arxiv.org/abs/1806.05185} {arXiv:1806.05185 [hep-ph]} \BibitemShut {NoStop}%
\bibitem [{\citenamefont {Berlin}\ \emph {et~al.}(2018)\citenamefont {Berlin}, \citenamefont {Gori}, \citenamefont {Schuster},\ and\ \citenamefont {Toro}}]{Berlin:2018pwi}%
  \BibitemOpen
  \bibfield  {author} {\bibinfo {author} {\bibfnamefont {Asher}\ \bibnamefont {Berlin}}, \bibinfo {author} {\bibfnamefont {Stefania}\ \bibnamefont {Gori}}, \bibinfo {author} {\bibfnamefont {Philip}\ \bibnamefont {Schuster}}, \ and\ \bibinfo {author} {\bibfnamefont {Natalia}\ \bibnamefont {Toro}},\ }\bibfield  {title} {\enquote {\bibinfo {title} {{Dark Sectors at the Fermilab SeaQuest Experiment}},}\ }\href {\doibase 10.1103/PhysRevD.98.035011} {\bibfield  {journal} {\bibinfo  {journal} {Phys. Rev. D}\ }\textbf {\bibinfo {volume} {98}},\ \bibinfo {pages} {035011} (\bibinfo {year} {2018})},\ \Eprint {http://arxiv.org/abs/1804.00661} {arXiv:1804.00661 [hep-ph]} \BibitemShut {NoStop}%
\bibitem [{\citenamefont {Batell}\ \emph {et~al.}(2021)\citenamefont {Batell}, \citenamefont {Berger}, \citenamefont {Darm\'e},\ and\ \citenamefont {Frugiuele}}]{Batell:2021ooj}%
  \BibitemOpen
  \bibfield  {author} {\bibinfo {author} {\bibfnamefont {Brian}\ \bibnamefont {Batell}}, \bibinfo {author} {\bibfnamefont {Joshua}\ \bibnamefont {Berger}}, \bibinfo {author} {\bibfnamefont {Luc}\ \bibnamefont {Darm\'e}}, \ and\ \bibinfo {author} {\bibfnamefont {Claudia}\ \bibnamefont {Frugiuele}},\ }\bibfield  {title} {\enquote {\bibinfo {title} {{Inelastic dark matter at the Fermilab Short Baseline Neutrino Program}},}\ }\href {\doibase 10.1103/PhysRevD.104.075026} {\bibfield  {journal} {\bibinfo  {journal} {Phys. Rev. D}\ }\textbf {\bibinfo {volume} {104}},\ \bibinfo {pages} {075026} (\bibinfo {year} {2021})},\ \Eprint {http://arxiv.org/abs/2106.04584} {arXiv:2106.04584 [hep-ph]} \BibitemShut {NoStop}%
\bibitem [{\citenamefont {Chen}\ \emph {et~al.}(2018)\citenamefont {Chen}, \citenamefont {Kozaczuk},\ and\ \citenamefont {Zhong}}]{Chen:2018vkr}%
  \BibitemOpen
  \bibfield  {author} {\bibinfo {author} {\bibfnamefont {Chien-Yi}\ \bibnamefont {Chen}}, \bibinfo {author} {\bibfnamefont {Jonathan}\ \bibnamefont {Kozaczuk}}, \ and\ \bibinfo {author} {\bibfnamefont {Yi-Ming}\ \bibnamefont {Zhong}},\ }\bibfield  {title} {\enquote {\bibinfo {title} {{Exploring leptophilic dark matter with NA64-$\mu$}},}\ }\href {\doibase 10.1007/JHEP10(2018)154} {\bibfield  {journal} {\bibinfo  {journal} {JHEP}\ }\textbf {\bibinfo {volume} {10}},\ \bibinfo {pages} {154} (\bibinfo {year} {2018})},\ \Eprint {http://arxiv.org/abs/1807.03790} {arXiv:1807.03790 [hep-ph]} \BibitemShut {NoStop}%
\bibitem [{\citenamefont {Batell}\ \emph {et~al.}(2018)\citenamefont {Batell}, \citenamefont {Freitas}, \citenamefont {Ismail},\ and\ \citenamefont {Mckeen}}]{Batell:2017kty}%
  \BibitemOpen
  \bibfield  {author} {\bibinfo {author} {\bibfnamefont {Brian}\ \bibnamefont {Batell}}, \bibinfo {author} {\bibfnamefont {Ayres}\ \bibnamefont {Freitas}}, \bibinfo {author} {\bibfnamefont {Ahmed}\ \bibnamefont {Ismail}}, \ and\ \bibinfo {author} {\bibfnamefont {David}\ \bibnamefont {Mckeen}},\ }\bibfield  {title} {\enquote {\bibinfo {title} {{Flavor-specific scalar mediators}},}\ }\href {\doibase 10.1103/PhysRevD.98.055026} {\bibfield  {journal} {\bibinfo  {journal} {Phys. Rev. D}\ }\textbf {\bibinfo {volume} {98}},\ \bibinfo {pages} {055026} (\bibinfo {year} {2018})},\ \Eprint {http://arxiv.org/abs/1712.10022} {arXiv:1712.10022 [hep-ph]} \BibitemShut {NoStop}%
\bibitem [{\citenamefont {Dreiner}\ \emph {et~al.}(2010)\citenamefont {Dreiner}, \citenamefont {Haber},\ and\ \citenamefont {Martin}}]{Dreiner:2008tw}%
  \BibitemOpen
  \bibfield  {author} {\bibinfo {author} {\bibfnamefont {Herbi~K.}\ \bibnamefont {Dreiner}}, \bibinfo {author} {\bibfnamefont {Howard~E.}\ \bibnamefont {Haber}}, \ and\ \bibinfo {author} {\bibfnamefont {Stephen~P.}\ \bibnamefont {Martin}},\ }\bibfield  {title} {\enquote {\bibinfo {title} {{Two-component spinor techniques and Feynman rules for quantum field theory and supersymmetry}},}\ }\href {\doibase 10.1016/j.physrep.2010.05.002} {\bibfield  {journal} {\bibinfo  {journal} {Phys. Rept.}\ }\textbf {\bibinfo {volume} {494}},\ \bibinfo {pages} {1--196} (\bibinfo {year} {2010})},\ \Eprint {http://arxiv.org/abs/0812.1594} {arXiv:0812.1594 [hep-ph]} \BibitemShut {NoStop}%
\bibitem [{\citenamefont {Dalla Valle~Garcia}(2025{\natexlab{b}})}]{DallaValleGarcia:2024zva}%
  \BibitemOpen
  \bibfield  {author} {\bibinfo {author} {\bibfnamefont {Giovani}\ \bibnamefont {Dalla Valle~Garcia}},\ }\bibfield  {title} {\enquote {\bibinfo {title} {{A minimalistic model for inelastic dark matter}},}\ }\href {\doibase 10.1016/j.physletb.2025.139320} {\bibfield  {journal} {\bibinfo  {journal} {Phys. Lett. B}\ }\textbf {\bibinfo {volume} {862}},\ \bibinfo {pages} {139320} (\bibinfo {year} {2025}{\natexlab{b}})},\ \Eprint {http://arxiv.org/abs/2411.02147} {arXiv:2411.02147 [hep-ph]} \BibitemShut {NoStop}%
\bibitem [{\citenamefont {Izaguirre}\ \emph {et~al.}(2017)\citenamefont {Izaguirre}, \citenamefont {Kahn}, \citenamefont {Krnjaic},\ and\ \citenamefont {Moschella}}]{Izaguirre:2017bqb}%
  \BibitemOpen
  \bibfield  {author} {\bibinfo {author} {\bibfnamefont {Eder}\ \bibnamefont {Izaguirre}}, \bibinfo {author} {\bibfnamefont {Yonatan}\ \bibnamefont {Kahn}}, \bibinfo {author} {\bibfnamefont {Gordan}\ \bibnamefont {Krnjaic}}, \ and\ \bibinfo {author} {\bibfnamefont {Matthew}\ \bibnamefont {Moschella}},\ }\bibfield  {title} {\enquote {\bibinfo {title} {{Testing Light Dark Matter Coannihilation With Fixed-Target Experiments}},}\ }\href {\doibase 10.1103/PhysRevD.96.055007} {\bibfield  {journal} {\bibinfo  {journal} {Phys. Rev. D}\ }\textbf {\bibinfo {volume} {96}},\ \bibinfo {pages} {055007} (\bibinfo {year} {2017})},\ \Eprint {http://arxiv.org/abs/1703.06881} {arXiv:1703.06881 [hep-ph]} \BibitemShut {NoStop}%
\bibitem [{\citenamefont {Krnjaic}(2016)}]{Krnjaic:2015mbs}%
  \BibitemOpen
  \bibfield  {author} {\bibinfo {author} {\bibfnamefont {Gordan}\ \bibnamefont {Krnjaic}},\ }\bibfield  {title} {\enquote {\bibinfo {title} {{Probing Light Thermal Dark-Matter With a Higgs Portal Mediator}},}\ }\href {\doibase 10.1103/PhysRevD.94.073009} {\bibfield  {journal} {\bibinfo  {journal} {Phys. Rev. D}\ }\textbf {\bibinfo {volume} {94}},\ \bibinfo {pages} {073009} (\bibinfo {year} {2016})},\ \Eprint {http://arxiv.org/abs/1512.04119} {arXiv:1512.04119 [hep-ph]} \BibitemShut {NoStop}%
\bibitem [{\citenamefont {Foguel}\ \emph {et~al.}(2024)\citenamefont {Foguel}, \citenamefont {Reimitz},\ and\ \citenamefont {Funchal}}]{Foguel:2024lca}%
  \BibitemOpen
  \bibfield  {author} {\bibinfo {author} {\bibfnamefont {Ana~Luisa}\ \bibnamefont {Foguel}}, \bibinfo {author} {\bibfnamefont {Peter}\ \bibnamefont {Reimitz}}, \ and\ \bibinfo {author} {\bibfnamefont {Renata~Zukanovich}\ \bibnamefont {Funchal}},\ }\bibfield  {title} {\enquote {\bibinfo {title} {{Unlocking the Inelastic Dark Matter Window with Vector Mediators}},}\ }\href@noop {} {\  (\bibinfo {year} {2024})},\ \Eprint {http://arxiv.org/abs/2410.00881} {arXiv:2410.00881 [hep-ph]} \BibitemShut {NoStop}%
\bibitem [{\citenamefont {Krnjaic}(2025)}]{Krnjaic:2025noj}%
  \BibitemOpen
  \bibfield  {author} {\bibinfo {author} {\bibfnamefont {Gordan}\ \bibnamefont {Krnjaic}},\ }\bibfield  {title} {\enquote {\bibinfo {title} {{Testing Thermal-Relic Dark Matter with a Dark Photon Mediator}},}\ }\href@noop {} {\  (\bibinfo {year} {2025})},\ \Eprint {http://arxiv.org/abs/2505.04626} {arXiv:2505.04626 [hep-ph]} \BibitemShut {NoStop}%
\bibitem [{\citenamefont {B{\'e}langer}\ \emph {et~al.}(2018)\citenamefont {B{\'e}langer}, \citenamefont {Boudjema}, \citenamefont {Goudelis}, \citenamefont {Pukhov},\ and\ \citenamefont {Zaldivar}}]{Belanger:2018ccd}%
  \BibitemOpen
  \bibfield  {author} {\bibinfo {author} {\bibfnamefont {Genevi{\`e}ve}\ \bibnamefont {B{\'e}langer}}, \bibinfo {author} {\bibfnamefont {Fawzi}\ \bibnamefont {Boudjema}}, \bibinfo {author} {\bibfnamefont {Andreas}\ \bibnamefont {Goudelis}}, \bibinfo {author} {\bibfnamefont {Alexander}\ \bibnamefont {Pukhov}}, \ and\ \bibinfo {author} {\bibfnamefont {Bryan}\ \bibnamefont {Zaldivar}},\ }\bibfield  {title} {\enquote {\bibinfo {title} {{micrOMEGAs5.0 : Freeze-in}},}\ }\href {\doibase 10.1016/j.cpc.2018.04.027} {\bibfield  {journal} {\bibinfo  {journal} {Comput. Phys. Commun.}\ }\textbf {\bibinfo {volume} {231}},\ \bibinfo {pages} {173--186} (\bibinfo {year} {2018})},\ \Eprint {http://arxiv.org/abs/1801.03509} {arXiv:1801.03509 [hep-ph]} \BibitemShut {NoStop}%
\bibitem [{\citenamefont {Krnjaic}\ \emph {et~al.}(2025)\citenamefont {Krnjaic}, \citenamefont {McKeen}, \citenamefont {Mizuta}, \citenamefont {Mohlabeng}, \citenamefont {Morrissey},\ and\ \citenamefont {Tuckler}}]{Krnjaic:2025zjl}%
  \BibitemOpen
  \bibfield  {author} {\bibinfo {author} {\bibfnamefont {Gordan}\ \bibnamefont {Krnjaic}}, \bibinfo {author} {\bibfnamefont {David}\ \bibnamefont {McKeen}}, \bibinfo {author} {\bibfnamefont {Riku}\ \bibnamefont {Mizuta}}, \bibinfo {author} {\bibfnamefont {Gopolang}\ \bibnamefont {Mohlabeng}}, \bibinfo {author} {\bibfnamefont {David~E.}\ \bibnamefont {Morrissey}}, \ and\ \bibinfo {author} {\bibfnamefont {Douglas}\ \bibnamefont {Tuckler}},\ }\bibfield  {title} {\enquote {\bibinfo {title} {{X-rays from Inelastic Dark Matter Freeze-in}},}\ }\href@noop {} {\  (\bibinfo {year} {2025})},\ \Eprint {http://arxiv.org/abs/2509.19428} {arXiv:2509.19428 [hep-ph]} \BibitemShut {NoStop}%
\bibitem [{\citenamefont {Heeba}\ \emph {et~al.}(2023)\citenamefont {Heeba}, \citenamefont {Lin},\ and\ \citenamefont {Schutz}}]{Heeba:2023bik}%
  \BibitemOpen
  \bibfield  {author} {\bibinfo {author} {\bibfnamefont {Saniya}\ \bibnamefont {Heeba}}, \bibinfo {author} {\bibfnamefont {Tongyan}\ \bibnamefont {Lin}}, \ and\ \bibinfo {author} {\bibfnamefont {Katelin}\ \bibnamefont {Schutz}},\ }\bibfield  {title} {\enquote {\bibinfo {title} {{Inelastic freeze-in}},}\ }\href {\doibase 10.1103/PhysRevD.108.095016} {\bibfield  {journal} {\bibinfo  {journal} {Phys. Rev. D}\ }\textbf {\bibinfo {volume} {108}},\ \bibinfo {pages} {095016} (\bibinfo {year} {2023})},\ \Eprint {http://arxiv.org/abs/2304.06072} {arXiv:2304.06072 [hep-ph]} \BibitemShut {NoStop}%
\bibitem [{\citenamefont {D'Agnolo}\ and\ \citenamefont {Ruderman}(2015)}]{DAgnolo:2015ujb}%
  \BibitemOpen
  \bibfield  {author} {\bibinfo {author} {\bibfnamefont {Raffaele~Tito}\ \bibnamefont {D'Agnolo}}\ and\ \bibinfo {author} {\bibfnamefont {Joshua~T.}\ \bibnamefont {Ruderman}},\ }\bibfield  {title} {\enquote {\bibinfo {title} {{Light Dark Matter from Forbidden Channels}},}\ }\href {\doibase 10.1103/PhysRevLett.115.061301} {\bibfield  {journal} {\bibinfo  {journal} {Phys. Rev. Lett.}\ }\textbf {\bibinfo {volume} {115}},\ \bibinfo {pages} {061301} (\bibinfo {year} {2015})},\ \Eprint {http://arxiv.org/abs/1505.07107} {arXiv:1505.07107 [hep-ph]} \BibitemShut {NoStop}%
\bibitem [{\citenamefont {Fitzpatrick}\ \emph {et~al.}(2022)\citenamefont {Fitzpatrick}, \citenamefont {Liu}, \citenamefont {Slatyer},\ and\ \citenamefont {Tsai}}]{Fitzpatrick:2020vba}%
  \BibitemOpen
  \bibfield  {author} {\bibinfo {author} {\bibfnamefont {Patrick~J.}\ \bibnamefont {Fitzpatrick}}, \bibinfo {author} {\bibfnamefont {Hongwan}\ \bibnamefont {Liu}}, \bibinfo {author} {\bibfnamefont {Tracy~R.}\ \bibnamefont {Slatyer}}, \ and\ \bibinfo {author} {\bibfnamefont {Yu-Dai}\ \bibnamefont {Tsai}},\ }\bibfield  {title} {\enquote {\bibinfo {title} {{New pathways to the relic abundance of vector-portal dark matter}},}\ }\href {\doibase 10.1103/PhysRevD.106.083517} {\bibfield  {journal} {\bibinfo  {journal} {Phys. Rev. D}\ }\textbf {\bibinfo {volume} {106}},\ \bibinfo {pages} {083517} (\bibinfo {year} {2022})},\ \Eprint {http://arxiv.org/abs/2011.01240} {arXiv:2011.01240 [hep-ph]} \BibitemShut {NoStop}%
\bibitem [{\citenamefont {Husdal}(2016)}]{Husdal:2016haj}%
  \BibitemOpen
  \bibfield  {author} {\bibinfo {author} {\bibfnamefont {Lars}\ \bibnamefont {Husdal}},\ }\bibfield  {title} {\enquote {\bibinfo {title} {{On Effective Degrees of Freedom in the Early Universe}},}\ }\href {\doibase 10.3390/galaxies4040078} {\bibfield  {journal} {\bibinfo  {journal} {Galaxies}\ }\textbf {\bibinfo {volume} {4}},\ \bibinfo {pages} {78} (\bibinfo {year} {2016})},\ \Eprint {http://arxiv.org/abs/1609.04979} {arXiv:1609.04979 [astro-ph.CO]} \BibitemShut {NoStop}%
\bibitem [{\citenamefont {Kolb}(2019)}]{Kolb:1990vq}%
  \BibitemOpen
  \bibfield  {author} {\bibinfo {author} {\bibfnamefont {Edward~W.}\ \bibnamefont {Kolb}},\ }\href {\doibase 10.1201/9780429492860} {\emph {\bibinfo {title} {{The Early Universe}}}},\ Vol.~\bibinfo {volume} {69}\ (\bibinfo  {publisher} {Taylor and Francis},\ \bibinfo {year} {2019})\BibitemShut {NoStop}%
\bibitem [{\citenamefont {Gninenko}\ \emph {et~al.}(2015)\citenamefont {Gninenko}, \citenamefont {Krasnikov},\ and\ \citenamefont {Matveev}}]{Gninenko:2014pea}%
  \BibitemOpen
  \bibfield  {author} {\bibinfo {author} {\bibfnamefont {S.~N.}\ \bibnamefont {Gninenko}}, \bibinfo {author} {\bibfnamefont {N.~V.}\ \bibnamefont {Krasnikov}}, \ and\ \bibinfo {author} {\bibfnamefont {V.~A.}\ \bibnamefont {Matveev}},\ }\bibfield  {title} {\enquote {\bibinfo {title} {{Muon g-2 and searches for a new leptophobic sub-GeV dark boson in a missing-energy experiment at CERN}},}\ }\href {\doibase 10.1103/PhysRevD.91.095015} {\bibfield  {journal} {\bibinfo  {journal} {Phys. Rev. D}\ }\textbf {\bibinfo {volume} {91}},\ \bibinfo {pages} {095015} (\bibinfo {year} {2015})},\ \Eprint {http://arxiv.org/abs/1412.1400} {arXiv:1412.1400 [hep-ph]} \BibitemShut {NoStop}%
\bibitem [{\citenamefont {Gninenko}\ and\ \citenamefont {Krasnikov}(2018)}]{Gninenko:2018tlp}%
  \BibitemOpen
  \bibfield  {author} {\bibinfo {author} {\bibfnamefont {S.~N.}\ \bibnamefont {Gninenko}}\ and\ \bibinfo {author} {\bibfnamefont {N.~V.}\ \bibnamefont {Krasnikov}},\ }\bibfield  {title} {\enquote {\bibinfo {title} {{Probing the muon $g_\mu$ - 2 anomaly, $L_\mu - L_\tau$ gauge boson and Dark Matter in dark photon experiments}},}\ }\href {\doibase 10.1016/j.physletb.2018.06.043} {\bibfield  {journal} {\bibinfo  {journal} {Phys. Lett. B}\ }\textbf {\bibinfo {volume} {783}},\ \bibinfo {pages} {24--28} (\bibinfo {year} {2018})},\ \Eprint {http://arxiv.org/abs/1801.10448} {arXiv:1801.10448 [hep-ph]} \BibitemShut {NoStop}%
\bibitem [{\citenamefont {Andreev}\ \emph {et~al.}(2024{\natexlab{b}})\citenamefont {Andreev} \emph {et~al.}}]{NA64:2024nwj}%
  \BibitemOpen
  \bibfield  {author} {\bibinfo {author} {\bibfnamefont {Yu.~M.}\ \bibnamefont {Andreev}} \emph {et~al.} (\bibinfo {collaboration} {NA64}),\ }\bibfield  {title} {\enquote {\bibinfo {title} {{Shedding light on dark sectors with high-energy muons at the NA64 experiment at the CERN SPS}},}\ }\href {\doibase 10.1103/PhysRevD.110.112015} {\bibfield  {journal} {\bibinfo  {journal} {Phys. Rev. D}\ }\textbf {\bibinfo {volume} {110}},\ \bibinfo {pages} {112015} (\bibinfo {year} {2024}{\natexlab{b}})},\ \Eprint {http://arxiv.org/abs/2409.10128} {arXiv:2409.10128 [hep-ex]} \BibitemShut {NoStop}%
\bibitem [{\citenamefont {Voronchikhin}\ and\ \citenamefont {Kirpichnikov}(2025)}]{Voronchikhin:2024vfu}%
  \BibitemOpen
  \bibfield  {author} {\bibinfo {author} {\bibfnamefont {I.~V.}\ \bibnamefont {Voronchikhin}}\ and\ \bibinfo {author} {\bibfnamefont {D.~V.}\ \bibnamefont {Kirpichnikov}},\ }\bibfield  {title} {\enquote {\bibinfo {title} {{Implication of the Weizsacker-Williams approximation for the dark matter mediator production}},}\ }\href {\doibase 10.1103/PhysRevD.111.035034} {\bibfield  {journal} {\bibinfo  {journal} {Phys. Rev. D}\ }\textbf {\bibinfo {volume} {111}},\ \bibinfo {pages} {035034} (\bibinfo {year} {2025})},\ \Eprint {http://arxiv.org/abs/2409.12748} {arXiv:2409.12748 [hep-ph]} \BibitemShut {NoStop}%
\bibitem [{\citenamefont {Tsai}\ and\ \citenamefont {Whitis}(1966)}]{Tsai:1966js}%
  \BibitemOpen
  \bibfield  {author} {\bibinfo {author} {\bibfnamefont {Yung-Su}\ \bibnamefont {Tsai}}\ and\ \bibinfo {author} {\bibfnamefont {Van}\ \bibnamefont {Whitis}},\ }\bibfield  {title} {\enquote {\bibinfo {title} {{THICK TARGET BREMSSTRAHLUNG AND TARGET CONSIDERATION FOR SECONDARY PARTICLE PRODUCTION BY ELECTRONS}},}\ }\href {\doibase 10.1103/PhysRev.149.1248} {\bibfield  {journal} {\bibinfo  {journal} {Phys. Rev.}\ }\textbf {\bibinfo {volume} {149}},\ \bibinfo {pages} {1248--1257} (\bibinfo {year} {1966})}\BibitemShut {NoStop}%
\bibitem [{\citenamefont {Tsai}(1986)}]{Tsai:1986tx}%
  \BibitemOpen
  \bibfield  {author} {\bibinfo {author} {\bibfnamefont {Yung-Su}\ \bibnamefont {Tsai}},\ }\bibfield  {title} {\enquote {\bibinfo {title} {{AXION BREMSSTRAHLUNG BY AN ELECTRON BEAM}},}\ }\href {\doibase 10.1103/PhysRevD.34.1326} {\bibfield  {journal} {\bibinfo  {journal} {Phys. Rev. D}\ }\textbf {\bibinfo {volume} {34}},\ \bibinfo {pages} {1326} (\bibinfo {year} {1986})}\BibitemShut {NoStop}%
\bibitem [{\citenamefont {Andreas}\ \emph {et~al.}(2012)\citenamefont {Andreas}, \citenamefont {Niebuhr},\ and\ \citenamefont {Ringwald}}]{Andreas:2012mt}%
  \BibitemOpen
  \bibfield  {author} {\bibinfo {author} {\bibfnamefont {Sarah}\ \bibnamefont {Andreas}}, \bibinfo {author} {\bibfnamefont {Carsten}\ \bibnamefont {Niebuhr}}, \ and\ \bibinfo {author} {\bibfnamefont {Andreas}\ \bibnamefont {Ringwald}},\ }\bibfield  {title} {\enquote {\bibinfo {title} {{New Limits on Hidden Photons from Past Electron Beam Dumps}},}\ }\href {\doibase 10.1103/PhysRevD.86.095019} {\bibfield  {journal} {\bibinfo  {journal} {Phys. Rev. D}\ }\textbf {\bibinfo {volume} {86}},\ \bibinfo {pages} {095019} (\bibinfo {year} {2012})},\ \Eprint {http://arxiv.org/abs/1209.6083} {arXiv:1209.6083 [hep-ph]} \BibitemShut {NoStop}%
\bibitem [{\citenamefont {Liu}\ \emph {et~al.}(2017)\citenamefont {Liu}, \citenamefont {McKeen},\ and\ \citenamefont {Miller}}]{Liu:2016mqv}%
  \BibitemOpen
  \bibfield  {author} {\bibinfo {author} {\bibfnamefont {Yu-Sheng}\ \bibnamefont {Liu}}, \bibinfo {author} {\bibfnamefont {David}\ \bibnamefont {McKeen}}, \ and\ \bibinfo {author} {\bibfnamefont {Gerald~A.}\ \bibnamefont {Miller}},\ }\bibfield  {title} {\enquote {\bibinfo {title} {{Validity of the Weizs\"acker-Williams approximation and the analysis of beam dump experiments: Production of a new scalar boson}},}\ }\href {\doibase 10.1103/PhysRevD.95.036010} {\bibfield  {journal} {\bibinfo  {journal} {Phys. Rev. D}\ }\textbf {\bibinfo {volume} {95}},\ \bibinfo {pages} {036010} (\bibinfo {year} {2017})},\ \Eprint {http://arxiv.org/abs/1609.06781} {arXiv:1609.06781 [hep-ph]} \BibitemShut {NoStop}%
\bibitem [{\citenamefont {Liu}\ and\ \citenamefont {Miller}(2017)}]{Liu:2017htz}%
  \BibitemOpen
  \bibfield  {author} {\bibinfo {author} {\bibfnamefont {Yu-Sheng}\ \bibnamefont {Liu}}\ and\ \bibinfo {author} {\bibfnamefont {Gerald~A.}\ \bibnamefont {Miller}},\ }\bibfield  {title} {\enquote {\bibinfo {title} {{Validity of the Weizs\"acker-Williams approximation and the analysis of beam dump experiments: Production of an axion, a dark photon, or a new axial-vector boson}},}\ }\href {\doibase 10.1103/PhysRevD.96.016004} {\bibfield  {journal} {\bibinfo  {journal} {Phys. Rev. D}\ }\textbf {\bibinfo {volume} {96}},\ \bibinfo {pages} {016004} (\bibinfo {year} {2017})},\ \Eprint {http://arxiv.org/abs/1705.01633} {arXiv:1705.01633 [hep-ph]} \BibitemShut {NoStop}%
\bibitem [{\citenamefont {Bjorken}\ \emph {et~al.}(2009)\citenamefont {Bjorken}, \citenamefont {Essig}, \citenamefont {Schuster},\ and\ \citenamefont {Toro}}]{Bjorken:2009mm}%
  \BibitemOpen
  \bibfield  {author} {\bibinfo {author} {\bibfnamefont {James~D.}\ \bibnamefont {Bjorken}}, \bibinfo {author} {\bibfnamefont {Rouven}\ \bibnamefont {Essig}}, \bibinfo {author} {\bibfnamefont {Philip}\ \bibnamefont {Schuster}}, \ and\ \bibinfo {author} {\bibfnamefont {Natalia}\ \bibnamefont {Toro}},\ }\bibfield  {title} {\enquote {\bibinfo {title} {{New Fixed-Target Experiments to Search for Dark Gauge Forces}},}\ }\href {\doibase 10.1103/PhysRevD.80.075018} {\bibfield  {journal} {\bibinfo  {journal} {Phys. Rev. D}\ }\textbf {\bibinfo {volume} {80}},\ \bibinfo {pages} {075018} (\bibinfo {year} {2009})},\ \Eprint {http://arxiv.org/abs/0906.0580} {arXiv:0906.0580 [hep-ph]} \BibitemShut {NoStop}%
\bibitem [{\citenamefont {Andreev}\ \emph {et~al.}(2023)\citenamefont {Andreev} \emph {et~al.}}]{NA64:2023wbi}%
  \BibitemOpen
  \bibfield  {author} {\bibinfo {author} {\bibfnamefont {Yu.~M.}\ \bibnamefont {Andreev}} \emph {et~al.} (\bibinfo {collaboration} {NA64}),\ }\bibfield  {title} {\enquote {\bibinfo {title} {{Search for Light Dark Matter with NA64 at CERN}},}\ }\href {\doibase 10.1103/PhysRevLett.131.161801} {\bibfield  {journal} {\bibinfo  {journal} {Phys. Rev. Lett.}\ }\textbf {\bibinfo {volume} {131}},\ \bibinfo {pages} {161801} (\bibinfo {year} {2023})},\ \Eprint {http://arxiv.org/abs/2307.02404} {arXiv:2307.02404 [hep-ex]} \BibitemShut {NoStop}%
\bibitem [{\citenamefont {Gninenko}(2014)}]{Gninenko:2013rka}%
  \BibitemOpen
  \bibfield  {author} {\bibinfo {author} {\bibfnamefont {S.~N.}\ \bibnamefont {Gninenko}},\ }\bibfield  {title} {\enquote {\bibinfo {title} {{Search for MeV dark photons in a light-shining-through-walls experiment at CERN}},}\ }\href {\doibase 10.1103/PhysRevD.89.075008} {\bibfield  {journal} {\bibinfo  {journal} {Phys. Rev. D}\ }\textbf {\bibinfo {volume} {89}},\ \bibinfo {pages} {075008} (\bibinfo {year} {2014})},\ \Eprint {http://arxiv.org/abs/1308.6521} {arXiv:1308.6521 [hep-ph]} \BibitemShut {NoStop}%
\bibitem [{\citenamefont {Dworkin}\ \emph {et~al.}(1986)\citenamefont {Dworkin} \emph {et~al.}}]{Dworkin:1986tk}%
  \BibitemOpen
  \bibfield  {author} {\bibinfo {author} {\bibfnamefont {J.~S.}\ \bibnamefont {Dworkin}} \emph {et~al.},\ }\bibfield  {title} {\enquote {\bibinfo {title} {{Electron Identification using a Synchrotron Radiation Detector}},}\ }\href {\doibase 10.1016/0168-9002(86)91325-2} {\bibfield  {journal} {\bibinfo  {journal} {Nucl. Instrum. Meth. A}\ }\textbf {\bibinfo {volume} {247}},\ \bibinfo {pages} {412--419} (\bibinfo {year} {1986})}\BibitemShut {NoStop}%
\bibitem [{NA6(2018)}]{NA64:2018iqr}%
  \BibitemOpen
  \bibfield  {title} {\enquote {\bibinfo {title} {{Addendum to the Proposal P348: Search for dark sector particles weakly coupled to muon with NA64 {\ensuremath{\mu}}}},}\ }\href@noop {} {\  (\bibinfo {year} {2018})}\BibitemShut {NoStop}%
\bibitem [{\citenamefont {Davoudiasl}\ and\ \citenamefont {Marciano}(2015)}]{Davoudiasl:2015hxa}%
  \BibitemOpen
  \bibfield  {author} {\bibinfo {author} {\bibfnamefont {Hooman}\ \bibnamefont {Davoudiasl}}\ and\ \bibinfo {author} {\bibfnamefont {William~J.}\ \bibnamefont {Marciano}},\ }\bibfield  {title} {\enquote {\bibinfo {title} {{Running of the U(1) coupling in the dark sector}},}\ }\href {\doibase 10.1103/PhysRevD.92.035008} {\bibfield  {journal} {\bibinfo  {journal} {Phys. Rev. D}\ }\textbf {\bibinfo {volume} {92}},\ \bibinfo {pages} {035008} (\bibinfo {year} {2015})},\ \Eprint {http://arxiv.org/abs/1502.07383} {arXiv:1502.07383 [hep-ph]} \BibitemShut {NoStop}%
\bibitem [{\citenamefont {Giudice}\ \emph {et~al.}(2018)\citenamefont {Giudice}, \citenamefont {Kim}, \citenamefont {Park},\ and\ \citenamefont {Shin}}]{Giudice:2017zke}%
  \BibitemOpen
  \bibfield  {author} {\bibinfo {author} {\bibfnamefont {Gian~F.}\ \bibnamefont {Giudice}}, \bibinfo {author} {\bibfnamefont {Doojin}\ \bibnamefont {Kim}}, \bibinfo {author} {\bibfnamefont {Jong-Chul}\ \bibnamefont {Park}}, \ and\ \bibinfo {author} {\bibfnamefont {Seodong}\ \bibnamefont {Shin}},\ }\bibfield  {title} {\enquote {\bibinfo {title} {{Inelastic Boosted Dark Matter at Direct Detection Experiments}},}\ }\href {\doibase 10.1016/j.physletb.2018.03.043} {\bibfield  {journal} {\bibinfo  {journal} {Phys. Lett. B}\ }\textbf {\bibinfo {volume} {780}},\ \bibinfo {pages} {543--552} (\bibinfo {year} {2018})},\ \Eprint {http://arxiv.org/abs/1712.07126} {arXiv:1712.07126 [hep-ph]} \BibitemShut {NoStop}%
\bibitem [{\citenamefont {Lees}\ \emph {et~al.}(2017)\citenamefont {Lees} \emph {et~al.}}]{BaBar:2017tiz}%
  \BibitemOpen
  \bibfield  {author} {\bibinfo {author} {\bibfnamefont {J.~P.}\ \bibnamefont {Lees}} \emph {et~al.} (\bibinfo {collaboration} {BaBar}),\ }\bibfield  {title} {\enquote {\bibinfo {title} {{Search for Invisible Decays of a Dark Photon Produced in ${e}^{+}{e}^{-}$ Collisions at BaBar}},}\ }\href {\doibase 10.1103/PhysRevLett.119.131804} {\bibfield  {journal} {\bibinfo  {journal} {Phys. Rev. Lett.}\ }\textbf {\bibinfo {volume} {119}},\ \bibinfo {pages} {131804} (\bibinfo {year} {2017})},\ \Eprint {http://arxiv.org/abs/1702.03327} {arXiv:1702.03327 [hep-ex]} \BibitemShut {NoStop}%
\bibitem [{\citenamefont {Cortina~Gil}\ \emph {et~al.}(2021)\citenamefont {Cortina~Gil} \emph {et~al.}}]{NA62:2021bji}%
  \BibitemOpen
  \bibfield  {author} {\bibinfo {author} {\bibfnamefont {Eduardo}\ \bibnamefont {Cortina~Gil}} \emph {et~al.} (\bibinfo {collaboration} {NA62}),\ }\bibfield  {title} {\enquote {\bibinfo {title} {{Search for $K^+$ decays to a muon and invisible particles}},}\ }\href {\doibase 10.1016/j.physletb.2021.136259} {\bibfield  {journal} {\bibinfo  {journal} {Phys. Lett. B}\ }\textbf {\bibinfo {volume} {816}},\ \bibinfo {pages} {136259} (\bibinfo {year} {2021})},\ \Eprint {http://arxiv.org/abs/2101.12304} {arXiv:2101.12304 [hep-ex]} \BibitemShut {NoStop}%
\bibitem [{\citenamefont {Krnjaic}\ \emph {et~al.}(2020)\citenamefont {Krnjaic}, \citenamefont {Marques-Tavares}, \citenamefont {Redigolo},\ and\ \citenamefont {Tobioka}}]{Krnjaic:2019rsv}%
  \BibitemOpen
  \bibfield  {author} {\bibinfo {author} {\bibfnamefont {Gordan}\ \bibnamefont {Krnjaic}}, \bibinfo {author} {\bibfnamefont {Gustavo}\ \bibnamefont {Marques-Tavares}}, \bibinfo {author} {\bibfnamefont {Diego}\ \bibnamefont {Redigolo}}, \ and\ \bibinfo {author} {\bibfnamefont {Kohsaku}\ \bibnamefont {Tobioka}},\ }\bibfield  {title} {\enquote {\bibinfo {title} {{Probing Muonphilic Force Carriers and Dark Matter at Kaon Factories}},}\ }\href {\doibase 10.1103/PhysRevLett.124.041802} {\bibfield  {journal} {\bibinfo  {journal} {Phys. Rev. Lett.}\ }\textbf {\bibinfo {volume} {124}},\ \bibinfo {pages} {041802} (\bibinfo {year} {2020})},\ \Eprint {http://arxiv.org/abs/1902.07715} {arXiv:1902.07715 [hep-ph]} \BibitemShut {NoStop}%
\bibitem [{\citenamefont {Blinov}\ \emph {et~al.}(2024)\citenamefont {Blinov}, \citenamefont {Gori},\ and\ \citenamefont {Hamer}}]{Blinov:2024gcw}%
  \BibitemOpen
  \bibfield  {author} {\bibinfo {author} {\bibfnamefont {Nikita}\ \bibnamefont {Blinov}}, \bibinfo {author} {\bibfnamefont {Stefania}\ \bibnamefont {Gori}}, \ and\ \bibinfo {author} {\bibfnamefont {Nick}\ \bibnamefont {Hamer}},\ }\bibfield  {title} {\enquote {\bibinfo {title} {{Diphoton signals of muon-philic scalars at DarkQuest}},}\ }\href {\doibase 10.1103/PhysRevD.110.075006} {\bibfield  {journal} {\bibinfo  {journal} {Phys. Rev. D}\ }\textbf {\bibinfo {volume} {110}},\ \bibinfo {pages} {075006} (\bibinfo {year} {2024})},\ \Eprint {http://arxiv.org/abs/2405.17651} {arXiv:2405.17651 [hep-ph]} \BibitemShut {NoStop}%
\bibitem [{\citenamefont {Griest}\ and\ \citenamefont {Seckel}(1991)}]{Griest:1990kh}%
  \BibitemOpen
  \bibfield  {author} {\bibinfo {author} {\bibfnamefont {Kim}\ \bibnamefont {Griest}}\ and\ \bibinfo {author} {\bibfnamefont {David}\ \bibnamefont {Seckel}},\ }\bibfield  {title} {\enquote {\bibinfo {title} {{Three exceptions in the calculation of relic abundances}},}\ }\href {\doibase 10.1103/PhysRevD.43.3191} {\bibfield  {journal} {\bibinfo  {journal} {Phys. Rev. D}\ }\textbf {\bibinfo {volume} {43}},\ \bibinfo {pages} {3191--3203} (\bibinfo {year} {1991})}\BibitemShut {NoStop}%
\bibitem [{\citenamefont {Edsjo}\ and\ \citenamefont {Gondolo}(1997)}]{Edsjo:1997bg}%
  \BibitemOpen
  \bibfield  {author} {\bibinfo {author} {\bibfnamefont {Joakim}\ \bibnamefont {Edsjo}}\ and\ \bibinfo {author} {\bibfnamefont {Paolo}\ \bibnamefont {Gondolo}},\ }\bibfield  {title} {\enquote {\bibinfo {title} {{Neutralino relic density including coannihilations}},}\ }\href {\doibase 10.1103/PhysRevD.56.1879} {\bibfield  {journal} {\bibinfo  {journal} {Phys. Rev. D}\ }\textbf {\bibinfo {volume} {56}},\ \bibinfo {pages} {1879--1894} (\bibinfo {year} {1997})},\ \Eprint {http://arxiv.org/abs/hep-ph/9704361} {arXiv:hep-ph/9704361} \BibitemShut {NoStop}%
\bibitem [{\citenamefont {Gondolo}\ and\ \citenamefont {Gelmini}(1991)}]{Gondolo:1990dk}%
  \BibitemOpen
  \bibfield  {author} {\bibinfo {author} {\bibfnamefont {Paolo}\ \bibnamefont {Gondolo}}\ and\ \bibinfo {author} {\bibfnamefont {Graciela}\ \bibnamefont {Gelmini}},\ }\bibfield  {title} {\enquote {\bibinfo {title} {{Cosmic abundances of stable particles: Improved analysis}},}\ }\href {\doibase 10.1016/0550-3213(91)90438-4} {\bibfield  {journal} {\bibinfo  {journal} {Nucl. Phys. B}\ }\textbf {\bibinfo {volume} {360}},\ \bibinfo {pages} {145--179} (\bibinfo {year} {1991})}\BibitemShut {NoStop}%
\bibitem [{\citenamefont {Shtabovenko}\ \emph {et~al.}(2020)\citenamefont {Shtabovenko}, \citenamefont {Mertig},\ and\ \citenamefont {Orellana}}]{Shtabovenko:2020gxv}%
  \BibitemOpen
  \bibfield  {author} {\bibinfo {author} {\bibfnamefont {Vladyslav}\ \bibnamefont {Shtabovenko}}, \bibinfo {author} {\bibfnamefont {Rolf}\ \bibnamefont {Mertig}}, \ and\ \bibinfo {author} {\bibfnamefont {Frederik}\ \bibnamefont {Orellana}},\ }\bibfield  {title} {\enquote {\bibinfo {title} {{FeynCalc 9.3: New features and improvements}},}\ }\href {\doibase 10.1016/j.cpc.2020.107478} {\bibfield  {journal} {\bibinfo  {journal} {Comput. Phys. Commun.}\ }\textbf {\bibinfo {volume} {256}},\ \bibinfo {pages} {107478} (\bibinfo {year} {2020})},\ \Eprint {http://arxiv.org/abs/2001.04407} {arXiv:2001.04407 [hep-ph]} \BibitemShut {NoStop}%
\bibitem [{\citenamefont {Shtabovenko}\ \emph {et~al.}(2016)\citenamefont {Shtabovenko}, \citenamefont {Mertig},\ and\ \citenamefont {Orellana}}]{Shtabovenko:2016sxi}%
  \BibitemOpen
  \bibfield  {author} {\bibinfo {author} {\bibfnamefont {Vladyslav}\ \bibnamefont {Shtabovenko}}, \bibinfo {author} {\bibfnamefont {Rolf}\ \bibnamefont {Mertig}}, \ and\ \bibinfo {author} {\bibfnamefont {Frederik}\ \bibnamefont {Orellana}},\ }\bibfield  {title} {\enquote {\bibinfo {title} {{New Developments in FeynCalc 9.0}},}\ }\href {\doibase 10.1016/j.cpc.2016.06.008} {\bibfield  {journal} {\bibinfo  {journal} {Comput. Phys. Commun.}\ }\textbf {\bibinfo {volume} {207}},\ \bibinfo {pages} {432--444} (\bibinfo {year} {2016})},\ \Eprint {http://arxiv.org/abs/1601.01167} {arXiv:1601.01167 [hep-ph]} \BibitemShut {NoStop}%
\bibitem [{\citenamefont {Inc.}()}]{Mathematica}%
  \BibitemOpen
  \bibfield  {author} {\bibinfo {author} {\bibfnamefont {Wolfram~Research{,}}\ \bibnamefont {Inc.}},\ }\href {https://www.wolfram.com/mathematica} {\enquote {\bibinfo {title} {Mathematica, {V}ersion 13.1},}\ }\bibinfo {note} {Champaign, IL, 2022}\BibitemShut {NoStop}%
\bibitem [{\citenamefont {Hooper}(2019)}]{Hooper:2018kfv}%
  \BibitemOpen
  \bibfield  {author} {\bibinfo {author} {\bibfnamefont {Dan}\ \bibnamefont {Hooper}},\ }\bibfield  {title} {\enquote {\bibinfo {title} {{TASI Lectures on Indirect Searches For Dark Matter}},}\ }\href@noop {} {\bibfield  {journal} {\bibinfo  {journal} {PoS}\ }\textbf {\bibinfo {volume} {TASI2018}},\ \bibinfo {pages} {010} (\bibinfo {year} {2019})},\ \Eprint {http://arxiv.org/abs/1812.02029} {arXiv:1812.02029 [hep-ph]} \BibitemShut {NoStop}%
\bibitem [{\citenamefont {Choi}\ \emph {et~al.}(2017)\citenamefont {Choi}, \citenamefont {Lee},\ and\ \citenamefont {Seo}}]{Choi:2017mkk}%
  \BibitemOpen
  \bibfield  {author} {\bibinfo {author} {\bibfnamefont {Soo-Min}\ \bibnamefont {Choi}}, \bibinfo {author} {\bibfnamefont {Hyun~Min}\ \bibnamefont {Lee}}, \ and\ \bibinfo {author} {\bibfnamefont {Min-Seok}\ \bibnamefont {Seo}},\ }\bibfield  {title} {\enquote {\bibinfo {title} {{Cosmic abundances of SIMP dark matter}},}\ }\href {\doibase 10.1007/JHEP04(2017)154} {\bibfield  {journal} {\bibinfo  {journal} {JHEP}\ }\textbf {\bibinfo {volume} {04}},\ \bibinfo {pages} {154} (\bibinfo {year} {2017})},\ \Eprint {http://arxiv.org/abs/1702.07860} {arXiv:1702.07860 [hep-ph]} \BibitemShut {NoStop}%
\bibitem [{\citenamefont {Wells}(1994)}]{Wells:1994qy}%
  \BibitemOpen
  \bibfield  {author} {\bibinfo {author} {\bibfnamefont {James~D.}\ \bibnamefont {Wells}},\ }\bibfield  {title} {\enquote {\bibinfo {title} {{Annihilation cross-sections for relic densities in the low velocity limit}},}\ }\href@noop {} {\  (\bibinfo {year} {1994})},\ \Eprint {http://arxiv.org/abs/hep-ph/9404219} {arXiv:hep-ph/9404219} \BibitemShut {NoStop}%
\bibitem [{\citenamefont {Ellis}\ \emph {et~al.}(2000)\citenamefont {Ellis}, \citenamefont {Falk}, \citenamefont {Olive},\ and\ \citenamefont {Srednicki}}]{Ellis:1999mm}%
  \BibitemOpen
  \bibfield  {author} {\bibinfo {author} {\bibfnamefont {John~R.}\ \bibnamefont {Ellis}}, \bibinfo {author} {\bibfnamefont {Toby}\ \bibnamefont {Falk}}, \bibinfo {author} {\bibfnamefont {Keith~A.}\ \bibnamefont {Olive}}, \ and\ \bibinfo {author} {\bibfnamefont {Mark}\ \bibnamefont {Srednicki}},\ }\bibfield  {title} {\enquote {\bibinfo {title} {{Calculations of neutralino-stau coannihilation channels and the cosmologically relevant region of MSSM parameter space}},}\ }\href {\doibase 10.1016/S0927-6505(99)00104-8} {\bibfield  {journal} {\bibinfo  {journal} {Astropart. Phys.}\ }\textbf {\bibinfo {volume} {13}},\ \bibinfo {pages} {181--213} (\bibinfo {year} {2000})},\ \bibinfo {note} {[Erratum: Astropart.Phys. 15, 413--414 (2001)]},\ \Eprint {http://arxiv.org/abs/hep-ph/9905481} {arXiv:hep-ph/9905481} \BibitemShut {NoStop}%
\bibitem [{\citenamefont {Harigaya}\ \emph {et~al.}(2020)\citenamefont {Harigaya}, \citenamefont {Nakai},\ and\ \citenamefont {Suzuki}}]{Harigaya:2020ckz}%
  \BibitemOpen
  \bibfield  {author} {\bibinfo {author} {\bibfnamefont {Keisuke}\ \bibnamefont {Harigaya}}, \bibinfo {author} {\bibfnamefont {Yuichiro}\ \bibnamefont {Nakai}}, \ and\ \bibinfo {author} {\bibfnamefont {Motoo}\ \bibnamefont {Suzuki}},\ }\bibfield  {title} {\enquote {\bibinfo {title} {{Inelastic Dark Matter Electron Scattering and the XENON1T Excess}},}\ }\href {\doibase 10.1016/j.physletb.2020.135729} {\bibfield  {journal} {\bibinfo  {journal} {Phys. Lett. B}\ }\textbf {\bibinfo {volume} {809}},\ \bibinfo {pages} {135729} (\bibinfo {year} {2020})},\ \Eprint {http://arxiv.org/abs/2006.11938} {arXiv:2006.11938 [hep-ph]} \BibitemShut {NoStop}%
\bibitem [{\citenamefont {Wang}\ \emph {et~al.}(2025)\citenamefont {Wang}, \citenamefont {Yun}, \citenamefont {He},\ and\ \citenamefont {Meng}}]{Wang:2025uwh}%
  \BibitemOpen
  \bibfield  {author} {\bibinfo {author} {\bibfnamefont {Yu-Chen}\ \bibnamefont {Wang}}, \bibinfo {author} {\bibfnamefont {Youhui}\ \bibnamefont {Yun}}, \bibinfo {author} {\bibfnamefont {Hong-Jian}\ \bibnamefont {He}}, \ and\ \bibinfo {author} {\bibfnamefont {Yue}\ \bibnamefont {Meng}},\ }\bibfield  {title} {\enquote {\bibinfo {title} {{Search for Light Inelastic Dark Matter with Low-Energy Ionization Signatures in PandaX-4T}},}\ }\href@noop {} {\  (\bibinfo {year} {2025})},\ \Eprint {http://arxiv.org/abs/2508.13062} {arXiv:2508.13062 [hep-ph]} \BibitemShut {NoStop}%
\bibitem [{\citenamefont {Weiner}\ and\ \citenamefont {Yavin}(2012)}]{Weiner:2012cb}%
  \BibitemOpen
  \bibfield  {author} {\bibinfo {author} {\bibfnamefont {Neal}\ \bibnamefont {Weiner}}\ and\ \bibinfo {author} {\bibfnamefont {Itay}\ \bibnamefont {Yavin}},\ }\bibfield  {title} {\enquote {\bibinfo {title} {{How Dark Are Majorana WIMPs? Signals from MiDM and Rayleigh Dark Matter}},}\ }\href {\doibase 10.1103/PhysRevD.86.075021} {\bibfield  {journal} {\bibinfo  {journal} {Phys. Rev. D}\ }\textbf {\bibinfo {volume} {86}},\ \bibinfo {pages} {075021} (\bibinfo {year} {2012})},\ \Eprint {http://arxiv.org/abs/1206.2910} {arXiv:1206.2910 [hep-ph]} \BibitemShut {NoStop}%
\bibitem [{\citenamefont {Batell}\ \emph {et~al.}(2009)\citenamefont {Batell}, \citenamefont {Pospelov},\ and\ \citenamefont {Ritz}}]{Batell:2009vb}%
  \BibitemOpen
  \bibfield  {author} {\bibinfo {author} {\bibfnamefont {Brian}\ \bibnamefont {Batell}}, \bibinfo {author} {\bibfnamefont {Maxim}\ \bibnamefont {Pospelov}}, \ and\ \bibinfo {author} {\bibfnamefont {Adam}\ \bibnamefont {Ritz}},\ }\bibfield  {title} {\enquote {\bibinfo {title} {{Direct Detection of Multi-component Secluded WIMPs}},}\ }\href {\doibase 10.1103/PhysRevD.79.115019} {\bibfield  {journal} {\bibinfo  {journal} {Phys. Rev. D}\ }\textbf {\bibinfo {volume} {79}},\ \bibinfo {pages} {115019} (\bibinfo {year} {2009})},\ \Eprint {http://arxiv.org/abs/0903.3396} {arXiv:0903.3396 [hep-ph]} \BibitemShut {NoStop}%
\end{thebibliography}%

\end{document}